{}
{}

\documentclass{article}

\usepackage[utf8]{inputenc}
\usepackage{lmodern}
\usepackage[dvipsnames]{xcolor}
\usepackage{graphicx,caption}
\usepackage{caption}
\usepackage{subcaption}
\usepackage{cite}

\usepackage{comment}
\usepackage{microtype}

\usepackage{pgfplots}
\pgfplotsset{compat=1.18}

\usepackage{wrapfig}
\usepackage{capt-of}

\usepackage{geometry}
\usepackage{afterpage}
\usepackage{multirow}
\usepackage{tabularx}

\usepackage{bm}
\usepackage{mathtools,slashed}
\usepackage{amsmath, amssymb} 

\numberwithin{equation}{section}
\usepackage{mathrsfs}
\usepackage{physics}
\usepackage{bigints}

\usepackage{url}
\usepackage[bottom,hang,flushmargin]{footmisc}
\usepackage{fancyhdr}

\usepackage{subfiles}
\usepackage{enumitem}\setlist[enumerate]{font=\bfseries}

\usepackage{titlesec}
\usepackage[toc,page]{appendix}
\usepackage[breaklinks]{hyperref}
\usepackage{cleveref}

\titleformat*{\section}{\LARGE\bfseries}
\titleformat*{\subsection}{\Large\bfseries}
\titleformat*{\subsubsection}{\large\bfseries}

\newcommand{\td}[1]{\widetilde{#1}}
\newcommand{\ha}[1]{\widehat{#1}}

\newcommand{\dlangle}{\left\langle\!\left\langle}
\newcommand{\drangle}{\right\rangle\!\right\rangle}

\newcommand{\comma}{\text{,\space\space\space\space\space}}

\newcommand{\tn}[1]{\underline{\bm{#1}}}
\newcommand{\ol}[1]{\overline{#1}}

\newcommand{\ra}{\rightarrow}

\newcommand{\hg}{\mathfrak{g}}

\newcommand{\cn}{\text{cn}}
\newcommand{\dn}{\text{dn}}
\newcommand{\cd}{\text{cd}}
\newcommand{\sd}{\text{sd}}
\newcommand{\sn}{\text{sn}}

\newcommand{\ab}{\text{ab}}
\newcommand{\nc}{\text{nc}}
\newcommand{\dc}{\text{dc}}

\newcommand{\ad}{\text{ad}}
\newcommand{\SU}{\text{SU}}

\newcommand{\SLNC}{\text{SL}_{\C}(N)}
\newcommand{\slNC}{\mf{sl}_{\C}(N)}

\newcommand{\C}{\mathbb{C}}
\newcommand{\Z}{\mathbb{Z}}
\newcommand{\R}{\mathbb{R}}
\newcommand{\Pexp}{\text{P}\overleftarrow{\text{exp}}}

\newcommand{\mc}[1]{\mathcal{#1}}
\newcommand{\mf}[1]{\mathfrak{#1}}
\newcommand{\mb}[1]{\mathbb{#1}}
\newcommand{\ms}[1]{\mathscr{#1}}
\newcommand{\Ad}{\text{Ad}}
\newcommand{\tlim}{\text{tlim}}
\newcommand{\Lc}{\mathcal{L}}
\newcommand{\Rc}{\mathcal{R}}
\newcommand{\vp}{\varphi}
\newcommand{\p}{\partial}

\DeclareFontEncoding{LS1}{}{}
\DeclareFontSubstitution{LS1}{stix}{m}{n}
\DeclareSymbolFont{stixsymbols}{LS1}{stixscr}{m}{n}
\SetSymbolFont{stixsymbols}{bold}{LS1}{stixscr}{b}{n}
\DeclareMathSymbol{\kay}{\mathalpha}{stixsymbols}{"6B}

\makeatletter
\let\@keywords\@empty
\let\@subject\@empty
\providecommand{\keywords}[1]{\gdef\@keywords{#1}}
\providecommand{\subject}[1]{\gdef\@subject{#1}}
\def\thetitle{\@title}
\def\theauthor{\@author}
\def\thesubject{\@subject}
\def\thedate{\@date}
\def\thekeywords{\@keywords}
\makeatother
\AtBeginDocument{
\hypersetup{pdftitle={\thetitle}}
\hypersetup{pdfauthor={\theauthor}}
\hypersetup{pdfsubject={\thesubject}}
\hypersetup{pdfkeywords={\thekeywords}}}

\title{An Elliptic Integrable Deformation of the Principal Chiral Model}

\author{Sylvain Lacroix and Anders Wallberg}

\begin{document}
	
	\begin{titlepage}

		\begin{flushright}
			{\ttfamily CERN-TH-2023-205}
		\end{flushright}
		
		\begin{center}
			
			\vspace*{2cm}
			
			\begingroup\Large\bfseries
			An Elliptic Integrable Deformation\\ of the Principal Chiral Model
			\par\endgroup
			
			\vspace{1.cm}
			
			\begingroup
			Sylvain Lacroix$^{a,}$\footnote{E-mail:~sylvain.lacroix@eth-its.ethz.ch} and
			Anders Wallberg$^{b,c,d,}$\footnote{E-mail:~anders.heide.wallberg@cern.ch}
			\endgroup
			
			\vspace{1cm}
			
			\begingroup
			$^a$\it Institute for Theoretical Studies, ETH Z\"urich, \\ Clausiusstrasse 47, 8092 Z\"urich, Switzerland,\\~\\

			$^b$\it  Department of Theoretical Physics, CERN,\\
			
			1211 Meyrin, Switzerland\\~\\
			
			$^c$\it  Laboratory for Theoretical Fundamental Physics, EPFL, \\
			
			Rte de la Sorge, 1015 Lausanne, Switzerland\\~\\
			
			$^d$\it  Institut f\"ur Theoretische Physik, ETH Z\"urich,\\
			Wolfgang-Pauli-Strasse 27, 8093 Z\"urich, Switzerland\\
			\endgroup
			
		\end{center}
		
		\vspace{1cm}
		
		\begin{abstract}
			We introduce a new elliptic integrable $\sigma$-model in the form of a two-parameter deformation of the Principal Chiral Model on the group $\text{SL}_\R(N)$, generalising a construction of Cherednik for $N=2$ (up to reality conditions). We exhibit the Lax connection and $\Rc$-matrix of this theory, which depend meromorphically on a spectral parameter valued in the torus. Furthermore, we explain the origin of this model from an equivariant semi-holomorphic 4-dimensional Chern-Simons theory on the torus. This approach opens the way for the construction of a large class of elliptic integrable $\sigma$-models, with the deformed Principal Chiral Model as the simplest example.
			
		\end{abstract}
	\end{titlepage}
	
	\tableofcontents
	
	\thispagestyle{empty}
	\clearpage
	
	\setcounter{page}{1}
	\setcounter{footnote}{0} 
	
	\section{Introduction}\label{sec:Intro}
	
	Integrable field theories are highly symmetric models which admit an infinite number of integrals of motion, \textit{i.e.} quantities which are conserved under time evolution. The strong constraints on the dynamics of these theories which are imposed by the presence of this large amount of symmetry have allowed for the development of various methods to compute some of their physical observables exactly. Integrable field theories are thus valuable models if one wants to explore non-perturbative effects in physics. The study of these models is a well-established but still quite active domain of research, in particular in the context of two-dimensional field theories. In that case, integrability is generally defined, at the classical level, using the so-called Lax formalism.
	
	The key ingredient of this formalism is the Lax connection $\mathcal{L}(z)$, which is a Lie algebra-valued one-form on the space-time of the theory, built from the fields and depending meromorphically on an auxiliary complex variable $z\in\C$ called the spectral parameter. The main property required of this connection is that the equations of motion of the fields are equivalent to the flatness of the connection for all values of $z$. This property ensures that one can build an infinite tower of conserved charges from $\Lc(z)$, signifying the integrability of the theory. In addition to being conserved, we also want these charges to be pairwise Poisson-commuting in the Hamiltonian formulation: this imposes strong constraints on the Hamiltonian structure, in particular on the Poisson-bracket of (the spatial component of) $\mathcal{L}(z)$ with itself. One way to satisfy this constraint is if this bracket takes the form of a so-called \textit{non-ultralocal Maillet bracket}~\cite{maillet_kac-moody_1985,maillet_new_1986}. The latter depends on an \textit{$\Rc$-matrix} $\mathcal{R}(z_1,z_2)$ valued in the tensor product of two copies of the Lie algebra. $\mathcal{R}$ is a function of two spectral parameters $z_1,z_2\in\mathbb{C}$ and satisfies the \textit{classical Yang-Baxter equation (cYBE)}. For the class of integrable field theories that we will consider in this paper, the $\mathcal{R}$-matrix can, in turn, be expressed in terms of another \textit{seed $\Rc$-matrix} $\mathcal{R}^0(z_2-z_1)$ (which depends only on the difference of the spectral parameters) and the so-called \textit{twist function} $\varphi(z)$:
	\begin{equation}\label{eq:Rintro}
		\mathcal{R}(z_1,z_2)=\mathcal{R}^0(z_2-z_1)\,\varphi(z_2)^{-1}\,.
	\end{equation}
	Modulo some basic assumptions,  the seed $\Rc$-matrices $\mathcal{R}^0$ have been classified by Belavin and Drinfel'd in the seminal work~\cite{belavin_solutions_1982} and come in three classes: \textit{rational, trigonometric} and \textit{elliptic}. This paper will mostly be concerned with the third class, \textit{i.e.} with elliptic integrable field theories.\\
	
	We will focus our attention on a specific family of two-dimensional field theories, the \textit{$\sigma$-models}. These have been studied across many fields of physics and have applications in string theory, high-energy physics, holography, condensed matter systems and various other domains. Of particular importance in this context are the integrable $\sigma$-models, which although rare are quite valuable since they allow for the derivation of exact results, contributing greatly to the aforementioned applications. The search for integrable $\sigma$-models started in the seventies and is still an active domain of research, leading to the discovery of a large panorama of such models: for a recent review, we refer to the lecture notes~\cite{hoare_integrable_2022}. Most of the models in this panorama belong to the class of rational and trigonometric integrable field theories, according to the classification introduced above. Although there are a few known examples~\cite{cherednik_relativistically_1981,sfetsos_anisotropic_2015,costello_gauge_2019,bykov_sigma_2021,derryberry_lax_2021,bykov_cpn-1-model_2022}, the class of elliptic integrable $\sigma$-models has so far been explored less than the others and offers a natural subject for further developments. The goal of this paper is to report on some first progress in this direction: although the long-term aim is to develop methods for a systematic study of elliptic integrable $\sigma$-models (see later in the introduction as well as in the concluding section for more details on these approaches), we will focus here on the simplest example that falls within this scheme, which takes the form of an integrable elliptic deformation of the \textit{Principal Chiral Model (PCM)}.
	
	The undeformed PCM is the prototypical example of a rational integrable $\sigma$-model, which we now briefly review. The space-time $\Sigma$ of this theory is two-dimensional with coordinates $(t,x)$ (or equivalently, light-cone coordinates $x^\pm=t \pm x$). On this space-time lives a $G$-valued field $g(x^+,x^-)$, with $G$ a real, simple Lie group. The associated Lie algebra is denoted by $\hg$ and is equipped with an invariant non-degenerate bilinear form $\left\langle\cdot,\cdot\right\rangle:\hg\times\hg\ra \mathbb{R}$. The action of the theory is then given by
	\begin{equation}\label{eq:PCMIntro}
		S_{\text{PCM}}[g]  \equiv h\int_{\Sigma}\dd x^+ \, \dd x^-\,\left\langle j_+,j_-\right\rangle\,,
	\end{equation}
	where $j_\pm=g^{-1}\partial_\pm g$ is a current valued in $\hg$ and $h$ is a constant parameter. This action admits a global $G_L\times G_R$ symmetry which acts by mapping $g\mapsto u_Lgu_R$, with constant $u_{L/R}\in G$. Importantly, this model is integrable and many exact results have been derived concerning its classical solutions and its quantisation. To prove the integrability of the PCM, one has to find the associated Lax connection and $\mathcal{R}$-matrix. The Lax connection has been derived in~\cite{zakharov_relativistically_1978} and is given by
	\begin{equation*}
		\mathcal{L}_\pm(z)=\frac{j_\pm}{1\mp z}\,.
	\end{equation*}
	The spatial component of this Lax connection satisfies a Maillet bracket, as established in~\cite{maillet_hamiltonian_1986}, and the corresponding $\mathcal{R}$-matrix takes the form \eqref{eq:Rintro}, for a certain seed $\Rc^0(z)$ and an explicit twist function $\varphi(z)$. More precisely, the seed $\Rc$-matrix is the simplest solution of the cYBE in the Belavin-Drinfel'd classification, called the \textit{Yangian $\Rc$-matrix}. Both $\Rc^0(z)$ and $\vp(z)$ are rational functions of the spectral parameter, making the PCM part of the class of rational integrable field theories.\\
	
	Integrability is inherently quite sensitive and is thus generally broken when one deforms the theory. However, some integrable $\sigma$-models admit very specific deformations which preserve integrability. We refer to the review~\cite{hoare_integrable_2022} for an extensive discussion of this subject and its history. For the purpose of this introduction, we will restrict our attention to the integrable deformations of the PCM. The first known example was found by Cherednik in \cite{cherednik_relativistically_1981} and concerns only the case where the underlying Lie group $G$ is given by SU$(2)$: it corresponds to a deformation which breaks the $\SU (2)_R$ symmetry of the action \eqref{eq:PCMIntro} but preserves $\SU(2)_L$. Another integrable deformation of the SU$(2)$ PCM was found in the work~\cite{Fateev:1996ea} of Fateev and breaks both the left and right symmetries to their $\text{U}(1)$ Cartan subgroups. It was later extended by Lukyanov in~\cite{Lukyanov:2012zt} to include a Wess-Zumino term. For general Lie groups $G$, the construction of integrable deformations of the PCM started with the introduction of the so-called Yang-Baxter model~\cite{klimcik_yang-baxter_2002,klimcik_integrability_2009} by Klim\v{c}\'{i}k, which breaks the right symmetry to the Cartan subgroup of $G$, preserves the left one and coincides with a specific case of Cherednik's model when $G=\text{SU}(2)$. Similarly, the generalisations of the integrable deformations of Fateev and Lukyanov for higher rank compact groups were constructed in~\cite{klimcik_integrability_2009,klimcik_integrability_2014,Delduc:2017fib}. Finally, several of these theories are related through Poisson-Lie T-duality~\cite{Klimcik:1995ux,Klimcik:1995dy} to another type of integrable $\sigma$-models referred to as $\lambda$-deformations~\cite{Sfetsos:2013wia,sfetsos_generalised_2015,sfetsos_anisotropic_2015}. These various constructions result in a very rich family of integrable $\sigma$-models on deformed Lie groups.
 
 In the present paper, we will be interested in simple deformations of the PCM defined in terms of a constant linear \textit{deformation operator} $D:\hg\ra\hg$, with the action being given by 
	\begin{equation}\label{eq:DPCMaction}
		S_{\text{DPCM}}[g]\equiv \int_{\Sigma}\dd x^+ \, \dd x^-\,\left\langle j_+,D\left[j_-\right]\right\rangle\,.
	\end{equation}
	In this language, the undeformed PCM then corresponds to $D$ being proportional to the identity operator. This deformation preserves the left symmetry of the model but breaks the right one to the subgroup of $G$ which commutes with $D$. It is expected that a generic deformation of this type breaks integrability. However, it can be preserved for very specific choices of $D$: this includes the aforementioned Cherednik deformation~\cite{cherednik_relativistically_1981} for $G=\text{SU}(2)$ and Yang-Baxter deformations \cite{klimcik_yang-baxter_2002,klimcik_integrability_2009} for arbitrary Lie groups $G$. All of the latter belong to the class of rational or trigonometric integrable theories. Surprisingly, it was shown in the former work~\cite{cherednik_relativistically_1981} that for $\SU(2)$, every linear deformation $D$ leads to an integrable model: these can be rational, trigonometric or elliptic, with the most generic choice of operator producing elliptic integrable theories.  We will show in this paper that the Lax connection of the generic Cherednik deformation furthermore satisfies a Maillet bracket with an $\Rc$-matrix of the form \eqref{eq:Rintro}: more precisely, we will give an explicit expression of the twist function $\varphi(z)$ in terms of known elliptic functions and will show that the seed matrix $\mathcal{R}^0(z)$ is Sklyanin's elliptic $\Rc$-matrix on $\mathfrak{sl}_\C(2)$ \cite{sklyanin_complete_1979}.\\
	
	Given this, we wish to generalise Cherednik's result to find elliptic integrable deformations for groups of higher rank. It is already known how to generalise Sklyanin's $\mathcal{R}^0$-matrix to higher rank cases, specifically for the Lie algebra $\mathfrak{sl}_\mathbb{C}(N)$. These $\mathcal{R}^0$-matrices were constructed by Belavin \cite{belavin_discrete_1981} and depend on a specific basis $\{T_\alpha\}_{\alpha\in\mathbb{A}}$ of $\mathfrak{sl}_\mathbb{C}(N)$ (where $\mathbb{A}\equiv\mathbb{Z}_N\times\mathbb{Z}_N\setminus 0$), henceforth called the \textit{Belavin basis}, and a specific set of $N^2-1$ elliptic functions $\{r^\alpha(z)\}_{\alpha\in\mathbb{A}}$. To avoid getting into technicalities, we will not give the explicit definition of these objects here and refer to section \ref{sec:HigherRank} for details. For the present discussion, it suffices to know that the functions $\{r^\alpha(z)\}$ have a single pole at $z=0$ and are elliptic, in the sense that they are doubly periodic on the complex plane: the corresponding periods $\lambda_1$ and $\lambda_2$ are free parameters, which we take to be respectively in $\R$ and $i\R$ for future considerations of reality conditions. In these terms, Belavin's $\mathcal{R}^0$-matrix takes the form
	\begin{equation}\label{eq:EllipticRIntro}
		\mathcal{R}^0(z)=\sum_{\alpha\in\mathbb{A}} r^\alpha(z) \, T^\alpha\otimes T_\alpha\,,
	\end{equation}
	where $\{T^\alpha\}_{\alpha\in\mathbb{A}}$ is the dual basis of $\{T_\alpha\}_{\alpha\in\mathbb{A}}$. This matrix satisfies the cYBE for all values of $\lambda_1$ and $\lambda_2$, using various properties of $\{T_\alpha\}_{\alpha\in\mathbb{A}}$ and $\{r^\alpha(z)\}_{\alpha\in\mathbb{A}}$. In the limit where $\lambda_1\to \infty$ and $\lambda_2\to i\infty$, the functions $\{r^\alpha(z)\}$ lose their periodicity properties and all tend to the fraction $\frac 1z$: Belavin's $\Rc^0$-matrix then becomes the rational Yangian $\Rc^0$-matrix.
	
	The main result of this paper is the explicit construction of an elliptic integrable deformation of the PCM on $G=\text{SL}_{\mathbb{R}}(N)$ whose $\Rc$-matrix has Belavin's $\Rc^0$-matrix as seed. Remarkably, we will find that the Belavin basis and the functions $\{r^\alpha(z)\}$ are also the only necessary ingredients for us to be able to write down the action and Lax connection of this integrable deformation. For instance, the action takes the form \eqref{eq:DPCMaction}, with a specific choice of deformation operator $D$, acting diagonally on the Belavin basis. More precisely, it is given by
	\begin{equation}\label{eq:DIntro}
		S_{\text{DPCM}}[g]\equiv \int_{\Sigma}\dd x^+ \, \dd x^-\,\left\langle j_+,D\left[j_-\right]\right\rangle\,, \qquad \text{ with } \qquad
		D[T_\alpha]=-\rho\,\frac{r^{\alpha\prime}(1)}{r^\alpha(1)}\,T_\alpha\,.
	\end{equation}
	Here $\rho\in\mathbb{R}$ is a free parameter, which simply appears as a global factor in the action. In addition to $\rho$, the elliptic integrable model depends on the two periods $\lambda_1,\lambda_2$ of the functions $r^\alpha(z)$, which serve as deformation parameters: indeed, in the limit $\lambda_1\to \infty$ and $\lambda_2\to i\infty$, the above operator $D$ reduces to $\rho\,\mathbb{I}$, leading to the undeformed PCM. The deformed theory then depends on 3 continuous parameters: in the case $N=2$, this covers all deformation operators $D: \mathfrak{sl}_{\mathbb{R}}(2) \to \mathfrak{sl}_{\mathbb{R}}(2)$, up to adding total derivatives to the Lagrangian and redefining the field $g$. For $N>2$ however, not all operators $D$ can be obtained by this construction, showing that the integrable elliptic model corresponds to a very specific choice of deformation. We note that this deformed PCM is defined on the real form $\text{SL}_{\mathbb{R}}(N)$ of $\text{SL}_{\mathbb{C}}(N)$: in particular, this means that the target space of the $\sigma$-model is non-Euclidean and that its energy is not bounded below. It is thus natural to wonder whether one can define a similar model on other real forms of the same complex group, most importantly the unitary group SU$(N)$, which would lead to a compact Euclidean target space. To the best of our knowledge, it is not possible in these cases to find appropriate reality conditions on the parameters making the action real for $N>2$. In the case $N=2$, due to low-dimensional simplifications, the same choice of operator $D$ on $\mathfrak{sl}_{\C}(2)$ stabilises both the real forms $\mathfrak{su}(2)$ and $\mathfrak{sl}_{\R}(2)$, leading to the Cherednik model on SU(2) and a close cousin on SL$_{\mathbb{R}}(2)$.
	
	As claimed earlier, this deformed model is integrable. In particular, its equations of motion can be recast as the flatness of the Lax connection
	\begin{equation*}
		\mathcal{L}_\pm(z)=\sum_{\alpha\in\mathbb{A}}\frac{r^\alpha(z\pm1)}{r^\alpha(\pm1)}j^\alpha_{\pm}T_\alpha\,,
	\end{equation*}
	which has simple poles at $z=\pm 1$, as in the undeformed case. Finally, the spatial component of this connection satisfies a Maillet bracket with the Belavin $\mathcal{R}$-matrix \eqref{eq:EllipticRIntro} as seed and an explicit twist function, which we will not give here for simplicity (see section \ref{sec:HigherRank} for details).\\

	Finding the right choice of deformation operator and the right Lax connection for the elliptic deformation of the PCM would be an arduous task without a guiding principle. The key idea that we follow in this paper is to use a unifying approach to integrable field theories, which allows for a systematic construction of $\sigma$-models in a way that automatically ensures their integrability. Two such approaches are the \textit{4D Chern-Simons theory with disorder defects}~\cite{costello_gauge_2019} and the \textit{Affine Gaudin Models} formalism~\cite{levin_hitchin_2003,feigin_quantization_2009,vicedo_integrable_2019,delduc_assembling_2019}, which are deeply related~\cite{vicedo_4d_2021, levin_2d_2022} to one another. Both of them can be used to obtain the integrable elliptic deformation of the PCM: for conciseness, we will only present in this paper its derivation from the 4D Chern-Simons formalism. Although integrable field theories based on elliptic curves have already been considered in this approach~\cite{costello_gauge_2019, bykov_sigma_2021,derryberry_lax_2021,bykov_cpn-1-model_2022,levin_2d_2022}, the derivation of the deformed elliptic PCM requires a slight generalisation of the initial setup proposed in~\cite{costello_gauge_2019}, by adding a new ingredient called \textit{equivariance}.\footnote{This equivariant setup is similar to the one considered in the work~\cite{costello_gauge_2018,costello_gauge_2018-1} to obtain integrable lattice systems with a quantum elliptic Belavin $\Rc$-matrix from 4D Chern-Simons theory. The approach developed here can thus be seen as an extension of this idea to the case of 2D field theories from disorder defects instead of lattice systems.} The derivation of elliptic integrable $\sigma$-models from \textit{equivariant 4D Chern-Simons theory} and the example of the elliptic deformed PCM will be the main subject of the last part of this paper. The parallel construction from Affine Gaudin Models showcases the Hamiltonian structure underlying these theories, in particular their construction in terms of current algebras, and requires a generalisation of the works~\cite{feigin_quantization_2009,vicedo_integrable_2019,delduc_assembling_2019} to include elliptic models. This is a natural perspective of the present article (see the concluding section \ref{sec:conclusion} for more details) and will be the subject of a future paper~\cite{ToAppear:Gaudin}.
	We stress here that these unifying approaches are quite general: they can be used to construct a very large class of elliptic integrable $\sigma$-models, with the deformed PCM considered here being the simplest example. We will give a quick overview of the more general theories expected in this class in the conclusion.
	
	In this paper, we will focus on the construction and the study of the elliptic integrable deformed $\text{SL}_{\mathbb{R}}(N)$-PCM at the classical level. We refer to the concluding section \ref{sec:conclusion} for a brief discussion of some perspectives about its quantisation. In particular, the explicit proof of its renormalisability at 1-loop and the study of the corresponding RG-flow of its parameters will be the concern of another upcoming paper~\cite{ToAppear:RG}.\\
	
	The plan of the paper is as follows. In section \ref{sec:Basics}, we start by reviewing the basics of 2D integrable field theories, including the notions of Lax connection, $\mathcal{R}$-matrix and twist function. We then go through the undeformed PCM, focusing on its integrable structure. In section \ref{sec:su2}, we consider Cherednik's deformation of the $\text{SU}(2)$ PCM. We review its Lax connection, which is an elliptic function of the spectral parameter $z$, and then go on to discuss its non-ultralocal Maillet bracket. We then proceed with the higher rank case in section \ref{sec:HigherRank}. We first review the necessary information about elliptic functions and the algebra of the Belavin basis, before showcasing the new elliptic deformation. We discuss its integrable structure, examine its reality conditions and consider its so-called trigonometric limit, where one regains a $\sigma$-model belonging to the class of Yang-Baxter deformations. We end with section \ref{sec:4dcs}, where we present the construction of the model from equivariant 4D Chern-Simons theory. Finally, some generalities about elliptic functions and some technical computations are gathered in appendices \ref{sec:EllipticAppendix} and \ref{sec:YBAppendix}.

	\section[Integrable Field Theories and the Principal Chiral Model]{Integrable Field Theories and the PCM}\label{sec:Basics}
	
	\subsection{Integrable Field Theories}\label{subsec:int}
	
	Let us consider a classical field theory defined on a 2-dimensional space-time manifold $\Sigma$, with coordinates $(t,x)$. To fix the setup, we take $\Sigma$ to be a cylinder and ask that the fields of the theory be periodic under the identification $x\sim x+2\pi$. However, we note that all the results of the paper also apply to the case where $\Sigma$ is the plane and the fields decrease sufficiently fast as the spatial coordinate $x$ tends to infinity.
	
	\paragraph{Lax Connection and Conserved Charges:} We say that a theory admits a \textit{Lax connection} if there exists a pair of matrices $\mathcal{L}_t(z\,;t,x)$ and $\mathcal{L}_x(z\,;t,x)$, valued in a complex Lie algebra $\hg^\C$, depending meromorphically on an auxiliary complex parameter $z\in\mathbb{C}$ and built from the fundamental fields of the model in such a way that the equations of motion of these fields are equivalent to the zero curvature equation
	\begin{equation}\label{eq:ZCE}
		\partial_t \mathcal{L}_x(z\,;t,x) - \partial_x \mathcal{L}_t(z\,;t,x) + \bigl[  \mathcal{L}_t(z\,;t,x), \mathcal{L}_x(z\,;t,x) \bigr] = 0\,.
	\end{equation}
	Crucially, we ask that this equation holds for all values of the complex variable $z\in\mathbb{C}$, which we call the \textit{spectral parameter} of the theory.
	Given such a Lax connection, let us define the quantity
	\begin{equation}
		\mathcal{T}_n(z\,;t) \equiv \Tr\left[\Pexp \left( - \int_0^{2\pi} \Lc_x(z\,;t,x)\,\dd x \right)^n \,\right]\,.
	\end{equation}
	In this equation, we are considering the monodromy of the connection along the spatial circle, computed as a path-ordered exponential over $x\in[0,2\pi]$ and valued in a Lie group $G^\C$ with Lie algebra $\hg^\C$, and we are taking the trace of its $n$-th power in a given matrix representation of $G^\C$. The main property of this quantity is that, as a consequence of the zero curvature equation \eqref{eq:ZCE}, it is conserved along the time evolution, \textit{i.e.}
	\begin{equation}
		\partial_t \mathcal{T}_n(z\,;t) = 0\,.
	\end{equation}
	By construction, this equation holds for all values of the spectral parameter $z\in\C$: one can thus extract \textit{infinitely many conserved quantities} from $\mathcal{T}_n(z)$, for instance by considering its evaluation at different points in $\C$ or its power series expansion. In other words, $\mathcal{T}_n(z)$ serves as a generating function for an infinite number of conserved charges of the model, making it a key ingredient in its integrability. We note that these conserved charges are in general non-local combinations of the fields of the theory, due to the presence of a path-ordered exponential in the definition of $\mathcal{T}_n(z)$.
	
	\paragraph{Maillet Bracket and $\bm{\mathcal{R}}$-Matrix:} The hallmark of integrable field theories is the existence of an infinite number of conserved charges which Poisson-commute with one another in the Hamiltonian formulation of the theory. In the previous paragraph, we have seen how to construct generating functions $\mathcal{T}_n(z)$ of conserved charges when the theory admits a Lax connection: a natural question at this point is thus to find a condition under which these charges are pairwise Poisson-commuting. Since $\mathcal{T}_n(z)$ is built from the Lax matrix $\Lc_x(z)$ (the spatial component of the Lax connection), its Poisson algebra is directly related to the Poisson brackets obeyed by the different components of this matrix. Introducing a basis $\lbrace T_a \rbrace_{a=1}^{\text{dim}\,\hg}$ of $\hg$ and the corresponding decomposition $\Lc_x(z)=\Lc_x^a(z) \,T_a$ (where we sum over repeated indices), we encode all of these Poisson brackets in a unique object
	\begin{equation}
		\bigl\lbrace \Lc_x(z_1\,;t,x)_{\tn{1}}, \Lc_x(z_2\,;t,y)_{\tn{2}} \bigr\rbrace = \bigl\lbrace \Lc_x^a(z_1\,;t,x), \Lc_x^b(z_2\,;t,y) \bigr\rbrace \; T_a \otimes T_b\,,
	\end{equation}
	valued in the tensor product $\hg^\C \otimes \hg^\C$. In this equation, we consider components of the Lax matrix evaluated at different values $z_1,z_2\in\C$ of the spectral parameter and different values $x,y\in[0,2\pi]$ of the spatial coordinate, but at an equal time $t$, since we are working in the Hamiltonian formulation. Moreover, we use the standard tensorial notation, tracking the two tensor factors of $\hg^\C \otimes \hg^\C$ with indices $\tn{1}$ and $\tn{2}$. Similarly, if $X$ is a matrix in $\hg^\C$, we will denote by $X_{\tn{1}} = X \otimes \mathbb{I}$ and $X_{\tn{2}}=\mathbb{I} \otimes X$ its embedding in the first and second tensor factors respectively, where $\mathbb{I}$ represents the identity matrix.
	
	In the above notation, a sufficient condition for the pairwise Poisson commutation of the charges $\mathcal{T}_n(z)$ is the \textit{non-ultralocal Maillet bracket}~\cite{maillet_kac-moody_1985,maillet_new_1986}
	\begin{align}\label{eq:Maillet}
		\bigl\lbrace \Lc_x(z_1\,;t,x)_{\tn{1}}, \Lc_x(z_2\,;t,y)_{\tn{2}} \bigr\rbrace &=\Big\{ \bigl[ \Rc(z_1,z_2)_{\tn{12}}, \Lc_x(z_1\,;t,x)_{\tn{1}}]- \bigl[ \Rc(z_2,z_1)_{\tn{21}}, \Lc_x(z_2\,;t,y)_{\tn{2}}]\Big\} \,\delta(x-y)  \notag \\
		& \hspace{30pt} + \bigl( \Rc(z_1,z_2)_{\tn{12}} + \Rc(z_2,z_1)_{\tn{21}} \bigr) \,\delta'(x-y)\,.
	\end{align}
	Here, $\delta(x-y)$ is the spatial Dirac-distribution, $\delta'(x-y)$ is its derivative and
	\begin{equation}
		\Rc(z_1,z_2)_{\tn{12}} = r^{ab}(z_1,z_2)\, T_a \otimes T_b
	\end{equation}
	is a matrix in $\hg^\C \otimes \hg^\C$, depending meromorphically on the two spectral parameters $(z_1,z_2)\in\C^2$ and from which we build the ``transpose'' as $\Rc(z_2,z_1)_{\tn{21}} = r^{ab}(z_2,z_1)\, T_b \otimes T_a$. It is called the \textit{$\Rc$-matrix} of the theory and is a key characteristic of its integrable structure. In this paper, we will only encounter $\Rc$-matrices which are non-dynamical, in the sense that they do not depend on the fields of the model. 
	
	It is natural to wonder whether all matrices $\Rc(z_1,z_2)_{\tn{12}}$ in $\hg^\C \otimes \hg^\C$ can play the role of an $\Rc$-matrix. This is in fact far from being the case, as the Jacobi identity for the Maillet bracket \eqref{eq:Maillet} imposes strong constraints on $\Rc$. After a few manipulations, one finds that a sufficient purely algebraic condition ensuring this Jacobi identity is that the $\Rc$-matrix satisfies the \textit{classical Yang-Baxter equation (cYBE)}
	\begin{equation}\label{eq:CYBE}
		\bigl[  \Rc(z_1,z_2)_{\tn{12}}, \Rc(z_1,z_3)_{\tn{13}}] + \bigl[  \Rc(z_1,z_2)_{\tn{12}}, \Rc(z_2,z_3)_{\tn{23}}] + \bigl[  \Rc(z_3,z_2)_{\tn{32}}, \Rc(z_1,z_3)_{\tn{13}}] = 0 \,,
	\end{equation}
	which involves the evaluation of the $\Rc$-matrix at three different values $(z_1,z_2,z_3)\in\C^3$ of the spectral parameters and its various embeddings into the 3-fold tensor product $\hg^\C\otimes\hg^\C\otimes\hg^\C$.
	
	\paragraph{Belavin-Drinfel'd $\bm{\Rc^0}$-Matrices and Twist Functions:} In this paper, we will be particularly interested in a specific class of integrable field theories, whose $\Rc$-matrix takes the following form:
	\begin{equation}\label{eq:RTwist}
		\Rc(z_1,z_2)_{\tn{12}} = \Rc^0(z_2-z_1)_{\tn{12}}\,\vp(z_2)^{-1}\,.
	\end{equation}
	Here, $\Rc^0(z_2-z_1)_{\tn{12}}=-\Rc^0(z_1-z_2)_{\tn{21}}$ is a skew-symmetric \textit{seed $\Rc$-matrix} depending only on the difference of the spectral parameters and $\vp(z)$ is a meromorphic function which we will call the \textit{twist function}. A quick analysis of the cYBE \eqref{eq:CYBE} shows that the above matrix is a solution thereof if and only if $\Rc^0(z_2-z_1)_{\tn{12}}$ is also a solution, as the twist function only contributes to a global factor $\vp(z_2)^{-1}\vp(z_3)^{-1}$ on the left-hand side of \eqref{eq:CYBE}. We are thus left with finding skew-symmetric solutions $\Rc^0$ of the cYBE which depend only on the difference of the spectral parameters. Working with a simple Lie algebra $\hg^\C$ and modulo an additional non-degeneracy assumption, such $\Rc$-matrices have been classified in the seminal work~\cite{belavin_solutions_1982} of Belavin and Drinfel'd. They come in three classes, depending on the properties of their poles in the complex plane:
	\begin{itemize}
		\item \textit{the rational solutions}, whose set of poles $\Gamma_0 =\lbrace 0 \rbrace$ is composed of a single point at the origin ;
		\item \textit{the trigonometric solutions}, for which the poles $\Gamma_1$ form a 1-dimensional lattice in $\C$ ;
		\item \textit{the elliptic solutions}, for which the poles $\Gamma_2$ form a 2-dimensional lattice in $\C$.
	\end{itemize}
	This terminology is justified by the type of meromorphic functions appearing in the solutions $\Rc^0(z)$. For instance, in the first case, the entries of $\Rc^0(z)$ are rational fractions of $z$, which can equivalently be seen as meromorphic functions on the Riemann sphere $\mathbb{CP}^1= \C \sqcup \lbrace \infty \rbrace$. In the second case, $\Rc^0(z)$ is in fact periodic under shifts of $z$ by elements of a sublattice $\Lambda_1=\mathbb{Z}\lambda$ of the 1d-lattice of poles $\Gamma_1$, \textit{i.e.} we have $\Rc^0(z+\lambda)=\Rc^0(z)$: typical examples of such functions are the trigonometric ones $\cos\left(\frac{2\pi z}\lambda\right)$, $\sin\left(\frac{2\pi z}\lambda\right)$ and their relatives. One can see such trigonometric functions as defined on the quotient $\C/\Lambda_1$, which is isomorphic to a cylinder. Finally, in the third case, $\Rc^0(z)$ is doubly periodic under shifts by a sublattice $\Lambda_2=\mathbb{Z} \lambda_1 + \mathbb{Z} \lambda_2$ of the 2d-lattice of poles $\Gamma_2$: such functions are called elliptic, as they pass to the quotient $\C/\Lambda_2$, which is identified with an elliptic curve / torus. We note that these elliptic solutions exist only for the Lie algebra $\hg^\C = \mathfrak{sl}_\C(N)$.
	
	Following this terminology, we will distinguish in this paper between three different classes of integrable field theories with twist function: the rational, trigonometric and elliptic ones. In addition to requiring that the seed $\Rc$-matrix $\Rc^0(z)$ is in the corresponding family of Belavin-Drinfel'd solutions, we further ask in this classification that the Lax connection $\Lc_\mu(z)$ and the twist function $\vp(z)$ are also rational, trigonometric or elliptic, with the same periods as $\Rc^0(z)$.
	
	\subsection{The Principal Chiral Model}
	\label{subsec:PCM}
	We now review the Principal Chiral Model (PCM), which is the prototypical example of a rational integrable $\sigma$-model.
	
	\paragraph{Field and Action:} The PCM is formulated in terms of a $G$-valued field $g(t,x)$, where $G$ is a simple real Lie group. Introducing the light-cone coordinates $x^\pm = t \pm x$ and their derivatives $\p_\pm = \frac{1}{2}(\p_t \pm \p_x)$, we define the \textit{Maurer-Cartan currents} $j_\pm = g^{-1} \p_\pm g$, which are valued in the Lie algebra $\hg$ of $G$. Since $\hg$ is simple, it is equipped with a non-degenerate invariant \textit{bilinear form} $\langle\cdot,\cdot\rangle$, which is proportional to the Killing form. Here, we normalise it so that $\langle X,Y\rangle = -\Tr(XY)$, where the trace is taken in the fundamental representation of $\hg$: the minus sign has been introduced so that $\langle\cdot,\cdot\rangle$ is positive definite if $G$ is a compact group. In these terms, the action of the PCM is defined as the functional
	\begin{equation}\label{eq:PCM}
		S_{\text{PCM}}[g]  \equiv h\int_{\Sigma}\dd x^+ \, \dd x^-\,\left\langle j_+,j_-\right\rangle\,,
	\end{equation}
	where $h\in\R$ is a constant parameter.
	
	\paragraph{Equations of Motion and Lax Connection:} Varying the action \eqref{eq:PCM} with respect to $g$, we obtain the equations of motion of the PCM, which take the form of a local conservation law
	\begin{equation}
		\p_+ j_- + \p_- j_+ = 0
	\end{equation}
	for the current $j_\pm$. In addition, the latter also satisfies the \textit{Maurer-Cartan identity}
	\begin{equation}
		\p_+ j_- - \p_- j_+ + [j_+,j_-] = 0\,,\label{eq:jflat}
	\end{equation}
	which is true off-shell and is a direct consequence of the definition $j_\pm = g^{-1}\p_\pm g$. One easily checks that the two equations above for $j_\pm$ are equivalent to the flatness of the Lax connection defined by~\cite{zakharov_relativistically_1978}
	\begin{equation}\label{eq:PCMlax}
		\Lc_\pm(z) \equiv \frac{j_\pm}{1 \mp z}\,.
	\end{equation}
	As explained in the previous subsection, this is the first step towards proving the integrability of the PCM, as it allows the construction of an infinite number of conserved charges from the monodromy of the spatial component $\Lc_x(z)=\Lc_+(z) - \Lc_-(z)$. We note that $\Lc_\pm(z)$ is a rational function of the spectral parameter $z$, with poles at $z=\pm 1$.
	
	\paragraph{$\bm{\Rc}$-matrix and Twist Function:} Passing to the Hamiltonian formulation, one can compute the Poisson bracket of $\Lc_x(z)$ with itself. This was done in~\cite{maillet_hamiltonian_1986}, where it was proven that this bracket takes the Maillet non-ultralocal form \eqref{eq:Maillet}, thus ensuring the Poisson-commutation of the aforementioned conserved charges. More precisely, one finds that the $\Rc$-matrix appearing in this Maillet bracket takes the factorised form \eqref{eq:RTwist}, in terms of a seed $\Rc$-matrix and a twist function given by
	\begin{equation}\label{eq:RforPCM}
		\Rc^0(z)_{\tn{12}} = \frac{\mathcal{C}_{\tn{12}}}{z} \qquad \text{ and } \qquad \vp(z) = h\,\frac{1-z^2}{z^2}\,.
	\end{equation}
	Here, $\mathcal{C}$ is the so-called \textit{split quadratic Casimir} of $\hg$, which is defined as
	\begin{equation}
		\mathcal{C}_{\tn{12}} = T^a \otimes T_a \; \in \hg \otimes \hg\,,
	\end{equation}
	with $\lbrace T_a \rbrace_{a=1}^{\text{dim}\,\hg}$ a basis of $\hg$ and $\lbrace T^a \rbrace_{a=1}^{\text{dim}\,\hg}$ its dual with respect to $\langle\cdot,\cdot\rangle$. One checks that this object is in fact independent of the choice of basis and satisfies the \textit{Casimir identity}
	\begin{equation}
		\bigl[\mathcal{C}_{\tn{12}}, X_{\tn{1}} + X_{\tn{2}} \bigr] = 0 \,, \qquad \forall\,X\in\hg\,,
	\end{equation}
	as a consequence of the ad-invariance of $\langle\cdot,\cdot\rangle$. Using this identity and a few algebraic manipulations, one finds that the skew-symmetric matrix $\Rc^0(z_2-z_1)_{\tn{12}}$ satisfies the cYBE \eqref{eq:CYBE}. It is the simplest solution of the cYBE in the Belavin-Drinfel'd classification and is called the Yangian $\Rc$-matrix; it belongs to the class of rational solutions. This achieves the identification of the PCM as a rational integrable field theory with twist function.
	
	\section{Elliptic Deformed SU(2) Principal Chiral Model}\label{sec:su2}
	We will now focus our attention on deformed versions of the Principal Chiral Model which belong to the class of elliptic integrable systems. Before presenting new results for higher-rank groups, we will first review the $G=\text{SU}(2)$ case. That the SU(2) PCM has a deformation that admits a flat, elliptic Lax connection has been known for some time due to the work \cite{cherednik_relativistically_1981} of Cherednik (see also~\cite[Subsection 3.1]{hoare_integrable_2022} for a review). In this paper, we will show that it also has a Maillet bracket with twist function and this will serve as an inspiration for the higher rank cases constructed later. 
	
	\subsection{Outline of the SU(2) Deformed PCM}
	
	Recall the PCM action \eqref{eq:PCM} and specialise to the case where the field $g$ is valued in the group SU(2). In this section, we will study deformations of this model given by
	\begin{equation}
		S_{\text{DPCM}}[g]=\int_\Sigma \dd x^+ \, \dd x^-\,\left\langle j_+,D[j_-]\right\rangle\,,\label{eq:DPCM2.1}
	\end{equation}
	where $D$ is a constant linear operator on $\mathfrak{su}(2)$, which we call the \textit{deformation operator}. The undeformed PCM \eqref{eq:PCM} then corresponds to taking $D=h\,\mathbb{I}$, \textit{i.e.} proportional to the identity operator. We will refer to theories of the form \eqref{eq:DPCM2.1} as \textit{Deformed Principal Chiral Models (DPCM)}. To begin our analysis, we fix the following basis $\{T_a\}_{a=1}^3$ of $\mathfrak{su}(2)$:
	\begin{equation}
		T_a\equiv\frac{i}{\sqrt{2}}\sigma_a\,, \qquad \text{ satisfying } \qquad [T_a,T_b]=-\sqrt{2} \sum_{c=1}^3\epsilon_{abc}\,T_c\,,\qquad \left\langle T_a,T_b\right\rangle = \delta_{ab}\, .\label{eq:su2basis}
	\end{equation}
	Here, $\sigma_a$'s are the Pauli matrices and $\epsilon_{abc}$ is the skew-symmetric Levi-Civita tensor. Moreover, recall that $\langle\cdot,\cdot\rangle=-\Tr(\cdot)$ is the positive bilinear form on $\hg=\mathfrak{su}(2)$, as defined in the previous section: here, we chose the normalisation of $\{T_a\}_{a=1}^3$ such that it forms an orthonormal basis. Having fixed the basis, we will then express the deformation operator $D$ in terms of a matrix $D_{\phantom{b}a}^{b}$, such that $D[T_a]=\sum_b T_b\,D_{\phantom{b}a}^{b}$. One can show that for $G=\SU(2)$, a skew-symmetric term in $D_{\phantom{b}a}^{b}$ contributes with a total derivative in the action \eqref{eq:DPCM2.1} and we can thus focus on symmetric operators without loss of generality. Since any symmetric operator is diagonalisable in an orthonormal basis of $\mathfrak{su}(2)$ and every such basis is conjugate to $\{T_a\}_{a=1}^3$, we are free to consider only diagonal deformation operators $D[T_a]=D_aT_a$ (modulo a redefinition of $g$). This gives us three different classes of models: all the eigenvalues could be the same $D_1=D_2=D_3$; only two could match $D_1=D_2\not=D_3$; or they could all be different $D_1\not= D_2\not=D_3\not=D_1$. One might expect integrability to be broken away from the undeformed case $D_1=D_2=D_3$; however, \cite{cherednik_relativistically_1981} found that the DPCM is integrable for any choice of deformation matrix, with the three classes corresponding respectively to a rational, trigonometric and elliptic integrable model. 
	
	\subsection{Integrability of the Deformations}
	We will approach the three cases in a unified manner. For that, it will be useful to reparametrise $D_1, D_2$ and $D_3$ in terms of three variables $h,\nu$ and $m$. The elliptic case keeps all of these variables general; setting $m=1$, we obtain the trigonometric model; and taking the limit $\nu\ra 0$, we regain the rational, undeformed PCM. This is done by setting
	\begin{subequations}\label{eq:su2D}
		\begin{eqnarray}
			&\text{Rational: } \qquad D_1=D_2=D_3=h\,,\label{eq:RationalD}\\[5pt]
			&\text{Trigonometric: } \qquad
			D_1=D_2=h\,\comma D_3=h\,\sech^2(\nu)\,,\label{eq:TrigD}\\[5pt]
			& \text{Elliptic: } \qquad
			D_1=h\,\comma D_2=h\,\cd^2(\nu;m)\,\comma D_3=h\,\cn^2(\nu;m)\,.\label{eq:EllipticD}
		\end{eqnarray}
	\end{subequations}
	The elliptic parametrisation \eqref{eq:EllipticD} is given in terms of $\cn$ and $\cd$, which are part of the 12-member family of \textit{Jacobi elliptic functions}. These are doubly-periodic meromorphic functions on the complex plane, with periods $4K(m)$ and $4iK(1-m)$, where $K(m)$ is the \textit{complete elliptic integral of the first kind}. Alternatively, one may view them as being defined on the \textit{complex torus}
	\begin{equation}
		\mathbb{T}\equiv \mathbb{C}/\left\{4n_1K(m)+4in_2K(1-m),\, n_1,n_2\in\mathbb{Z}\right\}\,.
	\end{equation}
	In the limit $m\ra 1^-$, one finds $K(m)\ra +\infty$. Thus, the real periodicity is lost (alternatively, one can say that the torus decompactifies to a cylinder) and the Jacobi elliptic functions degenerate to trigonometric (or hyperbolic) functions. For instance, $\cd(z;1)=1$ and $\cn(z;1)=\sech(z)$, so that one regains the trigonometric parametrisation \eqref{eq:TrigD}. We note that $\sech^2(\nu) \leq 1$ for $\nu$ real: this thus corresponds to a regime where $D_1=D_2 \geq D_3$. One can parametrise the regime $D_1=D_2 \leq D_3$ by considering $\nu$ as purely imaginary. Alternatively, one may also consider the limit $m\to 0^+$ where instead the imaginary periodicity is lost and the functions take the form $\cd(\nu;0)=\cn(\nu;0)=\cos(\nu)$: for $\nu$ real, this describes a trigonometric regime with $D_2=D_3\leq D_1$. We refer to Appendix \ref{app:Jacobi} for a more thorough review of Jacobi elliptic functions.
	
	\paragraph{Lax Connection:}
	Recall that the Lax connection \eqref{eq:PCMlax} for the undeformed model was linear in the Maurer-Cartan currents $j_\pm$. This also holds for the deformed models on SU(2) but we lose the isotropy along the three directions $\{ T_a \}_{a=1}^3$. More precisely, introducing the decomposition $j_\pm = \sum_a j_\pm^a T_a$, the Lax connection for all three cases takes the form
	\begin{equation}
		\mathcal{L}_\pm(z)=\sum_{a=1}^3 \frac{s^a(\nu)}{s^a(\nu[1\mp z])}\,j^a_\pm T_a\,,
	\end{equation}
	where $s^a(z)$ is a meromorphic function of $z$ with a zero at $z=0$. Following \cite{cherednik_relativistically_1981} (see also~\cite[Subsection 3.1]{hoare_integrable_2022}), one can choose the functions $s^a(z)$ such that $\mathcal{L}_\pm(z)$ is on-shell flat. We will focus on the elliptic deformation, seeing the trigonometric one as the $m\ra 1$ limit. A solution is then given by
	\begin{equation}
		s^1(z)=\text{sc}(z;m)\,,\qquad s^2(z)=\sd( z;m)\,,\qquad s^3(z)=\sn(z;m)\,,
	\end{equation}
	where $\text{sc},\sd$ and $\sn$ are three additional Jacobi elliptic functions (see Appendix \ref{app:Jacobi} for a review of these functions and their properties). Taking the suitable limits $m\ra 1$ and $\nu\ra 0$, the Lax connections for the three classes then take the form
	\begin{subequations}
		\begin{eqnarray}
			\hspace{-15pt}	&\text{Rational: }\qquad\displaystyle{\mathcal{L}_\pm(z)=\frac{1}{1\mp z}\Big\{j^1_\pm T_1+ j^2_\pm T_2+ j^3_\pm T_3\Big\}}\,,\label{eq:RationalL}\\[5pt]
			\hspace{-15pt}	&\text{Trigonometric: }\qquad\displaystyle{\mathcal{L}_\pm(z)=\frac{\sinh(\nu)}{\sinh\left(\nu[1\mp z]\right)}\Big\{j^1_\pm T_1+j^2_\pm T_2\Big\}+\frac{\tanh(\nu)}{\tanh(\nu[1\mp z])}j^3_\pm T_3}\,,\label{eq:TrigL}\\[5pt]
			\hspace{-15pt}	&\text{Elliptic: }\qquad\displaystyle{\mathcal{L}_\pm(z)=\frac{\text{sc}(\nu;m)}{\text{sc}(\nu[1\mp z];m)}j^1_\pm T_1+\frac{\text{sd}(\nu;m)}{\sd(\nu[1\mp z];m)}j^2_\pm T_2
				+\frac{\text{sn}(\nu;m)}{\sn(\nu[1\mp z];m)}j^3_\pm T_3}\,.\label{eq:EllipticL}
		\end{eqnarray}
	\end{subequations}
	
	\paragraph{$\bm{\mathcal{R}}$-Matrix:}
	In addition to its flatness, we also want this elliptic Lax connection to satisfy a non-ultralocal Maillet bracket. We prove this by computing the Poisson bracket of $\mathcal{L}_x(z)=\mathcal{L}_+(z)-\mathcal{L}_-(z)$ with itself and finding that it takes the form of a Maillet bracket \eqref{eq:Maillet} with an $\mathcal{R}$-matrix that is again in factorised form \eqref{eq:RTwist}.\footnote{In the work~\cite{sfetsos_anisotropic_2015}, the Maillet bracket of the so-called anisotropic SU(2) $\lambda$-model has been computed. In a specific limit, this theory coincides with the non-abelian T-dual of the Cherednik model. Since non-abelian T-duality is a canonical transformation~\cite{lozano_non-abelian_1995, alvarez_target_1996, sfetsos_non--abelian_1996}, the Maillet bracket of~\cite{sfetsos_anisotropic_2015} in this limit should coincide with the one presented here. The explicit comparison of these results as well as the inclusion of the anisotropic $\lambda$-deformation in the formalism of the present paper would require further work. In particular, the Lax connection of~\cite{sfetsos_anisotropic_2015} involves square roots of rational functions of the spectral parameter: we expect that an appropriate change of variables makes this Lax connection explicitly elliptic, written in terms of Jacobi functions. A more thorough analysis of these aspects forms a natural perspective of the present work (see also the concluding section \ref{sec:conclusion}).} The seed matrix $\Rc^0(z)$ is the Sklyanin $\Rc$-matrix \cite{sklyanin_complete_1979}, which in our conventions reads
	\begin{equation}
		\mathcal{R}^0(z)=\sum_{a=1}^3\frac{\nu}{s^a(\nu z)}T_a\otimes T_a\,.
	\end{equation}
	Similarly, the twist function can be written uniformly as
	\begin{equation}
		\varphi(z)=\frac{D_a\nu }{s^a(\nu)}\left[\prod_{b=1}^3 \frac{s^b(\nu)}{s^a(\nu)}\right] \frac{s^a(\nu)^2-s^a(\nu z)^2}{s^a(\nu z)^2}\,,\label{eq:su2uniformtwist}
	\end{equation}
	which is equally true for $a=1,2$ or $3$. Once again, we regain the undeformed case \eqref{eq:RforPCM} in the rational limit $\nu\to 0$ of these expressions. In proving that the Poisson bracket of $\mathcal{L}_x$ with itself takes the form of a Maillet bracket (as well as proving that the above expression for the twist function is indeed the same for all choices of $a$), one needs to use the following \textit{generalized power rule} of the $s^a$ functions:
	\begin{equation}
		s^a(z)^2-s^a(w)^2=s^a(z-w)s^a(z+w)\left(1-C^a(m)s^a(z)^2s^a(w)^2\right)\,.
	\end{equation}
	Here, $C^a(m)$ are $z$-independent constants given by
	\begin{equation}
		C^1(m)=1-m\,,\qquad C^2(m)=-m(1-m)\,,\qquad C^3(m)=m\,.
	\end{equation}
	Writing out the $\mathcal{R}^0$-matrix for all three cases explicitly, we find the expressions:
	\begin{subequations}
		\begin{eqnarray}
			\hspace{-5pt}	&\text{Rational: }\qquad \displaystyle{\mathcal{R}^0(z)=\frac{1}{z}\Big\{T_1\otimes T_1+T_2\otimes T_2+T_3\otimes T_3\Big\}}\,,
			\\[6pt]&
			\hspace{-5pt}	\text{Trigonometric: }\qquad \displaystyle{\mathcal{R}^0(z)=\frac{\nu}{\sinh\left(\nu z\right)}\Big\{T_1\otimes T_1+T_2\otimes T_2\Big\}+\frac{\nu}{\tanh(\nu z)}T_3\otimes T_3}\,,
			\\[5pt]&
			\hspace{-5pt}	\text{Elliptic: }\qquad \displaystyle{\mathcal{R}^0(z)=\frac{\nu}{\text{sc}(\nu z;m)}T_1\otimes T_1+\frac{\nu}{\sd(\nu z;m)}T_2 \otimes T_2
				+\frac{\nu}{\sn(\nu z;m)}T_3\otimes T_3}\label{eq:Rsu2}\,.
		\end{eqnarray}
	\end{subequations}
	Similarly, fixing the choice $a=1$ in \eqref{eq:su2uniformtwist}, we find the following expressions for the twist functions:
	\begin{subequations}
		\begin{eqnarray}
			\hspace{-3pt}	&\text{Rational: }\qquad \displaystyle{\varphi(z)=h\frac{1-z^2}{z^2}}\,,\label{eq:RationalPhi}
			\\[8pt]&
			\hspace{-3pt}	\text{Trigonometric: }\qquad \displaystyle{\varphi(z)=\frac{h\nu}{\sinh(\nu)\cosh(\nu)}\frac{\sinh^2(\nu)-\sinh^2(\nu z)}{\sinh^2(\nu z)}}\,,\label{eq:TrigPhi}
			\\[5pt]&
			\hspace{-3pt}	\text{Elliptic: }\qquad  \displaystyle{\varphi(z)=\frac{h\nu}{\nc(\nu;m)\dc(\nu;m)\text{sc}(\nu;m)}\frac{\text{sc}^2(\nu;m)-\text{sc}^2(\nu z;m)}{\text{sc}^2(\nu z;m)}}\,.\label{eq:EllipticPhi}
		\end{eqnarray}
	\end{subequations}
	This proves that the SU(2) DPCM is integrable for arbitrary choices of linear deformation. This will serve as inspiration for the higher-rank case presented in the next section. In the trigonometric regime, these results match the ones initially found in~\cite{kawaguchi_hybrid_2012,kawaguchi_classical_2012}.
	
	\paragraph{Elliptic Deformations of the $\text{SL}_{\mathbb{R}}$(2) PCM:}
	We will briefly note that the above construction also allows for an elliptic deformation of the PCM on $G=\text{SL}_{\mathbb{R}}(2)$. This stems from the fact that the two groups have the same complexified Lie algebra $\mathfrak{sl}_{\mathbb{C}}(2)$ and will be explained in more detail in the next section. For this deformation, we keep the same values of the coefficients $D_a$ as in \eqref{eq:EllipticD}, but change the basis from \eqref{eq:su2basis} to
	\begin{equation}
		T_1=\frac{1}{\sqrt{2}}\sigma_1\,,\qquad T_2=\frac{-i}{\sqrt{2}}\sigma_2\,,\qquad T_3=\frac{1}{\sqrt{2}}\sigma_3\,,\label{eq:sl2basis}
	\end{equation}
	which generate $\text{SL}_{\mathbb{R}}(2)$.
	For $h,\nu\in\mathbb{R}$ and $m\in [0,1]$, the DPCM action \eqref{eq:DPCM2.1} is then real for $g \in \text{SL}_{\mathbb{R}}(2)$. It is furthermore elliptic integrable, with the Lax connection and seed $\mathcal{R}$-matrix taking the same form as \eqref{eq:EllipticL} and \eqref{eq:Rsu2}, respectively, with the $\text{SU}(2)$ generators \eqref{eq:su2basis} replaced with the $\text{SL}_\mathbb{R}(2)$ generators \eqref{eq:sl2basis} and with additional minus signs in the terms $T_1 \otimes T_1$ and $T_3 \otimes T_3$ of the $\Rc$-matrix \eqref{eq:Rsu2}. The twist function \eqref{eq:EllipticPhi} is the same in both cases.
	
	\section{Elliptic Deformed \texorpdfstring{$\bm{\text{SL}_\mathbb{C}(N)}$}{SLC(N)} Principal Chiral Model}\label{sec:HigherRank}
	
	We are now ready to consider elliptic integrable deformed PCMs for higher rank groups. This will be more complicated than the $\text{SU}(2)$ DPCM discussed in the previous section: in particular, in the higher-rank case, it will not be true that the elliptic integrable deformation corresponds to the most general linear deformation of the PCM action. A natural starting point is Belavin's famous elliptic $\mathcal{R}$-matrix on $\mathfrak{sl}_\mathbb{C}(N)$ \cite{belavin_discrete_1981}. We will first review the construction of this object, which requires us to discuss a certain family of elliptic functions and a certain basis for $\mathfrak{sl}_\mathbb{C}(N)$. We will then present a deformed PCM action which admits a flat Lax connection and a Maillet bracket with twist function, with Belavin's $\Rc$-matrix as seed. This is the main result of the paper. In the present section, we will summarise the main properties of this model without explaining its origin: we postpone this to the next section \ref{sec:4dcs}, where we will derive it from \textit{equivariant 4D Chern-Simons theory}.
	
	\subsection{Belavin's Elliptic \texorpdfstring{$\mathcal{R}$}{R}-Matrix}
	\label{subsec:Belavin}
	
	\paragraph{Elliptic Functions:} To describe Belavin's $\Rc$-matrix, we will have to first review some elementary properties of certain elliptic functions, which are doubly periodic functions on the complex plane: further technical details are relegated to appendix \ref{subsec:ralphafamily}. A typical example is the \textit{Weierstrass $\wp$-function}, which is commonly parametrised by its half-periods $\ell_1$ and $\ell_2$ (collected together in a 2-vector $\bm{\ell}=(\ell_1,\ell_2)$). It is defined as\footnote{The second term in the sum ensures its convergence.}
	\begin{equation}
		\wp(z;\ell_1,\ell_2)\equiv\frac{1}{z^2}+\sum_{\substack{\bm{n}\in \mathbb{Z}^2\\\bm{n}\neq(0,0)} }\left\{\frac{1}{\left(z-2\bm{n}\cdot\bm{\ell}\right)^2}-\frac{1}{(2\bm{n}\cdot\bm{\ell})^2} \right\}\,.\label{eq:defofwp}
	\end{equation}
	It is elliptic by construction, \textit{i.e.} it satisfies
	\begin{equation}
		\wp(z+2\ell_i;\ell_1,\ell_2)=\wp(z;\ell_1,\ell_2)\,.
	\end{equation}
	This means that the function $\wp(z;\ell_1,\ell_2)$ is invariant under shifts by elements of the 2d-lattice
	\begin{equation}
		\Gamma\equiv\left\{2\bm{n}\cdot\bm{\ell}\,,\, \bm{n}\in\mathbb{Z}\times\mathbb{Z}\right\}\,.\label{eq:SmallLattice}
	\end{equation}
	The Weierstrass $\wp$-function can thus equivalently be seen as a function on the quotient $\C/\Gamma$, which is an elliptic curve / complex torus. Schematically, it can be thought of as the elliptic version of $1/z^2$. More precisely, it is the unique meromorphic function on $\C/\Gamma$ with a double pole at $z=0$ (with coefficient $1$) and no other singularities. To ease the reading, we will typically not write the half-period dependency of the $\wp$-function explicitly from now on.
	
	We can use the Weierstrass $\wp$-function to further define the \textit{Weierstrass $\zeta$- and $\sigma$-functions}, which are uniquely characterised by
	\begin{equation}
		\frac{\dd \zeta(z)}{\dd z}=-\wp(z)\,,\qquad \frac{\dd\log(\sigma(z))}{\dd z}=\zeta(z)\,,
	\end{equation}
	together with
	\begin{equation}
		\zeta(z) = \frac{1}{z} + O(z)\,,\qquad  \sigma(z)= z + O(z^2)\,.
	\end{equation}
	These are not doubly-periodic: instead, they inherit pseudo-elliptic properties from $\wp$ given by
	\begin{equation}
		\zeta(z+2\ell_i)=\zeta(z)+2L_i\,,\qquad \sigma(z+2\ell_i)=-\exp\Big(2L_i\left[z+\ell_i\right]\Big)\sigma(z)\,,\label{eq:zetasigmatransform}
	\end{equation}
	where $\bm{L}=(L_1,L_2)\equiv(\zeta(\ell_1),\zeta(\ell_2))$.\\
	
	For any $\alpha\in \mathbb{C}^2$, we can further use $\bm{\ell}$ and $\bm{L}$, as well as the two-dimensional scalar-valued ``cross-product'' 
	\begin{equation}
		(a_1,a_2)\times (b_1,b_2) \equiv a_1b_2-b_1a_2\,,\label{eq:crossproduct}
	\end{equation}
	to define numbers $q_\alpha$ and $Q_\alpha$ as
	\begin{equation}
		q_\alpha\equiv \frac{2}{N}\alpha\times \bm{\ell}
		\,,\qquad Q_\alpha\equiv \frac{2}{N}\alpha\times \bm{L}\,.\label{eq:qalphadef}
	\end{equation}
	Here, $N$ is an integer that will later be related to the chosen Lie group $\text{SL}_\mathbb{C}(N)$. Using $q_\alpha$ and $Q_\alpha$, we can consider the following combination of $\sigma$-functions: 
	\begin{equation}
		r^\alpha(z)\equiv\exp(-Q_\alpha z)\frac{\sigma(z+q_\alpha)}{\sigma(z)\sigma(q_\alpha)}\,,\label{eq:defofralpha}
	\end{equation}
	which is well defined for $q_\alpha\not\in\Gamma$. Up to an exponential pre-factor, this is the so-called \textit{Kronecker function}.
	The ratio of $\sigma$-functions and the specific exponential in the function above is chosen such that we get two interesting properties\footnote{For the proof of these, see appendix \ref{subsec:ralphafamily}.}. 
	First, note that \textit{Fay's identity} \cite{fay_theta_1973} implies that the family of functions $\{r^\alpha\}$ satisfies the following relation
	\begin{equation}
		r^\alpha(z_1)r^\beta(z_2)=r^{\alpha+\beta}(z_1)r^\beta(z_2-z_1)+r^{\alpha}(z_1-z_2)r^{\alpha+\beta}(z_2)\,,\label{eq:rcompatible}
	\end{equation}
	whenever both sides are well-defined (\textit{i.e.} when $q_\alpha$, $q_\beta$ and $q_{\alpha+\beta}$ are not in $\Gamma$). Second of all, $r^\alpha$ is not periodic under shifts by $2\ell_i$, since it is built out of the non-elliptic $\sigma$-function. Instead, the property \eqref{eq:zetasigmatransform} of the latter implies that 
	\begin{equation}\label{eq:shiftr}
		r^\alpha(z+2\bm{n}\cdot\bm{\ell})=\exp(2\pi i\frac{\bm{n}\cdot \alpha}{N})r^\alpha(z)\,.
	\end{equation}
	
	So far, we have seen $\alpha=(\alpha_1,\alpha_2)$ as a pair of (almost) arbitrary complex numbers. Let us now restrict to $\alpha_1$ and $\alpha_2$ being real integers. Using \eqref{eq:zetasigmatransform}, one checks that the function $r^\alpha(z)$ is invariant under a shift of these integers by $N$, so that we can equivalently label $r^\alpha$ by the equivalence class of $(\alpha_1,\alpha_2)$ in the cyclic group $\mathbb{A}_0 \equiv \mathbb{Z}_N \times \mathbb{Z}_N$.\footnote{By a slight abuse of notation, we will still denote this equivalence class as $\alpha=(\alpha_1,\alpha_2)$.} Moreover, since $r^\alpha(z)$ is not well defined for $\alpha=(0,0)$, this naturally leads us to consider the family of $N^2-1$ functions $\{r^\alpha(z)\}_{\alpha\in\mathbb{A}}$, where $\mathbb{A}\equiv \mathbb{Z}_N\times\mathbb{Z}_N\setminus\{(0,0)\}$. For these functions, the quasi-periodicity property \eqref{eq:shiftr} implies
	\begin{equation}
		r^\alpha(z+2\bm{n}\cdot\bm{\ell})=\xi^{\bm{n}\cdot\alpha}\,r^\alpha(z)\,,\label{eq:equivralpha}
	\end{equation}
	where $\xi=\exp(2\pi i/N)$ is a $N^{\text{th}}$-root of unity. Thus, even though $\{r^\alpha(z)\}_{\alpha\in\mathbb{A}}$ are not periodic under shifts of $z$ by vectors in $\Gamma$, they are periodic under shifts by the sub-lattice
	\begin{equation}
		\Lambda\equiv\left\{ 2N\bm{n}\cdot\bm{\ell}\,,\,\bm{n}\in\mathbb{Z}\times\mathbb{Z}\right\}\,.\label{eq:LargeLattice}
	\end{equation}
	Therefore, $\{r^\alpha\}_{\alpha\in\mathbb{A}}$ defines a family of elliptic functions on the torus $\mathbb{T}=\C/\Lambda$. The property \eqref{eq:rcompatible} with $\alpha,\beta,\alpha+\beta\in\mathbb{A}$ will be referred to as \textit{$\mathbb{A}$-compatibility} of the family $\{r^\alpha\}_{\alpha\in\mathbb{A}}$.
	
	\paragraph{The Belavin Basis:}
	The above property \eqref{eq:rcompatible} is important for the construction of Belavin's elliptic $\mathcal{R}$-matrix. Another key ingredient is a particular basis for $\mathfrak{sl}_\mathbb{C}(N)$, which we will call the \textit{Belavin basis}. First consider the following two $N\times N$ matrices $\Xi_1$ and $\Xi_2$:
	\begin{equation}
		\Xi_1\equiv\begin{pmatrix}
			0&1&&&\\
			0&0&\ddots&&\\
			&\ddots&\ddots&\ddots&\\
			&&\ddots&0&1\\
			1&&&0&0
		\end{pmatrix},\qquad
		\Xi_2\equiv\begin{pmatrix}
			1&0&&&\\
			0&\xi&\ddots&&\\
			&\ddots&\ddots&\ddots&\\
			\phantom{\xi^{N-2}}&\phantom{\xi^{N-2}}&\ddots&\xi^{N-2}&0\\
			&&&0&\xi^{N-1}
		\end{pmatrix},\label{eq:defofXi}
	\end{equation}
	where again $\xi=\exp(2\pi i/N)$. These matrices satisfy the algebraic relations
	\begin{equation}\label{eq:Xialgebra}
		\Xi_1^N = \Xi_2^N = \mathbb{I} \qquad \text{ and } \qquad \Xi_1 \cdot \Xi_2 = \xi\,\Xi_2 \cdot \Xi_1\,.
	\end{equation}
	In particular, both $\Xi_i$'s are cyclic of order $N$. This implies that for  $\alpha=(\alpha_1,\alpha_2)\in\mathbb{A}_0=\mathbb{Z}_N\times\mathbb{Z}_N$, the following matrices are well defined\footnote{The mixing of the $\alpha_i$ label and the $\Xi_i$ label as well as the minus sign will come in handy later.}:
	\begin{equation}
		T_\alpha\equiv\frac{1}{\sqrt{N}}\Xi_1^{-\alpha_2}\Xi_2^{\alpha_1}\,.\label{eq:defofT}
	\end{equation}
	The Belavin basis \cite{belavin_discrete_1981} is composed by the $N^2-1$ matrices $\{T_\alpha\}_{\alpha\in\mathbb{A}}$ with $\mathbb{A}=\mathbb{A}_0\setminus \{(0,0)\}$, which span $\mathfrak{sl}_\mathbb{C}(N)$.
	The commutation relations in this basis are quite simple, being given by
	\begin{equation}
		\left[T_\alpha,T_\beta\right]=f_{\alpha\beta}\,T_{\alpha+\beta} \qquad \text{ with } \qquad f_{\alpha\beta}=\frac{1}{\sqrt{N}}\left(\xi^{\alpha_1\beta_2}-\xi^{\beta_1\alpha_2}\right)\,.\label{eq:Tcommutator}
	\end{equation}
	Furthermore, recall the invariant bilinear form $\left\langle\cdot,\cdot\right\rangle=-\Tr(\cdot)$ defined in section \ref{subsec:PCM}. We will use $\{T^\alpha\}_{\alpha\in\mathbb{A}}$ with upper indices to denote the dual of $\{T_\alpha\}_{\alpha\in\mathbb{A}}$ with respect to this form. Explicitly, one then has
	\begin{equation}
		\left\langle T_\alpha,T_\beta\right\rangle =-\delta_{\alpha+\beta,0}\,\xi^{-\beta_1\beta_2} \qquad \text{ and } \qquad T^\alpha=-\xi^{\alpha_1\alpha_2}T_{-\alpha}\,. \label{eq:slcnkappa}
	\end{equation}
	
	\paragraph{Belavin's $\bm{\mathcal{R}}$-Matrix:} In \cite{belavin_discrete_1981}, Belavin combined the family of functions $\{r^\alpha\}_{\alpha\in\mathbb{A}}$ with the basis $\{T_\alpha\}_{\alpha\in\mathbb{A}}$ to define the skew-symmetric $\Rc$-matrix
	\begin{equation}
		\mathcal{R}^0(z)=\sum_{\alpha\in\mathbb{A}}r^\alpha(z)\,T^\alpha\otimes T_\alpha\,.\label{eq:HighRankR}
	\end{equation}
	Combining the commutation relations \eqref{eq:Tcommutator} of the Belavin basis $\{T_\alpha\}$ together with the $\mathbb{A}$- compatibility \eqref{eq:rcompatible} of the elliptic functions $\{r^\alpha\}$, one finds that $\Rc^0(z_2-z_1)$ satisfies the cYBE \eqref{eq:CYBE}. In the Belavin-Drinfel'd classification discussed in section \ref{subsec:int}, this $\mathcal{R}$-matrix belongs to the elliptic family of solutions of the cYBE. In agreement with the general analysis behind this classification, $\Rc^0(z)$ has poles along a 2d-lattice, which here is given by $\Gamma=2\ell_1\mathbb{Z}+2\ell_2\mathbb{Z}$, and is periodic under shifts by the sublattice $\Lambda=2N\ell_1\mathbb{Z}+2N\ell_2\mathbb{Z}$ of $\Gamma$.
	
	In the limit where $\ell_1$ and $\ell_2$ become infinite, the elliptic functions $r^\alpha(z)$ all degenerate to the rational fraction $\frac{1}{z}$ and $\Rc^0(z)$ thus becomes the Yangian $\Rc$-matrix \eqref{eq:RforPCM}. Furthermore, if one sets $N=2$, one finds that \eqref{eq:HighRankR} matches the Sklyanin $\Rc$-matrix \eqref{eq:Rsu2} exactly. To do this, one has to convert from the Weierstrass-family of elliptic functions to the Jacobi-family: a sketch of this computation can be found in appendix \ref{subsec:Nto2}.
	
	\subsection{Higher Rank Elliptic DPCM}\label{sec:DPCM-HR}
	It is natural to guess that the Belavin $\mathcal{R}$-matrix is related to some elliptic integrable $\sigma$-model based on the Lie algebra $\mathfrak{sl}_\C(N)$. This is indeed true: one way to construct such a model is the equivariant 4D Chern-Simons theory, as explained in section \ref{sec:4dcs}. In the present subsection, we will outline the final results of this construction. 
	
	\paragraph{Action:} One can naturally expect the integrable model we are searching for to be a deformation of the PCM on a Lie group $G$ whose complexified Lie algebra is given by $\mathfrak{sl}_\mathbb{C}(N)$. Here, we will initially consider a field $g(t,x)$ valued in the complex Lie group $\text{SL}_\C(N)$ and work directly with a complexified theory. We postpone the discussion of reality conditions to the next subsection, in which we will find restrictions on the field $g$ and the parameters of the theory which ensure that its action is real. As in the $\text{SU}(2)$ case, we consider a deformation of the PCM action by introducing a constant, linear deformation operator $D : \mathfrak{sl}_\C(N) \to \mathfrak{sl}_\C(N)$:
	\begin{equation}
		S_{\text{DPCM}}[g]=\int_\Sigma \dd x^+ \, \dd x^-\,\left\langle j_+,D[j_-] \right\rangle\,.\label{eq:DPCMActionHighRank}
	\end{equation}
	As will be explained in section \ref{sec:4dcs}, the equivariant 4D Chern-Simons approach allows to find a specific choice of $D$ which preserves the integrability of the model. This operator is diagonal in the Belavin basis, \textit{i.e.} $D[T_\alpha]=D_\alpha T_\alpha$. The corresponding coefficients $\{D_\alpha\}_{\alpha\in\mathbb{A}}$ can be naturally expressed in terms of 4 variables: the two half-periods $\ell_1$ and $\ell_2$ appearing in \eqref{eq:defofwp}, alongside a point $\ha{z}\in\mathbb{C}$ and an overall scale $\rho$, which plays a similar role as the constant $h$ in the undeformed PCM \eqref{eq:PCM}. In terms of these four parameters, the $\{D_\alpha\}_{\alpha\in\mathbb{A}}$ take the form
	\begin{equation}
		D_\alpha=-\rho\frac{r^{\alpha\prime}(\ha{z})}{r^\alpha(\ha{z})}=\rho\,\Big\{Q_\alpha + \zeta(\ha{z};\ell_1,\ell_2)-\zeta(\ha{z}+q_\alpha;\ell_1,\ell_2)\Big\}\,,\label{eq:EllipticDeformationsHighRank}
	\end{equation}
	where we recall that the numbers $q_\alpha$ and $Q_\alpha$ are defined as in \eqref{eq:qalphadef}. The 4 variables $(\ell_1,\ell_2,\rho,\ha{z})$ will appear as a natural way of parametrising the twist function $\vp(z)$ of the theory. One should expect the model to be invariant under a rescaling of the spectral parameter: in terms of the aforementioned variables, this amounts to an overall dilation  $(\ell_1,\ell_2,\rho,\ha{z}) \mapsto (\lambda\ell_1,\lambda\ell_2,\lambda \rho,\lambda\ha{z})$. Accordingly, one checks from the scaling properties of the Weierstrass $\zeta$-function that the coefficients $D_\alpha$ in \eqref{eq:EllipticDeformationsHighRank} are indeed invariant under such a transformation. In the end, the model thus depends only on 3 physical parameters: an overall factor in the action and 2 deformation parameters. This is not enough to furnish the entire space of linear deformations, unlike the case of the SU(2) DPCM. This is in agreement with the expectation that the DPCM is not integrable for arbitrary choices of linear deformation operator $D$ for groups of higher rank.
	
	One natural way to fix the scaling freedom of the parameters is to set $\ha{z}=1$, keeping $(\ell_1,\ell_2,\rho)$ as free variables: this is the choice made in the introduction, yielding equation \eqref{eq:DIntro}. We will call this the \textit{fixed-zero parametrisation}, as $\ha{z}$ will be later identified as the zero of the twist function. In this parametrisation, $\ell_1$ and $\ell_2$ are naturally interpreted as deformation parameters: indeed, in the limit where they become infinite, one recovers the undeformed operator $D=\rho\,\mathbb{I}$. Another natural choice, which is commonly done in the literature on elliptic functions, is to fix the period $2\ell_1$ to be $1$ and parametrize the second one as $2\ell_2=\tau$: the theory is then described by the 3 variables $(\tau,\rho,\ha{z})$, where $\tau$ is interpreted as the moduli of the underlying complex torus. We will call this the \textit{modular parametrisation}. In this choice, the discussion of the undeformed limit is less transparent.
	
	Upon setting $N=2$, this model coincides with (the complexification of) the Cherednik model discussed in section \ref{sec:su2}. This identification requires finding the right map between the parameters $(\ell_1,\ell_2,\rho)$ used here and the ones $(h,\nu,m)$ used in the SU(2) case, as well as relating the Weierstrass elliptic functions to the Jacobi ones. The details of this identification can be found in Appendix \ref{subsec:Nto2}.
	
	\paragraph{Isometries:} Let us quickly comment on the global symmetries of the model \eqref{eq:DPCMActionHighRank}. The undeformed theory \eqref{eq:PCM} with $D=h\,\mathbb{I}$ has $\text{SL}_{\mathbb{C}}(N)_L \times \text{SL}_{\mathbb{C}}(N)_R$ isometries, acting on the field $g$ by left and right multiplications $g \mapsto u_L g u_R$ (where $u_{L,R}$ are constant matrices in $\text{SL}_{\mathbb{C}}(N)$). The introduction of a non-trivial deformation operator $D$ preserves the left isometry $\text{SL}_{\mathbb{C}}(N)_L$, while breaking the right one to $H_R$, where $H$ is the subgroup of $\text{SL}_{\mathbb{C}}(N)$ whose adjoint action commutes with $D$. For the elliptic operator \eqref{eq:EllipticDeformationsHighRank} and generic values of the parameters, we expect this subgroup to be trivial, so that the right isometry is completely broken. We refer to the concluding section \ref{sec:conclusion} for further discussions and perspectives on the symmetries of the elliptic integrable DPCM.
	
	\paragraph{Lax Connection:}
	To show that this model is integrable, we have to exhibit a Lax connection whose flatness encodes the equations of motion of the theory. The analysis in section \ref{sec:4dcs} shows that, upon decomposing the currents along the Belavin basis as $j_\pm= \sum_\alpha j^{\alpha}_\pm T_\alpha$, the Lax connection is linear in $j^\alpha_\pm$ and is given by
	\begin{equation}
		\mathcal{L}_\pm(z)=\sum_{\alpha\in\mathbb{A}}\frac{r^\alpha(z\mp\ha{z})}{r^\alpha(\mp \ha{z})}\,j^\alpha_\pm \,T_\alpha\,.\label{eq:HighRankLax}
	\end{equation}
	This connection is doubly-periodic with respect to shifts of $z$ by the lattice vectors in $\Lambda$ \eqref{eq:LargeLattice}, \textit{i.e.} 
	\begin{equation}
		\mathcal{L}_\pm(z+2N\bm{n}\cdot\bm{\ell })=\mathcal{L}_\pm(z)\,.
	\end{equation}
	It is thus elliptic and defines a meromorphic function on the torus $\mathbb{T}=\C/\Lambda$. Furthermore, the component $\mathcal{L}_\pm(z)$ has a simple pole at $\pm \ha{z}$ (and its translates by the lattice $\Gamma=2\ell_1\mathbb{Z}+2\ell_2\mathbb{Z}$). Using the scaling freedom to fix $\ha{z}=1$, this matches the pole of the undeformed Lax connection \eqref{eq:PCMlax}: $\Lc_\pm(z)$ then defines a natural elliptic generalisation of this connection.
	
	\paragraph{$\bm{\mathcal{R}}$-Matrix and Twist Function:}
	As expected, the spatial component of this Lax connection satisfies a Maillet bracket with twist function. The seed $\mathcal{R}^0$-matrix is of course the Belavin one \eqref{eq:HighRankR} while the twist function reads
	\begin{equation}
		\varphi(z)=\rho\,\big\{\wp(z)-\wp(\ha{z})\big\}\,.\label{eq:higherrnaktwist}
	\end{equation}
	Similarly to the Lax connection and $\mathcal{R}^0$-matrix, $\vp(z)$ is doubly periodic under shifts by lattice vectors in $\Lambda$. However, it is also invariant under shifts by the smaller lattice vectors in $\Gamma$:
	\begin{equation}
		\varphi(z+2\bm{n}\cdot\bm{\ell})=\varphi(z)\,.
	\end{equation}
	The origin of this will be explained in the next paragraph. 
	
	The twist function \eqref{eq:higherrnaktwist} has a double pole at $z=0$ (and its translates under shifts by the lattice $\Gamma$), with expansion $\frac{\rho}{z^2} + O(z^0)$. Moreover, it has simple zeroes at $z=\pm \ha{z}$ (and their translates), which coincide with the poles of the Lax connection $\Lc_\pm(z)$. In fact, $\vp(z)$ can be characterised as the unique meromorphic elliptic function with periods $2\ell_i$ and this structure of poles and zeroes. This is what makes $(\ell_1,\ell_2,\rho,\ha{z})$ the natural parameters of the model.\footnote{One could also consider an extra parameter $\hat{p}$ specifying the location of the pole of $\varphi$. However, this also introduces an extra redundancy in the system given by the shift $(\hat{z},\hat{p})\mapsto (\hat{z}+z_0,\hat{p}+z_0)$. We can thus use this translation freedom to fix $\hat{p}=0$.} In the fixed-zero parameterisation $\ha{z}=1$, $\vp(z)$ is the unique elliptic function with a double pole at $z=0$ and simple zeroes at $z=\pm 1$: it is thus the most natural elliptic generalisation of the undeformed twist function \eqref{eq:RforPCM}.
	
	Finally, we note that the twist function \eqref{eq:higherrnaktwist} exactly matches the one \eqref{eq:EllipticPhi} given for the SU(2) elliptically deformed PCM, after an appropriate change of parameters and reformulation in terms of Jacobi elliptic functions (see Appendix \ref{subsec:Nto2} for details).
	
	\paragraph{Equivariance:} We will now describe a particular property of the Lax connection and $\mathcal{R}$-matrix: they are \textit{equivariant}. This plays a crucial role in the construction of the $\sigma$-model from 4D Chern-Simons theory in section \ref{sec:4dcs}. It stems from the fact that the group $\mathbb{A}_0=\mathbb{Z}_N\times\mathbb{Z}_N$ has a natural action on both the torus $\mathbb{C}/\Lambda$ and the Lie-algebra $\mathfrak{sl}_\mathbb{C}(N)$. 
	
	On the side of the torus $\C/\Lambda$, ${\alpha}=(\alpha_1,\alpha_2)\in\mathbb{A}_0$ act as translations by the vectors of the finer lattice $\Gamma$, according to
	\begin{equation}
		z\longmapsto z+2{\alpha}\cdot\bm{\ell}\,.\label{eq:equivactiononz}
	\end{equation}
	This defines an action of $\mathbb{A}_0=\mathbb{Z}_N\times\mathbb{Z}_N$, since performing $N$ successive translations by $2\ell_i$ produces a shift by $2N\ell_i\in\Lambda$, which thus acts as the identity on the quotient $\C/\Lambda$. In particular, the group $\mathbb{A}_0$ can then be identified with the lattice quotient $\Gamma/\Lambda$.
	
	On the Lie algebra side, ${\alpha}=(\alpha_1,\alpha_2)\in\mathbb{A}_0$ acts as the automorphism
	\begin{equation}
		\Ad_{\Xi_1}^{\alpha_1} \Ad_{\Xi_2}^{\alpha_2} \; : \; X \longmapsto \Xi_1^{\alpha_1}\Xi_2^{\alpha_2} X\, \Xi_2^{-\alpha_2}\Xi_1^{-\alpha_1}\,.
	\end{equation}
	Due to \eqref{eq:Xialgebra}, it is clear that $\Ad_{\Xi_1}$ and $\Ad_{\Xi_2}$ are two commuting automorphisms of order $N$ and thus define an action of $\mathbb{A}_0= \mathbb{Z}_N \times \mathbb{Z}_N$. The Lax connection $\mathcal{L}_\mu(z)$ of the model is equivariant as a function on $\mathbb{C}/\Lambda$ valued in $\mathfrak{sl}_\mathbb{C}(N)$, in the sense that the translational action of ${\alpha}=(\alpha_1,\alpha_2)\in\mathbb{A}_0$ on its argument $z$ amounts to the corresponding adjoint action on $\mathfrak{sl}_\mathbb{C}(N)$. Explicitly, this means that
	\begin{equation}\label{eq:equivL}
		\mathcal{L}_\mu(z+2{\alpha}\cdot\bm{\ell}) = \Ad_{\Xi_1}^{\alpha_1} \Ad_{\Xi_2}^{\alpha_2} \mathcal{L}_\mu(z)\,.
	\end{equation}
	The action of $\mathbb{A}_0$ on $\mathfrak{sl}_{\mathbb{C}}(N)$ defines a grading
	\begin{equation}
		\mathfrak{sl}_\mathbb{C}(N)=\bigoplus_{\alpha\in\mathbb{A}_0}\hg^{(\alpha)} \,,\qquad \bigl[\hg^{(\alpha)},\hg^{(\beta)}\bigr]\subset \hg^{(\alpha+\beta)}\,,
	\end{equation}
	where
	\begin{equation}\label{eq:Grading}
		\hg^{(\alpha)} = \bigl\lbrace X\in \mathfrak{sl}_{\mathbb{C}}(N)\,|\, \Ad_{\Xi_1}^{\beta_1} \Ad_{\Xi_2}^{\beta_2} X = \xi^{{\alpha}\cdot{\beta}} X \bigr\rbrace\,.
	\end{equation}
	From \eqref{eq:Xialgebra}, one easily checks that $\hg^{(\alpha)}=\mathbb{C}T_\alpha$ if $\alpha\neq (0,0)$ and $\hg^{(0,0)}=\lbrace 0 \rbrace$. The grading property $\left[\hg^{(\alpha)},\hg^{(\beta)}\right]\subset \hg^{(\alpha+\beta)}$ is then consistent with the commutation relation \eqref{eq:Tcommutator} of the Belavin basis. Introducing the decomposition $\mathcal{L}_\mu(z)=\sum_\beta\mathcal{L}_\mu^\beta(z)T_\beta$ along this basis, the equivariance property \eqref{eq:equivL} becomes
	\begin{equation}
		\mathcal{L}_\mu^\beta(z+2{\alpha}\cdot\bm{\ell}) = \xi^{{\alpha}\cdot{\beta}} \mathcal{L}_\mu^\beta(z)\,.
	\end{equation}
	That this holds is a simple consequence of the property \eqref{eq:equivralpha} of the functions $\lbrace r^\alpha \rbrace_{\alpha\in\mathbb{A}}$\footnote{Note that $r^\alpha$ is only well defined for $\alpha\in\mathbb{A}$ and we cannot allow for $\alpha=(0,0)$. This is not a problem, since $\hg^{(0,0)}$ is trivial.}. Let us finally mention that it was noted already in \cite{belavin_discrete_1981} that the  Belavin $\mathcal{R}$-matrix \eqref{eq:HighRankR} satisfies a similar equivariance property, namely
	\begin{equation}
		\mathcal{R}^0(z+2\bm{\alpha}\cdot\bm{\ell}) = \Bigl( \mathbb{I}  \otimes \Ad_{\Xi_1}^{\alpha_1} \Ad_{\Xi_2}^{\alpha_2} \Bigl) \mathcal{R}^0(z) = \Bigl( \Ad_{\Xi_1}^{-\alpha_1} \Ad_{\Xi_2}^{-\alpha_2} \otimes \mathbb{I} \Bigl) \mathcal{R}^0(z) \,.\label{eq:R0equiv}
	\end{equation}
	Meanwhile, as observed in the previous paragraph, the twist function is left invariant under the action of $\mathbb{A}_0$:
	\begin{equation}
		\varphi(z+2\alpha\cdot\bm{\ell})=\varphi(z)\,.\label{eq:phiequiv}
	\end{equation}
	Taken together, this ensures that the factorised $\mathcal{R}$-matrix \eqref{eq:RTwist} satisfies a similar property to \eqref{eq:R0equiv}.
	
	\paragraph{Modular Invariance:}
	Lastly, we will discuss the invariance of the action \eqref{eq:DPCMActionHighRank} under \textit{modular transformations}. We will fix the rescaling freedom by going to the modular parameterization, thus working with the parameters $(\tau,\rho,\ha{z})$. The group $\text{SL}_2(\mathbb{Z})$ has a very well-known action on the torus modulus $\tau$, given by Möbius transformations. We will supplement this action with a non-trivial transformation of the additional variables $\rho$ and $\ha{z}$. More precisely, we will define the action of $\gamma=\begin{pmatrix}
		a&b\\c&d
	\end{pmatrix}\in\text{SL}_2(\mathbb{Z})$ on the parameters to be given by
	\begin{equation}
		(\tau,\rho,\ha{z})\longmapsto \left(\frac{a\tau+b}{c\tau+d},\frac{\rho}{c\tau+d},\frac{\ha{z}}{c\tau+d}\right).\label{eq:modulartrans}
	\end{equation}
	Furthermore, $\gamma$ will also act on the Lie algebra $\mathfrak{sl}_{\mathbb{C}}(N)$, with its action on the Belavin generators $\{T_\alpha\}$ being given by
	\begin{equation}
		T_{\alpha}\longmapsto \sigma_\gamma[T_\alpha]\equiv\xi^{-\frac{bd}{2}\alpha_1^2-\frac{ac}{2}\alpha_2^2-bc\,\alpha_1\alpha_2}\,T_{\gamma'\alpha}\,, \qquad \text{ where } \quad \gamma'=\begin{pmatrix}
			d&c\\b&a
		\end{pmatrix}\label{eq:modularBelavin}
	\end{equation}
	and $\gamma'\alpha$ is the action of the matrix $\gamma'\in\text{SL}_2(\mathbb{Z})$ on the 2-vector $\alpha$, modulo $N$. One checks that this transformation $\sigma_\gamma$ is an automorphism of $\slNC$. We will also denote its uplift to the group $\SLNC$ by $\sigma_\gamma$. Moreover, a slightly tedious but straightforward computation shows that the above formula defines a proper action of $\text{SL}_2(\mathbb{Z})$, in the sense that $\sigma_{\gamma_1} \circ \sigma_{\gamma_2} = \sigma_{\gamma_1\cdot\gamma_2}$.
	
	The action \eqref{eq:modulartrans} is natural, since it is related to the theory of \textit{Jacobi forms}, see for example \cite{eichler_theory_1985} for an introduction. In particular, the Weierstrass $\wp$-function transforms as a Jacobi form of weight 2 and index 0, \textit{i.e.}
	\begin{align}
		\wp(z;\tau)\longmapsto \wp\left(\frac{z}{c\tau+d};\frac{a\tau+b}{c\tau+d}\right)=(c\tau+d)^2\wp(z;\tau)\,.
	\end{align}
	One then checks that the 1-form $\omega=\varphi(z)\dd z$, with the twist function $\varphi(z)$ being given by \eqref{eq:higherrnaktwist}, is left invariant under $\text{SL}_2(\mathbb{Z})$. Similarly, combining \eqref{eq:modulartrans} and the definition of $\gamma'$ in \eqref{eq:modularBelavin}, one checks that the components \eqref{eq:EllipticDeformationsHighRank} of the deformation operator in the Belavin basis transform as
	\begin{equation}
		D_\alpha\longmapsto D_{\gamma'\alpha}\,,
	\end{equation}
	which in turn implies that
	\vspace{-3pt}
	\begin{align}\label{eq:ModInv}
		S_{\text{DPCM}}(\tau,\rho,\ha{z})[g]=S_{\text{DPCM}}\left(\frac{a\tau+b}{c\tau+d},\frac{\rho}{c\tau+d},\frac{\ha{z}}{c\tau+d}\right)\bigl[\sigma_\gamma^{-1}(g)\bigr]\,.
	\end{align}
	Thus, by combining a $\text{SL}_2(\Z)$ Möbius transformation of $\tau$ with a transformation of the other variables $(\rho,\ha{z})$ and a non-trivial action on the $\SLNC$-valued field $g$, the DPCM is left invariant.
	
	\subsection{Reality Conditions}\label{subsec:Reality}
	The reader might worry about whether the action \eqref{eq:DPCMActionHighRank} is real. Initially, this question is moot since the field $g$ has so far been valued in the complex Lie group $\text{SL}_\C(N)$. To establish reality of the action, we need to fix a \textit{real form} of $\mathfrak{sl}_{\mathbb{C}}(N)$ and require that it is stabilised by the operator $D$: restricting the field $g$ to take values in the corresponding real group will then ensure the reality of the action \eqref{eq:DPCMActionHighRank}. There are two obvious candidates: $\mathfrak{sl}_{\mathbb{R}}(N)$ and $\mathfrak{su}(N)$. 
	
	We will first need to express these real forms in terms of the Belavin basis \eqref{eq:defofT}. First, notice that the complex conjugation of $T_\alpha$ yields
	\begin{equation}
		\ol{T_{\alpha}}=\frac{1}{\sqrt{N}}\ol{\left(\Xi_1^{-\alpha_2}\Xi_2^{\alpha_1}\right)}=\frac{1}{\sqrt{N}}\Xi_1^{-\alpha_2}\Xi_2^{-\alpha_1}=T_{\bar{\alpha}}\,,
	\end{equation}
	where we define $\ol{(\alpha_1,\alpha_2)}\equiv(-\alpha_1,\alpha_2)$. On the other hand, the transpose of $T_\alpha$ is given by
	\begin{equation}
		\left(T_\alpha\right)^{t}=\frac{1}{\sqrt{N}}\left(\Xi_1^{-\alpha_2}\Xi_2^{\alpha_1}\right)^t=\frac{1}{\sqrt{N}}\Xi_2^{\alpha_1}\Xi_1^{\alpha_2}=\frac{1}{\sqrt{N}}\xi^{-\alpha_1\alpha_2}\Xi_1^{\alpha_2}\Xi_2^{\alpha_1}=-T^{\ol{\alpha}}\,.
	\end{equation}
	Taken together, this implies
	\begin{equation}
		\left(T_\alpha\right)^{\dagger}=\left(\ol{T_\alpha}\right)^t=-T^\alpha\,,
	\end{equation}
	which means that generating families for the real forms are given by:
	\begin{equation*}
		\text{Generating family for $\mathfrak{sl}_{\mathbb{R}}(N)$: } \qquad T_{\alpha}^{\mathfrak{R}}\equiv\frac{T_\alpha+T_{\bar{\alpha}}}{\sqrt{2}}\,,\qquad T_\alpha^{\mathfrak{I}}\equiv i\frac{T_\alpha-T_{\bar{\alpha}}}{\sqrt{2}}\,, 
	\end{equation*}~\vspace{-7pt}
	\begin{equation*}
		\text{Generating family for $\mathfrak{su}(N)$: } \qquad T_{\alpha}^{+}\equiv \frac{T_\alpha+T^{\alpha}}{\sqrt{2}}\,,\qquad T_\alpha
		^-\equiv i\frac{T_\alpha-T^{\alpha}}{\sqrt{2}}\,.
	\end{equation*}
	These are not quite bases, since there are $N^2-1$ linear relations between the elements. For example, for $N=3$, a basis for $\mathfrak{sl}_\mathbb{R}(N)$ is given by
	\begin{equation}
		T^\mathfrak{R}_{(0,1)},\;T^\mathfrak{I}_{(0,1)},\;T^\mathfrak{R}_{(1,0)},\;T^\mathfrak{R}_{(1,1)},\;T^\mathfrak{I}_{(1,1)},\;T^\mathfrak{R}_{(2,0)},\;T^\mathfrak{R}_{(2,1)},\;T^\mathfrak{I}_{(2,1)},
	\end{equation}
	while for example $T^{\mathfrak{I}}_{(1,0)}$ vanishes and $T^{\mathfrak{R}}_{(1,2)}=T^{\mathfrak{R}}_{(1,1)}$.
	
	We will then ask that $D$ stabilises one of these real forms. To ensure this, we have to impose certain reality conditions on the parameters of the theory. In analogy to the $\mathfrak{su}(2)$ case, we will demand that
	\begin{equation}
		\ell_1,\rho,\ha{z}\in\mathbb{R}\,,\qquad\ell_2\in i\mathbb{R}\,,\label{eq:realitycond}
	\end{equation}
	which further implies that $L_1\in\mathbb{R}$ and $L_2\in i\mathbb{R}$. Then notice that for this choice of reality conditions, the ingredients of the deformation \eqref{eq:EllipticDeformationsHighRank} satisfy $\ol{\zeta(z)}=\zeta(\ol{z})$, $\ol{q_\alpha}=q_{\bar{\alpha}}$  and $\ol{Q_\alpha}=Q_{\bar{\alpha}}$, which implies
	\begin{equation}
		\ol{D_\alpha}=\ol{\rho\,\Big\{Q_\alpha+\zeta(\ha{z})-\zeta(\ha{z}+q_\alpha)\}}=\rho\,\Big\{Q_{\bar{\alpha}}+\zeta(\ha{z})-\zeta(\ha{z}+q_{\bar{\alpha}})\Big\}=D_{\bar{\alpha}}\,.
	\end{equation}
	We thus obtain the following action of $D$ on the $\mathfrak{sl}_{\mathbb{R}}(N)$ generators:
	\begin{subequations}
		\begin{equation}
			D\left[T_\alpha^{\mathfrak{R}}\right]=\frac{D_\alpha T_\alpha+ D_{\bar{\alpha}}T_{\bar{\alpha}}}{\sqrt{2}}=\mathfrak{Re}(D_\alpha)\,T_\alpha^{\mathfrak{R}}+\mathfrak{Im}(D_\alpha)\,T_\alpha^{\mathfrak{I}}\,,
		\end{equation}
		\begin{equation}
			D\left[T_\alpha^{\mathfrak{I}}\right]=i\frac{D_\alpha T_\alpha-D_{\bar{\alpha}}T_{\ol{\alpha}}}{\sqrt{2}}=\mathfrak{Re}(D_\alpha)\,T_\alpha^{\mathfrak{I}}-\mathfrak{Im}(D_\alpha)\,T_\alpha^{\mathfrak{R}}\,,
		\end{equation}
	\end{subequations}
	which means that $D$ stabilises the real form $\mathfrak{sl}_{\mathbb{R}}(N)$. The action \eqref{eq:DPCMActionHighRank} is then real for $g$ in $\text{SL}_{\mathbb{R}}(N)$. In particular, the target space of this $\sigma$-model is non-compact and non-Euclidean. We have not found a similar reality condition to \eqref{eq:realitycond} that would make $D$ stabilise $\mathfrak{su}(N)$, so the appropriate real form of $\mathfrak{sl}_{\mathbb{C}}(N)$ seems to be $\mathfrak{sl}_{\mathbb{R}}(N)$. The fact that $\mathfrak{su}(2)$ also has a real action for our choice of reality conditions \eqref{eq:realitycond} is a low-dimensional anomaly: indeed, for $N=2$, all the coefficients $D_\alpha$ are real themselves once one imposes \eqref{eq:realitycond}. However, this does not hold in general.
	
	\subsection{The Trigonometric Limit and Yang-Baxter Deformations}\label{subsec:MainYB}
	Recall that the SU(2) elliptic integrable DPCM became a trigonometric integrable model in the limit $m\ra 1^-$. Similarly, one would expect the higher rank elliptic integrable model to become trigonometric when $\ell_1\ra+\infty$: in this limit, we lose periodicity along the real axis, so we would expect to recover trigonometric (or rather hyperbolic) functions. We are then left with a 1-parameter deformation of the PCM. Furthermore, we saw previously that our deformed model has a left $\text{SL}_\mathbb{R}(N)_{L}$-symmetry, while the right symmetry of the theory is (at least partially) broken. Models with these properties are already well-known in the literature: namely, the \textit{Yang-Baxter deformations}~\cite{klimcik_yang-baxter_2002,klimcik_integrability_2009}. Thus, one might wonder if the trigonometric limit $\ell_1\ra+\infty$ of the elliptic deformation could be viewed as a Yang-Baxter deformation of the PCM. Indeed, we will find this to be true. The full calculation is done in appendix \ref{sec:YBAppendix}, but we will highlight the results here.
	
	\paragraph{The Trigonometric Limit:}
	We wish to take the limit $\ell_1\ra+\infty$ of the model, after which we are naively left with three parameters $(\ell_2,\rho,\ha{z})$. As mentioned earlier, there is still redundancy in this parametrisation due to the freedom of rescaling the spectral parameter: here we will use it to fix $\ha{z}$ to $1$. Finally, for future convenience, we make the choice of parametrising the second half-period as $\ell_2=\frac{i\pi}{2\nu}$ with $\nu >0$, so that the model in the limit now depends only on the two parameters $\rho$ and $\nu$. For brevity, we will refer to the limiting procedure $(\ell_1,\ell_2,\rho,\ha{z}) \to \left(+\infty, \frac{i\pi}{2\nu},\rho, 1 \right)$ as the \textit{trigonometric limit} and will denote it as a mathematical operator ``$\tlim$''. One then checks that the trigonometric limits of the Weierstrass elliptic functions are given by
	\begin{equation}
		\tlim \;\, \zeta(z)=\nu\,\coth(\nu z)-\frac{\nu^2 z}{3}\,,\qquad \tlim \;\, \wp(z)=\frac{\nu^2}{\sinh^2(\nu z)}+\frac{\nu^2}{3}\,.\label{eq:tlimzetawp}
	\end{equation}
	Using this, one finds that the twist function \eqref{eq:higherrnaktwist} has the limit
	\begin{equation}
		\tlim \;\, \varphi(z)= \rho\left( \frac{\nu^2}{\sinh^2(\nu z)}-\frac{\nu^2}{\sinh^2(\nu)}\right)\,,
	\end{equation}
	exactly reproducing the trigonometric twist function \eqref{eq:TrigPhi} found for SU(2) upon identifying $\rho$ with $h\,\tanh(\nu)/\nu$. 
	
	For the deformation coefficients $D_{\alpha}=D_{(\alpha_1,\alpha_2)}$ in \eqref{eq:EllipticDeformationsHighRank}, the limit sharply distinguishes between the cases with $\alpha_2=0$ and the ones with $\alpha_2\neq 0$. Furthermore, one has to choose a consistent way to enumerate the elements $\alpha=(\alpha_1,\alpha_2)$ of $\mathbb{A} = \mathbb{Z}_N \times \mathbb{Z}_N \setminus \lbrace (0,0) \rbrace$: we choose to list $\alpha_1$ and $\alpha_2$ between $0$ and $N-1$. In the trigonometric limit, the calculation in appendix \ref{sec:YBAppendix} then finds
	\begin{subequations}\label{eq:TrigDalpha}
		\begin{equation}
			\alpha_2=0: \qquad \tlim\;\, D_{(\alpha_1,0)}=\rho\nu\left\{\coth(\nu) - \coth(\nu+i\pi\frac{\alpha_1}{N})\right\},\label{eq:TrigDalpha0}
		\end{equation}
		\begin{equation}
			\alpha_2\neq 0: \qquad \tlim\;\, D_{(\alpha_1,\alpha_2)}=\rho \nu\left\{\coth(\nu) - 1 + \frac{2\alpha_2}{N} \right\}.\label{eq:TrigDalphaneq0}
		\end{equation}
	\end{subequations}
	
	\paragraph{Yang-Baxter Deformations:}
	A Yang-Baxter model~\cite{klimcik_yang-baxter_2002,klimcik_integrability_2009} is defined by choosing the constant deformation operator in \eqref{eq:DPCMActionHighRank} to be of the form
	\begin{equation}
		\mathscr{D}=h\frac{1-\eta^2}{1-\eta\mathscr{R}}\,,
	\end{equation}
	with $h$ and $\eta$ real parameters and $\mathscr{R}$ a skew-symmetric operator on $\hg$ satisfying the \textit{modified classical Yang-Baxter equation (mcYBE)}:
	\begin{equation}
		\big[\mathscr{R}(X),\mathscr{R}(Y)\big]-\mathscr{R}\Big(\left[X,\mathscr{R}(Y)\right]+\left[\mathscr{R}(X),Y\right]\Big)=-c^2[X,Y]\,, \qquad \forall \, X,Y\in\hg\,,\label{eq:mybe}
	\end{equation}
	where $c\in\{0,1,i\}$ can be enforced by rescaling. Can we find parameters $h$ and $\eta$ and mcYBE operator $\mathscr{R}$ such that the Yang-Baxter deformation coincides with the trigonometric limit \eqref{eq:TrigDalpha}?
	
	Given that the real form we were considering in section \ref{subsec:Reality} was $\mathfrak{sl}_\mathbb{R}(N)$, which is the \textit{split real form} of $\mathfrak{sl}_\mathbb{C}(N)$, it is natural to work with $\mathscr{R}$ a solution of the \textit{split mcYBE}, \textit{i.e.} a solution of \eqref{eq:mybe} with $c=1$. We will focus on a particular type of solution, known as an \textit{inhomogeneous, extended Drinfel'd-Jimbo solution}~\cite{Drinfeld:1985rx,Jimbo:1985zk}. It uses the fact that the matrices $\lbrace T_{(\alpha_1,0)} \rbrace_{\alpha_1=1}^{N-1}$ with $\alpha_2=0$ are diagonal and thus generate a Cartan subalgebra of $\mathfrak{sl}_\mathbb{C}(N)$, while the matrices $T_{(\alpha_1,\alpha_2)}$ with $\alpha_2\neq 0$ can be built from the corresponding postive/negative root-vectors $E^{\pm}_{ij}$. The calculation in appendix \ref{sec:YBAppendix} then shows that if we let the operator $\mathscr{R}$ act as
	\begin{equation}
		\mathscr{R}\left[T_{(\alpha_1,0)}\right]=i\,\cot\left(\pi\frac{\alpha_1}{N}\right)T_{(\alpha_1,0)}\,,\qquad \mathscr{R}\left[E^\pm_{ij}\right]=\mp E^{\pm}_{ij}\,,
	\end{equation}
	and let the constants $\eta$ and $h$ be given by
	\begin{equation}
		\eta=\tanh(\nu)\,,\qquad h= \frac{\rho\,\nu}{\tanh(\nu)}\,,
	\end{equation}
	then the Yang-Baxter-deformation matches the trigonometric limit of the elliptic deformation \eqref{eq:TrigDalpha}, up to a total derivative term in the Lagrangian. Note that these identifications, as well as  $\ell_1\ra+\infty$ and $\ell_2\ra \frac{i\pi}{2\nu}$ for $\nu>0$, are compatible with the reality conditions imposed in section \ref{subsec:Reality}; thus, the Yang-Baxter deformation also preserves the real form $\mathfrak{sl}_\mathbb{R}(N)$.
	
	\section{Construction From Equivariant 4D Chern-Simons}\label{sec:4dcs}
	
	In section \ref{sec:HigherRank}, we have presented an integrable elliptic deformation of the PCM on $\text{SL}_\R(N)$: in particular, we exhibited its action and Lax connection. Although the equivalence of the equations of motion derived from this action with the flatness of the Lax connection can be checked a posteriori, guessing the right form of these objects ex nihilo would in general be a quite arduous task. In fact, the expressions reviewed in section \ref{sec:HigherRank} were obtained using \textit{equivariant 4D Chern-Simons theory (4DCS)}, which offers a systematic framework for constructing elliptic integrable $\sigma$-models. This is the subject of this section.
	
	The relation between 4DCS theory~\cite{nekrassov_four_1996,costello_supersymmetric_2013} and integrable 2D $\sigma$-models has been initially proposed in~\cite{costello_gauge_2019} (see for instance~\cite{lacroix_4-dimensional_2022} for a review and a more complete list of references). In this section, we consider a slight generalisaton of the setup of~\cite{costello_gauge_2019}, by including equivariance as a built-in property of the 4DCS theory: this is what will allow us to derive the elliptic integrable DPCM. This property of equivariance can in principle be implemented for a 4DCS theory based on an arbitrary Riemann surface $\ha{\C}$ and Lie algebra $\hg^\C$, as long as both of these objects are equipped with an action of an abelian group $\mathbb{A}_0$ by automorphisms. Since we are interested in recovering elliptic integrable field theories with Belavin's $\Rc$-matrix as seed, we will focus in this section on the case where $\ha{\C}$ is a torus $\C/\Lambda$, $\hg^\C$ is $\mathfrak{sl}_\C(N)$ and $\mathbb{A}_0$ is the equivariance group $\mathbb{Z}_N \times \mathbb{Z}_N$ of this $\Rc$-matrix (see equation \eqref{eq:R0equiv} and the surrounding discussion).\footnote{Another example of equivariant 4DCS theory is the one considered in~\cite{schmidtt_symmetric_2021} to discuss $\lambda$-models on semi-symmetric spaces, for which $\ha{\C}=\mathbb{CP}^1$ and $\mathbb{A}_0=\mathbb{Z}_4$.} The 4DCS theory obtained this way is similar to the one considered in~\cite{costello_gauge_2018,costello_gauge_2018-1} to construct lattice models with Belavin's quantum $\Rc$-matrix: the results of this section can thus be seen as a generalisation of this construction from lattice systems to integrable $\sigma$-models (using so called disorder defects).\footnote{Here, we will consider a complexified 4DCS theory, which will then produce complexified integrable 2D $\sigma$-models. To obtain real ones, one would have to add reality conditions to the 4D theory, as done for instance in~\cite{delduc_unifying_2020}. For simplicity, we will not discuss these aspects here. In particular, we will then only obtain the  complexified DPCM on $\text{SL}_\C(N)$: we refer to the subsection \ref{subsec:Reality} for a discussion of the reality conditions of this model.}
	
	\subsection{The Basics of Equivariant 4D Chern-Simons Theory}\label{sec:Basics4d}
	The idea behind 4DCS is to enlarge the 2D space-time $\Sigma$ of the $\sigma$-model to a 4D manifold $\Sigma\times\ha{\mathbb{C}}$, with $\ha{\mathbb{C}}$ some Riemann surface. We will use coordinates $x^M=(t,x,z,\ol{z})$ on this 4D manifold, where $(t,x)$ parametrise $\Sigma$ and $z$ is a holomorphic coordinate on $\ha{\mathbb{C}}$. The latter will eventually be identified with the spectral parameter: in the present case, we thus take the Riemann surface $\ha{\mathbb{C}}$ to be the torus $\mathbb{T}=\mathbb{C}/\Lambda$. On the manifold $\Sigma\times\mathbb{T}$ lives a gauge field $\mathcal{A}$, taking its values in a Lie algebra $\hg^\C$, which we choose to be $\mathfrak{sl}_\mathbb{C}(N)$ for the remainder of this analysis. 
	The gauge field is a $1$-form and thus belongs to the space $\Omega^1(\Sigma\times\mathbb{T},\mathfrak{sl}_\mathbb{C}(N))$ (for $V$ a vector space, we denote by $\Omega(M,V)$ the space of $V$-valued differential forms on $M$ and by $\Omega^p(M,V)$ its subspace composed by forms of degree $p$).
	
	The space $\Omega(\Sigma\times\mathbb{T},\mathfrak{sl}_\mathbb{C}(N))$ is naturally endowed with a \textit{wedge-product} $\wedge$ (inherited from the wedge-product on $\Omega(\Sigma\times\mathbb{T},\mathbb{C})$ and the commutator on $\mathfrak{sl}_\mathbb{C}(N)$) and a \textit{pairing} $\dlangle\cdot,\cdot\drangle$ (inherited from the wedge-product on $\Omega(\Sigma\times\mathbb{T},\mathbb{C})$ and the bilinear form $\langle\cdot,\cdot\rangle$ on $\mathfrak{sl}_\mathbb{C}(N)$). For instance, for 1-forms $X = X_M \, \dd x^M$ and $Y = Y_M \, \dd x^M$ in $\Omega^1(\Sigma\times\mathbb{T},\mathfrak{sl}_\mathbb{C}(N))$, they are given by
	\begin{align}
		X\wedge Y=\frac{1}{2}\left[X_M,Y_N\right]\dd x^M\wedge\dd x^N\,,\qquad \dlangle X,Y\drangle=\left\langle X_M,Y_N\right\rangle\,\dd x^M\wedge \dd x^N\,,
	\end{align}
	respectively valued in $\Omega^2(\Sigma\times\mathbb{T},\mathfrak{sl}_\mathbb{C}(N))$ and $\Omega^2(\Sigma\times\mathbb{T},\mathbb{C})$.
	Using these, one can construct the \textit{curvature} of $\mathcal{A}$
	\begin{align}
		\mathcal{F}\left[\mathcal{A}\right]\equiv\dd\mathcal{A}+\mathcal{A}\wedge\mathcal{A} \qquad \text{in} \qquad \Omega^2\Big(\Sigma\times\mathbb{T},\mathfrak{sl}_\mathbb{C}(N)\Big)
	\end{align}
	and the \textit{Chern-Simons 3-form}
	\begin{align}
		\mathcal{CS}\left[\mathcal{A}\right]\equiv\dlangle\mathcal{A},\dd\mathcal{A}+\frac{2}{3}\mathcal{A}\wedge\mathcal{A}\drangle \qquad \text{in} \qquad \Omega^3\Big(\Sigma\times\mathbb{T},\mathbb{C}\Big)\,.
	\end{align}
	
	\paragraph{The 4D Action:}
	The action of the 4DCS theory is given by~\cite{nekrassov_four_1996,costello_supersymmetric_2013}
	\begin{equation}
		S_{\text{4DCS}}[\mathcal{A}] = \frac{i}{4\pi N^2}\int_{\Sigma\times\mathbb{T}} \omega\wedge\mathcal{CS}[\mathcal{A}]\,,\label{eq:4DCS Action}
	\end{equation}
	where the prefactor, depending on the Lie algebra $\mathfrak{sl}_\mathbb{C}(N)$ in which $\mathcal{A}$ is valued, has been introduced for future convenience. Here, $\omega$ is a meromorphic 1-form on $\mathbb{T}$ valued in $\mb{CP}^1$, which is data defining the model, \textit{i.e.} it is not dynamical and encodes the parameters of the theory. Having chosen a complex coordinate $z$ on $\mathbb{T}$, we can write this form as
	\begin{align}
		\omega=\varphi(z)\,\dd z\,,\label{eq:omegaandphi}
	\end{align}
	where $\vp(z)$ is a meromorphic elliptic function. The 4DCS theory under consideration will eventually reduce to an elliptic integrable 2D field theory: in that context, the function $\vp(z)$ appearing in $\omega$ will then be identified with the twist function of this model, as defined in section \ref{sec:Basics} from the point of view of Maillet brackets.\footnote{This relation between $\omega$ and the twist function has been proven in~\cite{vicedo_4d_2021} for the rational 4DCS theory, \textit{i.e.} when the underlying Riemann surface $\ha{\mathbb{C}}$ is the Riemann sphere $\mathbb{CP}^1$. That this relation also holds for the elliptic case $\ha{\mathbb{C}}=\mathbb{T}$ considered here is a conjecture (which holds for the elliptic DPCM).}
	Written in coordinates, the action \eqref{eq:4DCS Action} is then given by:
	\begin{align}\label{eq:4daction-comp}
		S_{\text{4DCS}}[\mathcal{A}]=\frac{i}{4\pi N^2}\int_{\Sigma}\dd t\,\dd x\int_{\mathbb{T}}\dd z\,\dd\ol{z}\,\varphi(z)\sum_{L,M,N\in\{t,x,\ol{z}\}}\epsilon^{LMN} \left\langle\mathcal{A}_L,\partial_M\mathcal{A}_N+\frac{1}{3}\left[\mathcal{A}_M,\mathcal{A}_N\right]\right\rangle\,,
	\end{align}
	where the $z$-component of the gauge field and the $z$-derivatives have decoupled due to the wedge product with $\omega=\vp(z)\dd z$.
	
	\paragraph{Equivariance:}
	As explained in subsection \ref{sec:DPCM-HR}, the elliptic DPCM possesses equivariance properties with respect to the abelian group $\mathbb{A}_0 = \mathbb{Z}_N \times \mathbb{Z}_N$. This arises naturally in its construction from 4DCS theory: indeed, we will consider here a variant of this theory with additional built-in equivariance properties, similar to that of~\cite[Section 10]{costello_gauge_2018-1}. These will take the form of restrictions on the allowed 1-forms $\omega$ and the gauge field $\mathcal{A}$. Recall from subsection \ref{sec:DPCM-HR} that an element $\alpha=(\alpha_1,\alpha_2)$ of the group $\mathbb{A}_0 = \mathbb{Z}_N \times \mathbb{Z}_N$ acts on the torus $\mathbb{T}=\C/\Lambda$ by the translation $ z \mapsto z + 2\alpha\cdot\bm{\ell}$. On the other hand, it acts on the Lie algebra $\mathfrak{sl}_\mathbb{C}(N)$ by the automorphism $X \mapsto \Ad_{\Xi_1}^{\alpha_1}\,\Ad_{\Xi_2}^{\alpha_2}\,X$. Here, we will impose that the meromorphic 1-form $\omega$ and the gauge field $\mathcal{A}$ of the 4DCS theory are equivariant under the action of $\mathbb{A}_0$. Concretely, this means that we ask
	\begin{equation}\label{eq:EquivCS}
		\omega(z + 2\alpha\cdot\bm{\ell}) = \omega(z) \qquad \text{ and } \qquad \mathcal{A}(t,x,z + 2\alpha\cdot\bm{\ell}) = \Ad_{\Xi_1}^{\alpha_1}\,\Ad_{\Xi_2}^{\alpha_2}\,\mathcal{A}(t,x,z)\,.
	\end{equation}
	Note that we make $\mathbb{A}_0$ act trivially on the space $\mathbb{CP}^1$ in which $\omega$ is valued, so that the $\mathbb{A}_0$-equivariance of $\omega$ simply coincides with its invariance under translations $z\mapsto z+2\alpha\cdot\bm{\ell}$. In particular, this means that $\omega$ is not only doubly-periodic with periods $2N\ell_i$ but in fact also with the smaller periods $2\ell_i$. In other words, it can seen as a meromorphic 1-form on the torus $\C/\Gamma$, where $\Gamma=2\ell_1\mathbb{Z}+2\ell_2\mathbb{Z}$ (this torus can be interpreted as the quotient of $\mathbb{T}=\C/\Lambda$ by the action of $\mathbb{A}_0$).
	
	\paragraph{Analytic Structure of $\bm{\omega}$:} Up to an overall constant, the meromorphic 1-form $\omega$ is characterised by the position and the order of its poles and zeroes in $\mathbb{T}$. Note that if $z\in\mathbb{T}$ is a pole/zero of $\omega$ of order $m$, then all the translates $z+2\alpha\cdot\bm{\ell}$, $\alpha\in\mathbb{A}_0$, are also poles/zeroes of the same order, due to the equivariance property \eqref{eq:EquivCS}. The sets of poles and zeroes are thus separated into $\mathbb{A}_0$-orbits, each composed of $|\mathbb{A}_0|=N^2$ translated points: we make a choice of representative for each of these orbits and denote by $\ha{\mathbb{P}}$ the set of representatives of the poles and by $\ha{\mathbb{Z}}$ the set of representatives of the zeroes.
	
	For simplicity, we will suppose here that all poles have multiplicity two while all zeroes are simple. In principle, more complicated analytical structures of $\omega$ are also allowed and should lead to a richer family of elliptic integrable $\sigma$-models: we choose to restrict to this simpler setup to lighten the discussion and as it will be enough to produce the integrable DPCM of section \ref{sec:HigherRank}. We refer to the concluding section \ref{sec:conclusion} for a brief discussion of perspectives on this more general family.
	
	Let us denote by $M=|\ha{\mathbb{P}}|$ the size of $\ha{\mathbb{P}}$, \textit{i.e.} the number of poles of $\omega$ up to the action of $\mathbb{A}_0$. By the Riemann-Hurwitz formula, we then find that the number of zeroes (up to the action of $\mathbb{A}_0$) is $|\ha{\mathbb{Z}}|=2M$. For reasons to be explained later, we separate these zeroes into two disjoint subsets $\ha{\mathbb{Z}}^+$ and $\ha{\mathbb{Z}}^-$ of equal size $M$. We then introduce notations for these poles and zeroes by letting
	\begin{equation}\label{eq:PolesAndZeroes}
		\ha{\mathbb{P}} = \bigl\lbrace \ha{p}_1,\dots,\ha{p}_M \bigr\rbrace \qquad \text{ and } \qquad \ha{\mathbb{Z}}^\pm = \bigl\lbrace \ha{z}_1^\pm,\dots,\ha{z}_M^\pm \bigr\rbrace\,.
	\end{equation}
	
	\paragraph{Permissible Field Configurations:}
	We will now describe the space of allowed configurations of the gauge field $\mathcal{A}$, by listing a number of constraints that it has to satisfy. First, we expand $\mathcal{A}$ as
	\begin{align}
		\mathcal{A}=\mathcal{A}_M\, \dd x^M=\mathcal{A}_+\,\dd x^+ + \mathcal{A}_-\,\dd x^- + \mathcal{A}_z\,\dd z + \mathcal{A}_{\bar{z}}\,\dd \ol{z}\,,
	\end{align}
	where we used the light-cone coordinates $x^\pm = t \pm x$ on $\Sigma$.
	\begin{enumerate}
		\item As observed in equation \eqref{eq:4daction-comp}, the component $\mathcal{A}_z$ drops out of the 4DCS action completely. We are thus free to shift this component by any function on $\Sigma\times\mathbb{T}$ without changing the theory. Here, we will fix this freedom by simply restricting to field configurations for which
		\begin{align}
			\mathcal{A}_{{z}}=0\,.\label{eq:Azis0}
		\end{align}
		
		\item Recall from equation \eqref{eq:EquivCS} that we require the gauge field to be equivariant under the action of $\mathbb{A}_0$. We also consider this as a restriction on the allowed field configurations for the components of $\mathcal{A}$, which we recall here for completeness:
		\begin{equation}\label{eq:Aisequiv}
			\mathcal{A}_M (t,x,z + 2\alpha\cdot\bm{\ell}) = \Ad_{\Xi_1}^{\alpha_1}\,\Ad_{\Xi_2}^{\alpha_2}\,\mathcal{A}_M(t,x,z)\,.
		\end{equation}
		
		\item Recall that $\omega$ has poles of order two at the points $\ha{\mathbb{P}} = \bigl\lbrace \ha{p}_1,\dots,\ha{p}_M \bigr\rbrace$. We will impose \textit{boundary conditions} on the gauge field at these points, namely that
		\begin{align}
			\mathcal{A}_\pm(t,x,\ha{p}_r)=0\,. \label{eq:boundaryconditions}
		\end{align}
		By the equivariance property \eqref{eq:Aisequiv}, this means that $\mathcal{A}_\pm$ vanishes as well at all the translates $\ha{p}_r + 2\alpha\cdot\bm{\ell}$ with $\alpha\in\mathbb{A}_0$, which are also double poles of $\omega$. This boundary condition ensures that the Lagrangian density of the 4DCS theory is locally integrable in the neighbourhood of these poles and thus that the action is finite. In addition, as we will see in the next subsection, these boundary conditions also play a crucial role in the derivation of the equations of motion of the 4DCS theory and in its invariance under certain gauge transformations.
		
		\item Likewise, the components $\mathcal{A}_\pm$ of the gauge field will be allowed to have poles, called \textit{disorder defects}~\cite{costello_gauge_2019}. To ensure finiteness of the Lagrangian density, these singularities will be located at the zeroes of $\omega$, whose representatives under the action of $\mathbb{A}_0$ form the set $\ha{\mathbb{Z}}$. This is where the separation of these zeroes into two disjoint subsets $\ha{\mathbb{Z}}^\pm=\bigl\lbrace \ha{z}_1^\pm,\dots,\ha{z}_M^\pm \bigr\rbrace$ made earlier enters the game: this will correspond to a distribution of the corresponding singularities of the gauge field along its light-cone components $\mathcal{A}_\pm$\footnote{More generally, for each zero $\hat{z}$ of $\omega$, one could allow for a singularity of the gauge field at $\hat{z}$ along an arbitrary spatio-temporal component, \textit{i.e.} a linear combination of the form $c_+\,\mathcal{A}_+ + c_-\mathcal{A}_-$. Here, we make the choice of distributing each singularity along one of the light-cone component $\mathcal{A}_+$ or $\mathcal{A}_-$: this will in fact ensure the relativistic invariance of the resulting integrable 2D $\sigma$-model (see e.g. \cite{costello_gauge_2019,delduc_unifying_2020}). Moreover, one finds that the consistency of the construction requires having as many singularities in the component $\mathcal{A}_+$ as in the one $\mathcal{A}_-$, which is why we chose the subsets $\ha{\mathbb{Z}}^+$ and $\ha{\mathbb{Z}}^-$ to be of the same size.}. More precisely, we allow $\mathcal{A}_\pm$ to have singularities of order one\footnote{We restrict to singularities of order one since we supposed that the zeroes of $\omega$ are simple.} at the points $\ha{z}_r^\pm$, such that
		\begin{align}
			\bigl|z-\ha{z}_r^\pm \bigr|\, \mathcal{A}_\pm(t,x,z) \quad \text{is bounded around } z=\ha{z}_r^\pm\,.\label{eq:zerosofomega}
		\end{align}
		By equivariance \eqref{eq:Aisequiv}, this permits similar singularities at the $\mathbb{A}_0$-translates $\ha{z}^\pm_r + 2\alpha\cdot\bm{\ell}$, which are also zeroes of $\omega$. Outside of these points, we ask that $\mathcal{A}_\pm$ is regular (in particular, we note that the component $\mathcal{A}_\pm$ is regular at the points $\ha{z}^\mp_r \in \ha{\mathbb{Z}}^\mp$). Moreover, we suppose that the component $\mathcal{A}_{\bar{z}}$ is regular everywhere.
		
		\item Let us finally discuss boundary conditions of the gauge field with respect to the spatio-temporal coordinates $(t,x)\in\Sigma$. In the case where $\Sigma$ is the cylinder with periodic spatial coordinate $x\sim x+2\pi$, $\mathcal{A}$ is periodic in $x$ by definition and we further suppose that it decreases sufficiently fast when $t$ tends to infinity. If $\Sigma$ is the plane, with non-compact spatial coordinate $x$, we ask that $\mathcal{A}$ decreases sufficiently fast when either $t$ or $x$ tends to infinity.
		
	\end{enumerate}
	\subsection{Equations of Motion and Gauge Symmetry}
	
	\paragraph{Equations of Motion:} Having defined the field content and the action of the equivariant 4DCS theory, we are now ready to analyse its properties. We will first derive its equations of motion. We introduce a variation $\delta\mathcal{A}$ of the gauge field: it is a standard result that the variation of the Chern-Simons 3-form is then given by
	\begin{align}
		\delta\mathcal{CS}[\mathcal{A}]=2\dlangle\delta\mathcal{A},\mathcal{F}[\mathcal{A}]\drangle+\dd\dlangle\delta\mathcal{A},\mathcal{A}\drangle\,.
	\end{align}
	In turn, the variation of the action reads
	\begin{align*}
		\delta S_{\text{4DCS}}[\mathcal{A}]=\frac{i}{4\pi N^2}\int_{\Sigma\times\mathbb{T}}\omega\wedge\delta\mathcal{CS}[\mathcal{A}]
	\end{align*}
	\begin{align}
		=\frac{i}{4\pi N^2} \left\lbrace 2\int_{\Sigma\times\mathbb{T}}\omega\wedge\dlangle\delta\mathcal{A},\mathcal{F}[\mathcal{A}]\drangle + \int_{\Sigma\times\mathbb{T}}\dd\omega\wedge\dlangle\delta\mathcal{A},\mathcal{A}\drangle -  \int_{\Sigma\times\mathbb{T}}\dd\Big(\omega\wedge\dlangle\delta\mathcal{A},\mathcal{A}\drangle \Big) \right\rbrace.\label{eq:SEOM}
	\end{align}
	The last term in this equation is a boundary one and can safely be discarded (due to the boundary conditions imposed on $\mathcal{A}$ at the spatio-temporal infinity of $\Sigma$ and the fact that the torus $\mathbb{T}$ has no boundary). Spelling out $\dd \omega$ in the middle term as
	\begin{align}
		\dd\omega= \partial_t\varphi(z)\,\dd t \wedge \dd z + \partial_x\varphi(z)\,\dd x \wedge \dd z+ \partial_{\bar{z}}\varphi(z)\,\dd \ol{z}\wedge\dd z\,,\label{eq:ddomega}
	\end{align}
	one might suspect that it vanishes, since $\varphi$ is only a function of $z$, and not of $t,x$ or $\ol{z}$. Accordingly, the first two terms in \eqref{eq:ddomega} are then zero. However, the third term vanishes everywhere \textit{except} for the poles of $\omega$. Indeed, a meromorphic function has a non-vanishing $\ol{z}$-derivative at its poles in the distributional sense, since in a local neighbourood of $z=\ha{p}$, one has
	\begin{align} 
		\partial_{\bar{z}}\left(\frac{1}{(z-\ha{p})^k}\right)=2i\pi\frac{(-1)^k}{(k-1)!}\partial_z^{k-1}\delta(z-\ha{p})\,.\label{eq:PolesandDelta}
	\end{align}
	Here $\delta(z-\ha{p})$ is the two-dimensional Dirac distribution with support at $z=\ha{p}$. Thus, the middle term in \eqref{eq:SEOM} cannot be automatically discarded and yields 2-dimensional defect contributions along $\Sigma$, localised at a discrete set of points in $\mathbb{T}$, namely the poles of $\omega$. We supposed that the latter are all of multiplicity two and are located at the points $\ha{\mathbb{P}} = \bigl\lbrace \ha{p}_1,\dots,\ha{p}_M \bigr\rbrace$, together with their translates under the action of $\mathbb{A}_0$. Introducing $\kay_{r,0}=\text{res}_{z=\hat{p}_r} \omega$ and $\kay_{r,1}=\text{res}_{z=\hat{p}_r} (z-\ha{p}_r)\omega$, we find that the contribution of the pole $\ha{p}_r$ to this defect term is proportional to
	\begin{equation}\label{eq:DefectEom}
		\int_{\Sigma \times \lbrace \hat{p}_r \rbrace} \Bigl( \kay_{r,0} \dlangle \delta \mathcal{A},\mathcal{A} \drangle + \kay_{r,1} \dlangle  \p_z\delta\mathcal{A},\mathcal{A} \drangle + \kay_{r,1} \dlangle \delta \mathcal{A},\p_z\mathcal{A} \drangle \Bigr)\,.
	\end{equation}
	To obtain this equation, we worked in a neighbourhood of $\ha{p}_r$, used the identity \eqref{eq:PolesandDelta} and performed the integral over $z$ against the distributions $\delta(z-\ha{p}_r)$ and $\p_z \delta(z-\ha{p}_r)$. This is where the boundary condition \eqref{eq:boundaryconditions} imposed on the gauge field plays a crucial role. Indeed, this condition means that the restriction of the 1-form $\mathcal{A}$ on $\Sigma \times \bigl\lbrace \ha{p}_r \bigr\rbrace$ vanishes. Moreover, the variation $\delta\mathcal{A}$ of the gauge field should preserve this boundary condition, so that the restriction of $\delta\mathcal{A}$ on $\Sigma \times \lbrace \ha{p}_r \rbrace$ is alzo zero. This ensures that the three terms in the above surface integral all vanish. A similar analysis holds for the poles of $\omega$ at the translated points $\ha{p}_r + 2\alpha\cdot\bm{\ell}$, $\alpha\in\mathbb{A}_0$, using the equivariance property \eqref{eq:Aisequiv} of the gauge field. To summarise, we conclude that the second term in equation \eqref{eq:SEOM} vanishes, thanks to the boundary conditions \eqref{eq:boundaryconditions} imposed on the gauge field at the poles of $\omega$. The variation of the action then simply reads
	\begin{align}
		\delta S_{\text{4DCS}}[\mathcal{A}]= \frac{i}{2\pi N^2}\int_{\Sigma\times\mathbb{T}}\omega\wedge\dlangle\delta\mathcal{A},\mathcal{F}[\mathcal{A}] \drangle\,.\label{eq:4DCSEOM}
	\end{align}
	Requiring this variation to vanish for arbitrary $\delta\mathcal{A}$ amounts to the equation of motion:
	\begin{align}\label{eq:EoM4d}
		\omega\wedge\mathcal{F}[\mathcal{A}]=0\,.
	\end{align}
	
	\paragraph{Gauge Symmetry:}
	We now turn to the discussion of the gauge symmetry of the equivariant 4DCS theory. Let us consider a standard gauge transformation of the connection $\mathcal{A}$, namely $u\mathcal{A}u^{-1}-(\dd u)u^{-1}$, where $u$ is a smooth function on $\Sigma\times\mathbb{T}$ valued in the group $\text{SL}_\mathbb{C}(N)$. In the present case, we will have to adapt this definition slightly and impose restrictions on $u$ to take into account the various constraints put on the configurations of $\mathcal{A}$ (see previous subsection). For instance, recall that we fixed $\mathcal{A}_z=0$, using the fact that this component completely decouples from the theory. In order to preserve this, we thus change the definition of the gauge transformation to  
	\begin{align}
		\mathcal{A}\longmapsto u\mathcal{A}u^{-1}-(\dd' u)u^{-1}\,,\label{eq:gaugetransformofA}
	\end{align}
	where we use a ``truncated differential'' $\dd'u = \p_t u \,\dd t +  \p_x u \,\dd x +  \p_{\bar{z}} u \,\dd \ol{z}$, with no $\dd z$--component. For compatibility with the equivariance property \eqref{eq:Aisequiv} of the gauge field, we further restrict to gauge parameters $u$ which are themselves equivariant, \textit{i.e.} which satisfy
	\begin{align}\label{eq:EquivGauge}
		u(t,x,z+2\alpha\cdot\bm{\ell}) = \Ad_{\Xi_1}^{\alpha_1}  \Ad_{\Xi_2}^{\alpha_2} u(t,x,z)\,.
	\end{align}
	Moreover, to preserve the boundary conditions \eqref{eq:boundaryconditions} imposed on $\mathcal{A}$, we ask that
	\begin{equation}\label{eq:BCgauge}
		u(t,x,\ha{p}_r) = \text{Id}
	\end{equation}
	at the poles of $\omega$ (note that the same condition holds at the $\mathbb{A}_0$-translates $\ha{p}_r + 2\alpha\cdot\bm{\ell}$ by equivariance of $u$). We finally require that $u$ converges to the identity at the space-time infinity of $\Sigma$. Together, these various conditions on $u$ ensure that the gauge transformations \eqref{eq:gaugetransformofA} preserve the space of permissible field configurations discussed in subsection \ref{sec:Basics4d}.\footnote{Note that the singularity condition \eqref{eq:zerosofomega} at the zeroes of $\omega$ is preserved without further restrictions on $u$ as we supposed that the latter is a smooth function on $\Sigma \times \mathbb{T}$ and thus in particular is regular at the zeroes.}\\
	
	Now that we have properly defined the gauge transformations, let us show that they are symmetries of the theory. The easiest invariance to check is that of the equation of motion \eqref{eq:EoM4d}. To this end, let us recall the standard result stating that the curvature of $\mathcal{A}$ is covariant under gauge transformations $\mathcal{A} \mapsto u\mathcal{A}u^{-1}-(\dd u)u^{-1}$, \textit{i.e.} it transforms as $\mathcal{F}[\mathcal{A}] \mapsto u\mathcal{F}[\mathcal{A}]u^{-1}$. Naively, we cannot apply this result directly in the present case, since we have slightly changed the definition of gauge transformations by replacing the differential $\dd$ by the truncated one $\dd'$, which does not include the $\dd z$-component -- see equation \eqref{eq:gaugetransformofA}. However, crucially, we can replace $\dd'$ by $\dd$ in any form which is taken in an exterior product with $\omega$, since the latter is proportional to $\dd z$. In particular, under the correct gauge transformation \eqref{eq:gaugetransformofA}, we thus find that
	\begin{equation}
		\omega \wedge \mathcal{F}[\mathcal{A}] \longmapsto u (\omega \wedge \mathcal{F}[\mathcal{A}])u^{-1}\,.
	\end{equation}
	Therefore, it is clear that the equation of motion \eqref{eq:EoM4d} is gauge invariant.\newpage
	
	The invariance of the action is a more subtle question. The behaviour of the Chern-Simons 3-form under gauge transformations is also a well-known result, which we can apply in the present case to the product $\omega \wedge \mathcal{CS}[\mathcal{A}]$, for similar reasons as for the curvature. One then finds
	\begin{equation}
		\omega \wedge \mathcal{CS}[\mathcal{A}] \longmapsto \omega \wedge \Bigl( \mathcal{CS}[\mathcal{A}] - \dd \dlangle \mathcal{A}, u^{-1} \dd u \drangle + \frac{1}{3} \dlangle u^{-1} \dd u, u^{-1} \dd u \wedge u^{-1} \dd u \drangle \Bigr)\,.
	\end{equation}
	Integrating this equation and discarding a boundary term, we obtain the variation of the action:
	\begin{equation}\label{eq:gaugeAction}
		S_{\text{4DCS}}[\mathcal{A}] \longmapsto S_{\text{4DCS}}[\mathcal{A}] + \frac{i}{4\pi N^2} \int_{\Sigma\times\mathbb{T}} \!\!\!\dd\omega \wedge \dlangle \mathcal{A}, u^{-1} \dd u \drangle + \frac{i}{12\pi N^2} \int_{\Sigma\times\mathbb{T}} \!\!\!\omega \wedge \dlangle u^{-1} \dd u, u^{-1} \dd u \wedge u^{-1} \dd u \drangle \,.
	\end{equation}
	We now want to argue that the second and third term in the right-hand side vanish. The second one is the easiest to treat. Indeed, as argued earlier when deriving the equations of motion, $\dd\omega$ is a distribution supported at the poles of $\omega$ (see equations \eqref{eq:ddomega}--\eqref{eq:PolesandDelta} and the surrounding discussion). The term under investigation is thus a two-dimensional defect term, localised on the surfaces $\Sigma \times \lbrace \ha{p} \rbrace$, where $\ha{p}$ are the poles of $\omega$. Recall that the latter are given by the points $\ha{\mathbb{P}} = \bigl\lbrace \ha{p}_1,\dots,\ha{p}_M \bigr\rbrace$ and their translates under the action of $\mathbb{A}_0$. Following an argument similar to the derivation of equation \eqref{eq:DefectEom} in the previous paragraph, one finds that the contribution of the pole $\ha{p}_r$ to this defect term is proportional to
	\begin{equation}
		\int_{\Sigma \times \lbrace \hat{p}_r \rbrace} \Bigl( \kay_{r,0} \dlangle \mathcal{A}, u^{-1} \dd u \drangle + \kay_{r,1} \dlangle  \p_z\mathcal{A}, u^{-1} \dd u \drangle + \kay_{r,1} \dlangle  \mathcal{A},\p_z (u^{-1} \dd u) \drangle \Bigr)\,.
	\end{equation}
	Recall that the gauge field $\mathcal{A}$ and gauge parameter $u$ satisfy the respective boundary conditions \eqref{eq:boundaryconditions} and \eqref{eq:BCgauge} at the point $z=\ha{p}_r$. This means that the restriction of the 1-forms $\mathcal{A}$ and $u^{-1}\dd u$ on $\Sigma \times \lbrace \ha{p}_r \rbrace$ vanish, so that the above surface integral is zero. Using equivariance and repeating the argument shows that the contributions from the translated poles $\ha{p}_r + 2\alpha\cdot\bm{\ell}$, $\alpha\in\mathbb{A}_0$, are also zero. In conclusion, we thus find that the second term in the gauge transformation \eqref{eq:gaugeAction} of the action vanishes.
	
	The final term in equation \eqref{eq:gaugeAction} is slightly more subtle to analyse: indeed it is a Wess-Zumino-like term and thus requires a careful treatment of topological and global issues. Here, for simplicity, we will only present a heuristic local argument of why this term does not contribute: we expect that this statement can be established rigorously using homotopical analysis as done in~\cite{benini_homotopical_2022} for the rational 4DCS theory and thus refer to this work for details. It is a well-known fact that the Cartan 3-form $\dlangle u^{-1} \dd u, u^{-1} \dd u \wedge u^{-1} \dd u \drangle$ is closed, \textit{i.e.} is annihilated by $\dd$. \textit{Locally}, one can thus write it as a closed form $\dd \Lambda$, with $\Lambda \in \Omega^2(\Sigma \times \mathbb{T}, \C)$. Performing an integration by part and discarding the boundary term, we are left with the integral of $\dd\omega \wedge \Lambda$. As before, this localises to two-dimensional integrals of $\Lambda$ and $\p_z\Lambda$ along the defects $\Sigma \times \lbrace \ha{p} \rbrace$, with $\ha{p}$ the poles of $\omega$. Using the boundary condition \eqref{eq:BCgauge}, one can argue that the restriction of $\Lambda$ and $\p_z\Lambda$ on these surfaces vanishes, hence closing our heuristic local argument.
	
	\subsection{Parametrisation of the Gauge Field and Lax Connection}\label{subsec:paramA}
	
	\paragraph{The Field $\bm{\ha{g}}$:} Having established the fundamentals of the equivariant 4DCS theory, we are now ready to analyse its properties further. For that, it will be useful to parameterise the gauge field in a slightly different way. In particular, we will suppose that the $\ol{z}$-component can be written as
	\begin{align}
		\mathcal{A}_{\bar{z}}=-\left(\partial_{\bar{z}}\,\ha{g}\right)\ha{g}^{\,-1} \label{eq:Aolzisg}
	\end{align}
	for some smooth field $\ha{g}(t,x,z,\ol{z})$ on $\Sigma\times\mathbb{T}$, valued in the group $\text{SL}_\mathbb{C}(N)$. The existence of such a parameterisation is non-trivial (see for instance~\cite[Remark 5.1]{vicedo_4d_2021}) and will be crucial for our construction. Working at a fixed point $(t,x)\in\Sigma$, we expect that the existence of $\ha{g}$ can be established using the results on rigid elliptic bundles of~\cite[Section 10.2]{costello_gauge_2018} and~\cite[Section 9]{costello_gauge_2018-1}.\footnote{As mentioned earlier, these references reconstruct the quantum Belavin $\Rc$-matrix from an equivariant 4DCS theory similar to the one considered here. A key property in this construction is that the adjoint holomorphic bundle defined by the connection $\partial_{\bar{z}} + \mathcal{A}_{\bar{z}}$ possesses a property called ``rigidity'', which can be formulated as the vanishing of a certain cohomology group of this bundle. This cohomology group in turn measures the obstructions in finding solutions of $(\partial_{\bar{z}} + \mathcal{A}_{\bar{z}})\hat{g}=0$, thus providing the desired result. We are grateful to K. Costello and B. Vicedo for useful discussions on this matter.} We stress that this analysis crucially relies on the equivariance property \eqref{eq:Aisequiv} imposed on the gauge field $\mathcal{A}_{\bar{z}}$: without this property, there would exist many field configurations for which the parametrisation \eqref{eq:Aolzisg} would not be possible.\footnote{A typical example, similar to the setup of~\cite[Section 15.10]{costello_gauge_2019}, is the case where the field $\mathcal{A}_{\bar{z}}$ is simply independent of $z$ and $\ol{z}$. In that case, one can easily construct solutions $\hat{g}(t,x,z,\ol{z})$ of \eqref{eq:Aolzisg} on $\Sigma \times \C$, but one checks that none of these solutions are invariant under shifts $z\mapsto z+2N\ell_i$ and thus reduce to $\Sigma \times \mathbb{T}$. Consistently with our claim in this paragraph, we note that such a field configuration for $\mathcal{A}_{\bar{z}}$, constant along $\mathbb{T}$, is forbidden by the equivariance condition \eqref{eq:Aisequiv}.} In addition to constructing the field $\ha{g}$ pointwise in $(t,x)\in\Sigma$, one should in principle also verify that these solutions can be recombined into a smooth function on the whole 4D manifold $\Sigma \times \mathbb{T}$. This is also a subtle problem (see for instance~\cite[Remark 5.1]{vicedo_4d_2021} for a related discussion), that we shall not address here for simplicity (if necessary, we will restrict to field configurations for which such a smooth extension exists). From now on, we will thus admit that $\mathcal{A}_{\bar{z}}$ can be rewritten as \eqref{eq:Aolzisg} in terms of a smooth field $\ha{g}$ on $\Sigma\times\mathbb{T}$.
	
	We note that the function $\ha{g}$ is not uniquely defined by \eqref{eq:Aolzisg}: indeed, $\ha{g}h$ is also a solution if and only if $h$ is a function on $\Sigma\times \mathbb{T}$ which depends holomorphically on $z$. Since $\mathbb{T}$ is a compact Riemann surface, this forces $h$ to be constant along $\mathbb{T}$ by Liouville's theorem. We thus conclude that $\ha{g}(t,x,z,\ol{z})$ is unique up to multiplication on the right by $h(t,x)$.\\
	
	Let us now discuss the equivariance properties of $\ha{g}$, starting from that  of $\mathcal{A}_{\bar{z}}$ \eqref{eq:Aisequiv}. We will show that one can always choose $\ha{g}$ such that
	\begin{equation}\label{eq:gEquiv}
		\ha{g}(t,x,z+2\ell_i) = \Xi_i\,\ha{g}(t,x,z)\,\Xi_i^{-1}\,.
	\end{equation}
	Moreover, with that condition, $\ha{g}$ is unique up to a discrete choice, namely the multiplication by a $N^{\text{th}}$-root of unity $\xi^k$, $k\in\lbrace 0,\dots,N-1\rbrace$. To prove these facts, let us start with any solution $\td{g}$ of \eqref{eq:Aolzisg} on $\Sigma\times \mathbb{T}$. We then define $\td{g}_i(z) = \Xi_i^{-1} \,\td{g}(z+2\ell_i)$. A direct computation shows that
	\begin{equation}
		\bigl(\partial_{\bar{z}}\, \td{g}_i(z) \bigr) \td{g}_i(z)^{-1} = \Xi_i^{-1} \bigl(\partial_{\bar{z}}\, \td{g}(z+2\ell_i) \bigr) \td{g}(z+2\ell_i)\,\Xi_i = \Ad_{\Xi_i}^{-1}\,\mathcal{A}_{\bar{z}}(z+2\ell_i) = \mathcal{A}_{\bar{z}}(z)\,,
	\end{equation}
	where in the last equality we used the equivariance property \eqref{eq:Aisequiv} of $\mathcal{A}_{\bar{z}}$. The function $\td{g}_i$ then satisfies the same equation \eqref{eq:Aolzisg} as $\td{g}$. As explained above, this means that there exists $\td{\Xi}_i(t,x)$ such that $\td{g}_i=\td{g}\,\td{\Xi}_i^{-1}$. From the definition of $\td{g}_i$, we thus get
	\begin{equation}
		\td{g}(t,x,z+2\ell_i) = \Xi_i\,\td{g}(t,x,z)\,\td{\Xi}_i(t,x)^{-1}\,.
	\end{equation}
	Using the fact that $\td{g}(z+2N\ell_i)=\td{g}(z)$ and writing up the compatibility of the two different ways to relate $\td{g}(z+2\ell_1+2\ell_2)$ to $\td{g}(z)$, we find that the matrices $\td{\Xi}_i(t,x)$ have to satisfy
	\begin{equation}
		\td{\Xi}_1^N = \td{\Xi}_2^N = \mathbb{I} \qquad \text{ and } \qquad \td{\Xi}_1 \cdot\td{\Xi}_2 = \xi\,\td{\Xi}_2 \cdot \td{\Xi}_1\,.
	\end{equation}
	These are exactly the relations \eqref{eq:Xialgebra} obeyed by $\Xi_i$. It is a standard result that the latter are the unique matrices obeying this algebra up to conjugation by $\text{SL}_\C(N)$. Thus, there exists a matrix $h(t,x)$ in $\text{SL}_\C(N)$ such that
	\begin{equation}
		\td{\Xi}_i(t,x) = h(t,x)^{-1}\, \Xi_i\, h(t,x)\,.
	\end{equation}
	To conclude, let us define $\ha{g}(t,x,z) = \td{g}(t,x,z)\,h(t,x)$. It is a solution of \eqref{eq:Aolzisg} by construction and one easily checks that it satisfies the required equivariance condition \eqref{eq:gEquiv}.
	
	Finally, let us discuss the freedom in finding a function $\ha{g}$ with these two properties \eqref{eq:Aolzisg} and \eqref{eq:gEquiv}. Any other solution is related to $\ha{g}$ by right multiplication by a matrix which commutes with both $\Xi_1$ and $\Xi_2$. One checks that the only such matrices in $\text{SL}_\C(N)$ are $\xi^k\,\mathbb{I}$, $k\in\lbrace 0,\dots,N-1\rbrace$, proving the claim made earlier. From now on, we will pick one arbitrary choice among these possibilities and will then work with a fixed $\ha{g}$ satisfying \eqref{eq:gEquiv} for the rest of the section.
	
	\paragraph{The Lax Connection $\bm{\mc{L}}$:}
	Now that we have rewritten the $\ol{z}$-component \eqref{eq:Aolzisg} of the gauge field in terms of $\ha{g}$, let us turn our attention to its components $\mathcal{A}_\pm$ along $\Sigma$. We define the following fields:
	\begin{equation}\label{eq:LA}
		\mathcal{L}_\pm \equiv \ha{g}^{\,-1} \mathcal{A}_\pm \ha{g} + \ha{g}^{\,-1} \p_\pm \ha{g}\,,
	\end{equation}
	valued in $\mathfrak{sl}_N(\C)$. The various components of the gauge field can then be written as
	\begin{align}
		\mathcal{A}_z = 0\,, \qquad \mathcal{A}_{\bar{z}} = \left(\partial_{\bar{z}}\,\ha{g}\right)\ha{g}^{\,-1}\,, \qquad \mathcal{\mathcal{A}}_\pm= \ha{g}\mathcal{L}_\pm \ha{g}^{\,-1}-\left(\partial_\pm\ha{g}\right)\ha{g}^{\,-1} \,.\label{eq:parametrisationofA}
	\end{align}
	We think of this as a reparametrisation of $\mathcal{A}$ in terms of three new fields $(\ha{g},\mathcal{L}_+,\mathcal{L}_-)$. Further down the line, we will identify $\mathcal{L} = \mathcal{L}_+\,\dd x^+ + \mathcal{L}_-\,\dd x^-$ with the Lax connection of the underlying 2D integrable $\sigma$-model, while $\ha{g}$ will encode the 2D field content.
	
	Recall that $\mathcal{A}_M$ and $\ha{g}$ satisfy the equivariance conditions \eqref{eq:Aisequiv} and \eqref{eq:gEquiv}. One easily checks that this induces the following equivariant behaviour for $\mathcal{L}_\pm$:
	\begin{equation}\label{eq:LEquiv}
		\mathcal{L}_\pm(t,x,z+2\ell_i)=\Ad_{\Xi_i}\,\mathcal{L}_\pm(t,x,z)\,.
	\end{equation}
	
	\paragraph{Gauge Transformations:} Furthermore, recall the gauge transformations \eqref{eq:gaugetransformofA} acting on $\mathcal{A}$, where the gauge parameter $u$ is a smooth function on $\Sigma\times\mathbb{T}$, valued in $\text{SL}_\mathbb{C}(N)$, which satisfies the conditions \eqref{eq:EquivGauge} and \eqref{eq:BCgauge}. In terms of our new parametrisation \eqref{eq:parametrisationofA}, these transformations simply act as
	\begin{equation}\label{eq:gLGauge}
		\ha{g} \longmapsto u\,\ha{g}\,, \quad \qquad \mathcal{L}_\pm \longmapsto \mathcal{L}_\pm\,.
	\end{equation}
	In particular, we see that $\mathcal{L}_\pm$ is gauge-invariant and thus contains only physical degrees of freedom. In contrast, $\ha{g}$ is not gauge-invariant. In fact, it would be tempting to say that none of the degrees of freedom contained in $\ha{g}$ are physical, since applying the gauge transformation by $u=\ha{g}^{-1}$ would set $\ha{g}$ to the identity and would send $\mathcal{A}$ to the connection $\mathcal{L} = \mathcal{L}_+\,\dd x^+ + \mathcal{L}_-\,\dd x^-$, which is entirely composed of physical fields. However, we stress that such a gauge transformation would be incompatible with the boundary condition \eqref{eq:BCgauge} imposed on $u$ and thus cannot be performed. More precisely, it will only be possible to attain such a gauge away from the poles of $\omega$. This fact will be crucial for the construction of the underlying 2D $\sigma$-model.
	
	\paragraph{Equations of Motion Revisited:} The equations of motion of the gauge field were given by \eqref{eq:EoM4d}. In the parametrisation \eqref{eq:parametrisationofA}, using the fact that $\omega\wedge\mathcal{F}[\mathcal{A}]=\ha{g}\,(\omega\wedge\mathcal{F}[\mathcal{L}])\,\ha{g}^{\,-1}$, this becomes
	\begin{align}
		\omega\wedge \mathcal{F}[\mathcal{L}]=0\,.
	\end{align}
	Recall that $\omega$ is proportional to $\dd z$ while $\mathcal{L}$ has non-zero components only along the directions $\dd x^\pm$. Separating the various components of the above equation, we then find
	\begin{align}
		\omega \wedge \mathcal{F}_{\Sigma}[\mathcal{L}] = 0 \qquad \text{ and} \qquad \omega \wedge \partial_{\bar{z}}\mathcal{L} = 0\,,\label{eq:4DCSLaxisflatandmero}
	\end{align}
	where
	\begin{equation}
		\mathcal{F}_{\Sigma}[\mathcal{L}] = \bigl( \p_+ \mathcal{L}_- - \p_- \mathcal{L}_+ + [\mathcal{L}_+,\mathcal{L}_-] \bigr) \dd x^+ \wedge \dd x^-
	\end{equation}
	is the curvature of $\mathcal{L}$ along $\Sigma$. Away from the zeroes of $\omega$, the first equation in \eqref{eq:4DCSLaxisflatandmero} means that $\mathcal{L}$ is a flat connection on $\Sigma$ while the second one implies that $\mathcal{L}$ depends holomorphically on $z$. These are exactly the properties of the Lax connection of a 2D integrable field theory on $\Sigma$, with spectral parameter $z$. This is the great strength of 4D Chern-Simons theory as a mean for generating such integrable models: the equations of motion are automatically written as the flatness condition of a Lax connection, depending holomorphically on the auxiliary complex parameter $z$ (which here is valued in $\mathbb{T}$ minus the zeroes of $\omega$).
	
	\paragraph{Solving for $\bm{\Lc_\pm(z)}$:} Let us stress however that the holomorphicity of $\Lc$ only holds \textit{away} from the zeroes of $\omega$. This is crucial: indeed, if $\Lc$ was holomorphic on the entire torus $\mathbb{T}$, it would have to be independent of $z$ by Liouville's theorem and could thus not be an appropriate Lax connection. The presence of the exterior product by $\omega$ in equation \eqref{eq:4DCSLaxisflatandmero} solves this problem, by allowing $\mathcal{L}$ to be meromorphic on $\mathbb{T}$ instead, with poles located at the zeroes of $\omega$. Indeed, recall from the identity \eqref{eq:PolesandDelta} that in a neighbourhood of $z=w$, the $\ol{z}$-derivative of the singularity $1/(z-w)^{k}$ is proportional to the distribution $\p_z^{k-1} \delta(z-w)$. Since we supposed that $\omega$ has only simple zeroes, the second equation in \eqref{eq:4DCSLaxisflatandmero} then allows for simple poles of $\Lc(z)$ at these points.
	
	This is in agreement with the condition \eqref{eq:zerosofomega} imposed on the gauge field configurations, which allowed $\mathcal{A}_\pm$ to have order one singularities exactly at these zeroes. Let us be a bit more precise. Recall from subsection \ref{sec:Basics4d} that $\ha{\mathbb{Z}}$ is a set of representatives of the zeroes of $\omega$ with respect to the action of the equivariance group $\mathbb{Z}_N\times\mathbb{Z}_N$. Furthermore, $\ha{\mathbb{Z}}$ was separated into two disjoint subsets $\ha{\mathbb{Z}}^\pm = \bigl\lbrace \ha{z}_1^\pm,\dots,\ha{z}_M^\pm \bigr\rbrace$, which corresponded to the allowed singularities of the light-cone components $\mathcal{A}_\pm$ of the gauge field in the condition \eqref{eq:zerosofomega}. Thus, we find that $\Lc_\pm(z)$ can have a simple pole at the points $\ha{z}_r^\pm$ in $\ha{\mathbb{Z}}^\pm$ but not at the points $\ha{z}_r^\mp$ in $\ha{\mathbb{Z}}^\mp$\footnote{As already mentioned around equation \eqref{eq:zerosofomega}, this separation of $\ha{\mathbb{Z}}^+$ and $\ha{\mathbb{Z}}^-$ has to do with the relativistic invariance of the resulting 2D $\sigma$-model.}. Concretely, this means that there exist 2D fields $U_{\pm,r} : \Sigma \to \mathfrak{sl}_\C(N)$ for $r\in\lbrace 1,\dots,M\rbrace$ such that
	\begin{equation}\label{eq:poleL}
		\Lc_\pm(z) = \frac{U_{\pm,r}}{z-\ha{z}_r^\pm} + O\bigl( (z-\ha{z}_r^\pm)^0 \bigr)\,.
	\end{equation}
	Combined with the equivariance property \eqref{eq:LEquiv}, this implies that $\Lc_\pm(z)$ also has a simple pole at the translates $\ha{z}^\pm_r + 2\alpha_1\ell_1+2\alpha_2\ell_2$, with residues $\Ad_{\Xi_1}^{\alpha_1}\Ad_{\Xi_2}^{\alpha_2}\,U_{\pm,r}$. Outside of these points, $\Lc_\pm(z)$ is finite and holomorphic.
	
	This analytical behaviour, together with the equivariance property \eqref{eq:LEquiv}, uniquely fixes $\mathcal{L}_\pm(z)$ in terms of the fields $U_{\pm,r}$. The corresponding expression for $\Lc_\pm(z)$ is most conveniently written in terms of the Belavin basis $\{T_\alpha\}$ of $\mathfrak{sl}_{\mathbb{C}}(N)$ and the family of functions $\{r^\alpha\}$ on $\mathbb{T}$, defined respectively in \eqref{eq:defofT} and \eqref{eq:defofralpha}. Namely, we find
	\begin{align}
		\mathcal{L}_\pm(z)=\sum_{r=1}^{M} \sum_{\alpha\in\mathbb{A}}r^\alpha\!\left(z-\ha{z}^\pm_r\right)U^\alpha_{r,\pm}\,T_\alpha\,,\label{eq:LintermsofU}
	\end{align}
	where $U^\alpha_{r,\pm}$ is the component of $U_{r,\pm}$ along the basis element $T_\alpha$.
	
	Let us argue that this is indeed the right expression for $\Lc_\pm(z)$ and that it is unique. For this paragraph only, we denote by $\td{\Lc}_\pm(z)$ the right-hand side of equation \eqref{eq:LintermsofU}: we then want to prove that the properties described above forces $\Lc_\pm(z)=\td{\Lc}_\pm(z)$. Recall that the functions $r^\alpha(z)$ satisfy the quasi-periodicity property \eqref{eq:equivralpha} under shifts by $2\ell_i$: combined with the behaviour \eqref{eq:Grading} of $T_\alpha$ under the adjoint action of $\Xi_i$, one easily checks that $\td{\Lc}_\pm(z)$ satisfies the equivariance property \eqref{eq:LEquiv}, as required. Moreover, one has $r^\alpha(z) \sim 1/z$ when $z\to 0$: we thus find that $\td{\Lc}_\pm(z)$ has a simple pole at the points $\ha{z}_r^\pm$, with the correct behaviour \eqref{eq:poleL}. The same holds at the translates $\ha{z}_r^\pm + 2\alpha\cdot\bm{\ell}$ by equivariance. One checks that these points are the only poles of $\td{\Lc}_\pm(z)$, as required of $\Lc_\pm(z)$. Thus, the function $\td{\Lc}_\pm(z)-\Lc_\pm(z)$ has no singularities and is holomorphic on the entire torus $\mathbb{T}$. It is then constant by Liouville's theorem: we finally conclude that it is in fact zero, as equivariance requires that it commutes with $\Xi_1$ and $\Xi_2$. We thus have $\Lc_\pm(z)=\td{\Lc}_\pm(z)$, completing the proof that \eqref{eq:LintermsofU} is the unique expression compatible with the various properties of $\Lc_\pm(z)$.
	
	To summarise, the second equation of motion in \eqref{eq:4DCSLaxisflatandmero} (and the various conditions imposed on the gauge field configurations) allows us to completely solve for $\Lc_\pm(z)$ in terms of the zeroes of $\omega$ and some two-dimensional fields $U_{\pm,r}$. Later on, we will see that the latter are related to the fundamental fields of the underlying $\sigma$-model.
	
	\subsection[Obtaining the Integrable 2D \texorpdfstring{$\sigma$}{sigma}-Model]{Obtaining the Integrable 2D \texorpdfstring{$\bm\sigma$}{sigma}-Model}
	
	What we have presented so far has all taken place on the 4D manifold $\Sigma\times\mathbb{T}$. However, we ultimately want a 2D integrable field theory on $\Sigma$ from this construction, with $\mathcal{L}$ as its Lax connection. This requires us to extract the relevant 2D degrees of freedom from the 4D fields, as well as writing down the action as a 2D integral. For that, we will follow the approach of~\cite{costello_gauge_2019} and~\cite{delduc_unifying_2020}.
	
	\paragraph{Extracting the 2D Fields:}
	The fundamental fields of the $\sigma$-model are constructed as the gauge-invariant degrees of freedom contained in the function $\ha{g}$. Recall that gauge transformations act on $\ha{g}$ by equation \eqref{eq:gLGauge} and that the gauge parameter $u$ has to satisfy the equivariance property \eqref{eq:EquivGauge} and the boundary condition \eqref{eq:BCgauge} at the points $\ha{\mathbb{P}} = \bigl\lbrace \ha{p}_1,\dots,\ha{p}_M \bigr\rbrace $. These points, together with their image under the equivariance group $\mathbb{Z}_N\times\mathbb{Z}_N$, form the poles of $\omega$. Outside of these poles, the gauge parameter $u$ does not have to satisfy any boundary conditions: the degrees of freedom contained in $\ha{g}$ in this region can thus be eliminated by well-chosen gauge transformations. Similarly, the boundary conditions only constrain the evaluation of $u$ at the poles and not that of its derivatives: this means that the derivatives of $\ha{g}$ at the poles can also be gauged away. In the end, we are thus led to consider the 2D fields
	\begin{equation}
		g_r(t,x) \equiv \ha{g}\left(t,x,z=\ha{p}_r\right)\,.\label{eq:fieldsfromhag}
	\end{equation}
	It is clear from the boundary condition \eqref{eq:BCgauge} that these are gauge-invariant. The other degrees of freedom in $\ha{g}$ that cannot be eliminated by gauge transformations are its evaluations at the translates $\ha{p}_r + 2\alpha\cdot\bm{\ell}$: by the equivariance property \eqref{eq:gEquiv} of $\ha{g}$, these are not independent of the fields $g_r$, as they are  related by
	\begin{equation}\label{eq:grOrbit}
		\ha{g}\left(t,x,z=\ha{p}_r+2\alpha\cdot\bm{\ell}\right) = \Ad_{\Xi_1}^{\alpha_1}\,\Ad_{\Xi_2}^{\alpha_2}\,g_r(t,x)\,.
	\end{equation}
	In the end, we thus conclude that the independent and gauge-invariant degrees of freedom contained in $\ha{g}$ are exactly the $M$ two-dimensional fields $g_r : \Sigma \to \text{SL}_\C(N)$.
	
	\paragraph{The Lax Connection in Terms of the 2D Fields:} Recall that $\Lc_\pm(z)$ also contains 2D gauge-invariant fields $U_{r,\pm} : \Sigma \to \mathfrak{sl}_\C(N)$. We will now argue that these are related to the fields $g_r$ constructed above and thus are not independent degrees of freedom. The main character of this argument is the boundary condition \eqref{eq:boundaryconditions} imposed on the gauge field at the point $\ha{p}_r$. Using this, the definition \eqref{eq:fieldsfromhag} of $g_r$ and evaluating the equation \eqref{eq:LA} at $z=\ha{p}_r$, we get
	\begin{equation}
		\Lc_\pm(\ha{p}_r) = j_{r,\pm}\,, \qquad \text{ where } \qquad j_{r,\pm} = g_r^{-1} \p_\pm g_r
	\end{equation}
	is the Maurer-Cartan current of $g_r$. By equation \eqref{eq:LintermsofU}, the above condition becomes
	\begin{equation}\label{eq:RelUj}
		\sum_{s=1}^{M} r^\alpha\!\left(\ha{p}_r-\ha{z}^\pm_s\right) U^\alpha_{s,\pm} = j^\alpha_{r,\pm}\,.
	\end{equation}
	This relation holds for all $r\in\lbrace 1,\dots,M \rbrace$ and all $\alpha\in\mathbb{A}$. We thus get $M(N^2-1)$ linear equations, relating the $M(N^2-1)$ fields $\lbrace U^\alpha_{s,\pm} \rbrace$ to the $M(N^2-1)$ current components $\lbrace j^\alpha_{r,\pm} \rbrace$\footnote{Crucially, there is an equal number $M$ of fields $U_{s,\pm}$ and currents $j_{r,\pm}$: tracing back the construction of these objects, this is due to the Riemann-Hurwitz formula, which ensures that there are as many double poles in $\omega$ as there are simple zeroes of chirality $\pm$ (see the discussion around \eqref{eq:PolesAndZeroes} for details).}.
	This linear system is invertible, which follows from well-known identities for elliptic Cauchy matrices (see for instance \cite{prokofev_elliptic_2023}). This means that the currents $\lbrace U_{s,\pm} \rbrace$ are not independent degrees of freedom but rather are built from $\lbrace g_r \rbrace$: we thus think of the latter as the fundamental fields of the theory. Moreover, the Lax connection \eqref{eq:LintermsofU} is then a complicated but specific linear combination $\Lc_\pm\bigl[\lbrace g_r \rbrace\bigr]$ of the components of the Maurer-Cartan currents $j_{r,\pm} = g_r^{-1} \p_\pm g_r$.
	
	\paragraph{The  $\bm{\sigma}$-Model Action:} We have now identified the field content $\lbrace g_r \rbrace$ of the underlying 2D integrable field theory and we have constructed a Lax connection in terms of these fields. The action of this theory is the 4DCS action \eqref{eq:4DCS Action}, once reexpressed in terms of the gauge-invariant independent degrees of freedom $\lbrace g_r \rbrace$ only. We now want to write this action as an explicit integral over the 2D manifold $\Sigma$, which thus requires us to integrate over $\mathbb{T}$. To achieve this, we will follow the approach of \cite{delduc_unifying_2020}: we refer to this work for details and only summarise the main steps and results here\footnote{While the analysis presented in \cite{delduc_unifying_2020} concerned a 4DCS theory on $\Sigma\times\mathbb{CP}^1$, one finds that the same construction also works for the elliptic equivariant 4DCS theory on $\Sigma\times\mathbb{T}$ considered here.}.
	
	We first insert the parametrisation \eqref{eq:parametrisationofA} of $\mathcal{A}$ in terms of $\mathcal{L}$ and $\ha{g}$ in the action \eqref{eq:4DCS Action}. We then go to a suitable choice of gauge, called \textit{archipelago gauge}, which allows us to fix $\mathcal{L}$ and $\ha{g}$ in terms of the fields $\{g_r\}$. The integration over $\mathbb{T}$ can then be performed explicitly, with the end result taking the form of the unifying action \cite[Equation (3.7)]{delduc_unifying_2020}. In the present case, this action reads
	\begin{align}\label{eq:unif}
		S_{\text{IFT}}[\{g_r\}]=\frac{1}{2N^2}\sum_{\hat{p}} \left\lbrace \int_{\Sigma} \bigl\langle\!\bigl\langle \text{res}_{z=\hat{p}}\left(\omega\wedge \mathcal{L}\right), \, \ha{g}^{\,-1}\dd_\Sigma \ha{g}\bigr|_{z=\hat{p}} \, \bigr\rangle\!\bigr\rangle - \text{res}_{z=\hat{p}}(\omega)\; \mathcal{I}_{\text{WZ}}\bigl[ \ha{g} \bigr|_{z=\hat{p}} \bigr] \right\rbrace\,,
	\end{align} 
	where the sum runs over all the poles $\ha{p}$ of $\omega$, the Lax connection $\Lc$ should be understood as its explicit expression $\Lc[\lbrace g_r \rbrace\bigr]$ in terms of the 2D fields and, for $g:\Sigma \to \text{SL}_\C(N)$ a group-valued 2D field, $\mathcal{I}_{\text{WZ}}\bigl[ g \bigr]$ denotes the \textit{Wess-Zumino term} of $g$~\cite{Wess:1971yu,Novikov:1982ei,Witten:1983ar}. The latter is defined as an integral over a 3D manifold $\Sigma_3$ whose boundary is the 2D surface $\p\Sigma_3=\Sigma$ and in terms of an extension $g:\Sigma_3 \to \text{SL}_\C(N)$, coinciding with the initial field $g$ on the boundary $\p\Sigma_3=\Sigma$:
	\begin{align}
		\mathcal{I}_{\text{WZ}}\bigl[ g \bigr] = \frac{1}{3} \int_{\Sigma_3} \dlangle g^{-1}\dd_{\Sigma_3} g , g^{-1}\dd_{\Sigma_3} g  \wedge g^{-1}\dd_{\Sigma_3} g \drangle\,.
	\end{align}
	The 3-form which is integrated in this expression is well-known to be closed and thus locally exact on $\Sigma_3$. At least locally, one can thus rewrite this 3D integral as a 2D one, depending only on the initial 2D field $g$ on $\Sigma=\p\Sigma_3$.\footnote{This is in general only local and the full treatment of the Wess-Zumino term involves various global and topological subtleties. We will not explain those here and refer to standard textbooks or references~\cite{Wess:1971yu,Novikov:1982ei,Witten:1983ar} for details.}\\
	
	As mentioned above, the sum in equation \eqref{eq:unif} runs over all the poles $\ha{p}$ of $\omega$. These poles are given by the points $\ha{\mathbb{P}}=\bigl\lbrace \ha{p}_1,\dots,\ha{p}_M \bigr\rbrace$ together with their images under the equivariance group $\mathbb{A}_0=\mathbb{Z}_N \times \mathbb{Z}_N$. At the pole $\ha{p}_r$, the evaluation $\ha{g}|_{z=\hat{p}_r}$ yields the 2D field $g_r$. At another point $\ha{p}_r + 2\alpha\cdot\bm{\ell}$ in the same $\mathbb{A}_0$-orbit, the evaluation of $\ha{g}$ is related to $g_r$ by an adjoint action -- see equation \eqref{eq:grOrbit} -- stemming from equivariance. Similarly, the residues of $\omega \wedge \Lc$ at different poles in the orbit of $\ha{p}_r$ are related to the residue at $\ha{p}_r$ by the equivariance \eqref{eq:EquivCS} and \eqref{eq:LEquiv} of $\omega$ and $\Lc$ respectively. In the end, using the invariance of the pairing $\dlangle\cdot,\cdot\drangle$ under adjoint transformations, one finds that all the poles in the $\mathbb{A}_0$-orbit of $\ha{p}_r$ give the same contribution in the sum \eqref{eq:unif}. We are then left with a sum over one representative $\ha{p}_r$ of each orbit. This was the reason for the introduction of the factor $N^2=|\mathbb{A}_0|$ in the equivariant 4DCS action \eqref{eq:4DCS Action}: it exactly cancels the overcounting due to equivariance. In the end, we get the action
	\begin{align} 
		S_{\text{IFT}}[\{g_r\}]=\frac{1}{2}\sum_{r=1}^M \left\lbrace \int_{\Sigma} \bigl\langle\!\bigl\langle \text{res}_{z=\hat{p}_r}\left(\omega\wedge \mathcal{L}\right), \, g_r^{-1}\dd_\Sigma g_r  \, \bigr\rangle\!\bigr\rangle - \text{res}_{z=\hat{p}_r}(\omega)\; \mathcal{I}_{\text{WZ}}\bigl[ g_r \bigr] \right\rbrace\,.\label{eq:IFTactionfrom4dcs}
	\end{align}
	We now have all the ingredients needed to define an integrable 2D field theory on $\Sigma$, with spectral parameter $z\in\mathbb{T}$. This theory depends on $M$ group-valued fields $\lbrace g_r \rbrace$ and is defined by the action \eqref{eq:IFTactionfrom4dcs}. By construction, the equations of motion of $\lbrace g_r \rbrace$ obtained by varying this action can be recast as the flatness of a Lax connection $\Lc_\pm[\lbrace g_r \rbrace\bigr]$ depending meromorphically on $z\in\mathbb{T}$. Moreover, we expect that the spatial component of this Lax connection satisfies a Maillet bracket with Belavin's $\Rc$-matrix as seed and a twist function determined by the choice of 1-form $\omega=\vp(z)\dd z$. This holds for the model with $M=1$, which, as we shall show in the next subsection, coincides with the elliptic integrable DPCM of section \ref{sec:HigherRank}. For more general cases, this conjecture is an elliptic generalisation of the work~\cite{vicedo_4d_2021}, which establishes that the rational integrable models obtained from 4DCS theory on $\Sigma \times \mathbb{CP}^1$ obey a Maillet bracket with twist function $\vp(z)$ and the Yangian $\Rc$-matrix as seed. \\
	
	Let us finally comment on the more precise nature and form of the elliptic integrable field theory obtained above. Recall from the discussion around equation \eqref{eq:RelUj} that the Lax connection $\Lc_\pm[\lbrace g_r \rbrace\bigr]$ is built linearly from the components of the Maurer-Cartan currents $j_{r,\pm}=g_r^{-1}\p_\pm g_r$. It is thus clear that the action \eqref{eq:IFTactionfrom4dcs} can eventually be brought to the form
	\begin{equation}
		S_{\text{IFT}}[\{g_r\}] = \int_{\Sigma}\dd x^+\dd x^- \sum_{r,s=1}^M D_{\alpha\beta}^{(rs)}\, j^\alpha_{r,+}\, j^\beta_{s,-} - \sum_{r=1}^M \text{res}_{z=\hat{p}_r}(\omega)\; \mathcal{I}_{\text{WZ}}\bigl[ g_r \bigr] \,,
	\end{equation}
	for some coefficients $D_{\alpha\beta}^{(rs)}$, built as complicated combinations of the parameters contained in $\omega$. This is exactly the action of a (complexified) $\sigma$-model with target space $\text{SL}_\C(N)^M$, built as $M$ copies of the group. Writing down the explicit metric and B-field of this $\sigma$-model would require the determination of the coefficients $D_{\alpha\beta}^{(rs)}$, which in turn would require solving the linear system \eqref{eq:RelUj}. We will perform this in the next subsection for the case $M=1$ and will show that the resulting $\sigma$-model is the elliptic integrable DPCM on $\text{SL}_\C(N)$ described in Section \ref{sec:HigherRank}. The model for arbitrary $M$ can be seen as a coupled version of this DPCM which preserves integrability: its explicit description is however beyond the scope of this paper.
	
	\subsection{Constructing the DPCM from 4DCS}
	
	Having presented the general approach, we are now ready to understand the construction of the $\text{SL}_\mathbb{C}(N)$-DPCM presented in section \ref{sec:HigherRank}, which corresponds to specialising to $M=1$ in the discussion above. The key ingredient is the choice of 1-form $\omega$:
	\begin{align}
		\omega_{\text{DPCM}}=\rho\,\big\{\wp(z)-\wp(\ha{z})\big\}\,\dd z\,,
	\end{align}
	which corresponds to the twist function \eqref{eq:higherrnaktwist} of the DPCM. Up to translation of $z$, this is the most general meromorphic 1-form on $\C/(2\ell_1\mathbb{Z}+2\ell_2\mathbb{Z})$ with only one double pole and two simple zeroes. Seen as a 1-form on $\mathbb{T}=\C/\Lambda$, $\Lambda=2N\ell_1\mathbb{Z}+2N\ell_2\mathbb{Z}$, $\omega$ then has $N^2$ poles, forming a single orbit of the equivariance group $\mathbb{Z}_N\times\mathbb{Z}_N$: we will choose its representative to be $\ha{p}_1=0$. Similarly, there are two orbit representatives of the zeroes, which we choose to be $\ha{z}^\pm_1=\pm\ha{z}$. This allows us to use the methods of the previous subsection to construct an integrable $\sigma$-model based on this data.
	
	\paragraph{Field Content and Lax Connection:} This model is described by a single $\text{SL}_{\mathbb{C}}(N)$-valued field $g\equiv\ha{g}(0)$ (to ease the notation, given $M=1$, we drop the index $r\in\lbrace 1,\dots,M\rbrace$ used in the previous subsection). The Lax connection \eqref{eq:LintermsofU} simply becomes
	\begin{align}
		\mathcal{L}_\pm(z)=\sum_{\alpha\in\mathbb{A}} r^\alpha(z\mp\ha{z})\,U^\alpha_\pm\, T_\alpha\,,
	\end{align}
	where similarly we dropped the subscript of the current $U_\pm$ for simplicity.
	To express $U_\pm^\alpha$ in terms of $g$, we use the relation \eqref{eq:RelUj}, which comes from the boundary condition of the gauge field. In the present case, it simply reads
	\begin{equation}
		r^\alpha( \mp \ha{z} ) U_\pm^\alpha = j_\pm^\alpha\,,
	\end{equation}
	where $j_\pm = g^{-1}\p_\pm g$. This is trivially solved for $U_\pm^\alpha$ and the Lax connection then becomes
	\begin{align}\label{eq:LaxDPCM4d}
		\mathcal{L}_\pm (z)=\sum_{\alpha\in\mathbb{A}}\frac{r^\alpha(z\mp\ha{z})}{r^\alpha(\mp \ha{z})}\,j^\alpha_\pm \,T_\alpha\,.
	\end{align}
	We recognise here the connection \eqref{eq:HighRankLax} introduced in Section \ref{sec:HigherRank}.
	
	\paragraph{Action:} The last thing we are missing is the action, for which we want to recover \eqref{eq:DPCMActionHighRank}--\eqref{eq:EllipticDeformationsHighRank}. This is obtained by specialising the general action \eqref{eq:IFTactionfrom4dcs} to $M=1$. One easily checks that $\omega$ has no residues, corresponding to the absence of a Wess-Zumino term in our DPCM action \eqref{eq:DPCMActionHighRank}. We are thus left with
	\begin{equation}
		S_{\text{DPCM}}[g] = \frac{1}{2} \int_{\Sigma} \bigl\langle\!\bigl\langle \text{res}_{z=0}\left(\omega\wedge \mathcal{L}\right), \, g^{-1}\dd_\Sigma g  \, \bigr\rangle\!\bigr\rangle\,.
	\end{equation}
	Using $\omega=\left( \dfrac{\rho}{z^2} + O(z^0) \right)\dd z$, we get
	\begin{equation}
		\text{res}_{z=0}\left(\omega\wedge \mathcal{L}_\pm\right) = \rho\,\Lc'_\pm(z=0) = \rho \sum_{\alpha\in\mathbb{A}} \frac{r^\alpha\null'(\mp \ha{z})}{r^\alpha(\mp \ha{z})} j_\pm^\alpha\,T_\alpha\,,
	\end{equation}
	where in the last equality we used \eqref{eq:LaxDPCM4d}. Finally, using the fact that $\left\langle T_\alpha,T_\beta\right\rangle$ is non-zero only for $\alpha=-\beta$ as well as the identity $r^{-\beta}(z)=-r^\beta(-z)$, we obtain
	\begin{equation}
		S_{\text{DPCM}}[g] = -\int_{\Sigma} \dd x^+ \, \dd x^-\,\sum_{\alpha,\beta\in\mathbb{A}}\,\rho\,\frac{r^\beta\null'(\ha{z})}{r^\beta(\ha{z})}\, \langle T_\alpha, T_\beta \rangle\,j_+^\alpha\,j_-^\beta\,.
	\end{equation}
	This coincides with the DPCM action \eqref{eq:DPCMActionHighRank}, providing we define the deformation operator $D$ by
	\begin{equation}
		D[T_\beta] = - \rho\,\frac{r^\beta\null'(\ha{z})}{r^\beta(\ha{z})}\,T_\beta\,,
	\end{equation}
	in agreement with equation \eqref{eq:EllipticDeformationsHighRank}. This ends the construction of the elliptic integrable DPCM on $\text{SL}_\C(N)$ from equivariant 4DCS theory.
	
	\section{Conclusion and Perspectives}
	\label{sec:conclusion}
	
	In this article, we have constructed a new integrable deformation of the PCM on $\text{SL}_\R(N)$ based on an elliptic spectral parameter, generalising a result of Cherednik~\cite{cherednik_relativistically_1981} for $N=2$. We have exhibited the action, Lax connection and twist function of this integrable $\sigma$-model and explained how this model was constructed from an equivariant 4D Chern-Simons theory on the torus. We conclude this paper by discussing various perspectives and potential directions for future developments.
	
	\paragraph{Panorama of Elliptic Integrable $\bm\sigma$-Models:} The deformed PCM constructed in this paper is the simplest model that can be obtained from equivariant 4D Chern-Simons theory on the torus. Indeed, it corresponds to a particularly simple choice of twist function, with one double pole and two simple zeroes. We expect that a rich panorama of elliptic integrable $\sigma$-models can be obtained by starting with more complicated twist functions. Before discussing this in more detail, let us mention that in parallel to the Lagrangian approach provided by 4D Chern-Simons theory, these models can also be constructed from a Hamiltonian perspective, using current Poisson algebras, based on appropriate generalisations of affine Gaudin models~\cite{feigin_quantization_2009,vicedo_integrable_2019,delduc_assembling_2019}: this will be the focus of an upcoming paper~\cite{ToAppear:Gaudin}. We further expect these theories to also be related to the Affine Higgs Bundles construction studied in~\cite{levin_hitchin_2003,levin_2d_2022}: it would be interesting to explore this in more detail\footnote{Most of the examples treated in~\cite{levin_hitchin_2003,levin_2d_2022}, see also~\cite{Zotov:2010kb}, are ultralocal theories, corresponding in the present language to constant twist functions. In contrast, the integrable $\sigma$-models discussed in this paper are inherently non-ultralocal and are described by non-constant twist functions $\varphi(z)$: we expect these models to be related to Affine Higgs Bundles with non-vanishing central charges, with the twist function $\varphi(z)$ corresponding to a non-constant choice for $k(z)$ in~\cite{levin_hitchin_2003} (or $\nu(z)$ in~\cite{levin_2d_2022}).}.

	The first natural generalisation of the elliptic DPCM consists of its integrable deformations, which can be obtained by splitting the double pole of its twist function into a pair of simple poles with opposite residues. From the point of view of 4D Chern-Simons theory, this should be accompanied by a modification of the boundary conditions at these poles, which can be chosen to be of Yang-Baxter-type or of $\lambda$-type, following~\cite{delduc_unifying_2020}. We expect these two choices to produce elliptic generalisations of the \textit{bi-Yang-Baxter model}~\cite{klimcik_integrability_2009, klimcik_integrability_2014} and the \textit{generalised $\lambda$-model}~\cite{sfetsos_generalised_2015} respectively. In the case where the underlying Lie group is SU(2), it would be interesting to compare the resulting theory with the anisotropic $\lambda$-model introduced in~\cite{sfetsos_anisotropic_2015}.
	
	One is also free to start with a twist function possessing $M$ double poles rather than just one, as was the case in the first part of Section \ref{sec:4dcs}. We expect this setup to lead to an elliptic generalisation of the integrable coupled model of~\cite{delduc_assembling_2019}, whose target space is formed by $M$ copies of the group $\text{SL}_\R(N)$. Moreover, this theory can be further deformed while preserving its integrability by splitting the double poles into pairs of simple poles.
	
	Finally, one could consider elliptic integrable $\sigma$-models based on arbitrary elliptic twist functions, with any pole structure and appropriate boundary conditions (adapting the rational construction of~\cite{benini_homotopical_2022}). These should form elliptic generalisations of the integrable $\mathcal{E}$-models considered in~\cite{lacroix_integrable_2021}. We expect these theories to be related by various Poisson-Lie T-dualities~\cite{Klimcik:1995ux,Klimcik:1995dy}.\\

    Another obvious direction to generalise the model presented in this paper would be to consider other choices of complex Lie group than $\text{SL}_\mathbb{C}(N)$. However, it is clear that such a theory would have to be quite different from the elliptic DPCM. In particular, recall that the Lax connection of the DPCM satisifies a Maillet bracket \eqref{eq:Maillet} with an $\mathcal{R}$-matrix \eqref{eq:RTwist}, stemming from a skew-symmetric seed $\Rc^0(z)$. As noted in subsection \ref{subsec:int}, Belavin and Drinfel'd proved in \cite{belavin_solutions_1982} that an elliptic, skew-symmetric, non-dynamical seed $\mc{R}$-matrix  exists only for the complex algebras $\mathfrak{sl}_\mathbb{C}(N)$. Similarly, from the point of view of the equivariant 4D Chern-Simons theory, we expect this question to be related with the existence of a field $\hat{g}$ (or alternatively, the rigidity property of the underlying elliptic bundle), which was a crucial step in generating an elliptic integrable $\sigma$-model from the 4D setup (see subsection \ref{subsec:paramA} for more details). This result was directly tied to the $\Z_N \times \Z_N$ equivariance condition imposed on the theory, which is only possible for the gauge group $\text{SL}_\mathbb{C}(N)$. One would then have to get around these restrictions to generate an integrable $\sigma$-model related to a different Lie group: for instance, using a dynamical or a non-skew-symmetric seed $\mc{R}$-matrix. The latter possibility is also naturally related to generating integrable $\sigma$-models whose target spaces are \textit{$\Z_T$-cosets} of Lie groups (which include for instance spheres and anti-de Sitter spaces) \cite{Young:2005jv,Ke:2011zzb,vicedo_integrable_2019}. It is not clear whether the approach followed in the present paper could be adapted to produce elliptic integrable $\sigma$-models on cosets. We note however that theories of this type on certain specific cosets can alternatively be constructed from another approach, using so-called $\beta\gamma$ order defects in 4D Chern-Simons theory (see for instance~\cite{costello_gauge_2019, bykov_sigma_2021, bykov_cpn-1-model_2022}).
	
	\paragraph{Elliptic Symmetries:} The undeformed PCM on a group $G$ has an isometry group $G_L \times G_R$, acting by left and right multiplication on the field. The elliptic integrable deformation constructed in the present paper for $G=\text{SL}_\R(N)$ breaks the right isometry $G_R$ while keeping the left-one $G_L$. In the trigonometric limit, we have argued in Subsection \ref{subsec:MainYB} that the model takes the form of a Yang-Baxter deformation. It is well known that such a deformation in fact does not fully break the right-isometry $G_R$ but rather deform it to a Poisson-Lie symmetry~\cite{kawaguchi_classical_2012,delduc_classical_2013}, \textit{i.e.} the semi-classical version of a quantum group. Moreover, this deformed symmetry belongs to a larger infinite algebraic structure, extracted from the monodromy of the Lax matrix and forming a semi-classical $q$-deformed affine algebra~\cite{kawaguchi_classical_2012,delduc_affine_2017}. It would be interesting to generalise this analysis to the elliptic DPCM and to determine its algebra of symmetries, which could for instance take the form of a semi-classical elliptic quantum group. A natural extension of this programme is the study of the symmetries of the more general elliptic integrable $\sigma$-models described in the previous paragraph.
	
	\paragraph{Renormalisation:} In this paper, we have focused on the classical aspects of the integrable elliptic DPCM. It would be interesting to also explore its quantum properties. A first step in this direction would be to study its renormalisation and determine the RG-flow of its parameters. The realisation of this programme at 1-loop, \textit{i.e.} at the first order in the quantum expansion, will be the subject of an upcoming publication~\cite{ToAppear:RG}.
	
	The renormalisation of integrable $\sigma$-models has attracted a lot of attention in the past few years. In particular, various results and conjectures have been proposed on the path towards a universal formulation of the 1-loop RG-flow of these models. For instance, it has been conjectured in~\cite{delduc_rg_2021} that the 1-loop renormalisation of rational integrable $\sigma$-models can be recast in a very simple way as a flow of their twist function, or equivalently of the 1-form $\omega$ appearing in the 4D Chern-Simons theory. This was proven for a large class of such models in the recent works~\cite{hassler_rg_2021,Hassler:2023xwn}. In parallel, a conjecture of Costello, reported in~\cite{derryberry_lax_2021}, states that this 1-loop RG-flow can be written compactly in terms of certain well-chosen integrals of the 1-form $\omega$: this conjecture is of a more geometrical nature and also applies to models based on higher-genus Riemann surfaces, including elliptic ones. In our future work~\cite{ToAppear:RG}, we will extend the conjecture of~\cite{delduc_rg_2021} to the elliptic case, detail its relation to the one of~\cite{derryberry_lax_2021}, and check that both of these conjectures hold true for the elliptic integrable DPCM considered in the present article. We expect these results to extend to all integrable $\sigma$-models that have the elliptic Belavin $\Rc$-matrix as seed. Some of the methods recently developed for the renormalisation of integrable $\sigma$-models might offer natural starting points to explore these more general cases, including for instance the ``universal-divergences'' approach of~\cite{levine_universal_2023,Levine:2023wvt} or the $\mathcal{E}$-models~\cite{Klimcik:1995ux,Klimcik:1995dy} techniques of~\cite{Valent:2009nv,sfetsos_renormalization_2010,hassler_rg_2021,Hassler:2023xwn}.

 \paragraph{Quantum integrability and the problem of non-ultralocality:}
	In addition to their renormalisation, it would be interesting to study other quantum aspects of elliptic integrable $\sigma$-models, such as their quantum integrability, their S-matrices and their spectra.\footnote{In the case of rank one, an approach to the quantisation of the elliptic Cherednik model was suggested in~\cite{Faddeev:1985qu}. Moreover, it was conjectured in~\cite{Appadu:2017bnv} that the scattering of the anisotropic SU(2) $\lambda$-model~\cite{sfetsos_anisotropic_2015} is governed by the Zamolodchikov elliptic S-matrix of~\cite{Zamolodchikov:1979ba} (as mentioned earlier in this conclusion, we expect this theory to belong to the class of elliptic $\sigma$-models obtained from equivariant 4D Chern-Simons on the torus).} We note that these $\sigma$-models will generally define non-unitary quantum field theories since their target space is built from the split group $\text{SL}_\R(N)$ and is thus non-Euclidean.

 The main obstacle in establishing the quantum integrability of these models is the so-called \textit{problem of non-ultralocality}, which we now briefly discuss. This problem originates from the presence of derivatives $\partial_x\delta(x-y)$ of the Dirac distribution in the Maillet bracket \eqref{eq:Maillet} obeyed by the Lax matrix $\Lc_x(z)$. When the $\Rc$-matrix is skew-symmetric, \textit{i.e.} when $\Rc(z_1,z_2)_{\tn{12}} = - \Rc(z_2,z_1)_{\tn{21}}$, the term containing $\partial_x\delta(x-y)$ in this bracket vanishes and the theory is said to be ultralocal. In that case, one can derive the Poisson algebra of the monodromy matrix of $\Lc_x(z)$ starting from this bracket and eventually quantise it\footnote{In practice this is often done by discretisation, \textit{i.e.} by putting the theory on a lattice.}, within the framework of the Quantum Inverse Scattering Method~\cite{Sklyanin:1979pfu}. This leads to the Yang-Baxter algebra
 \begin{eqnarray}\label{Eq:YBAlg}
     \mathsf{R} (z_1,z_2)_{\tn{12}}\, \mathsf{M}(z_1)_{\tn{1}}\, \mathsf{M}(z_2)_{\tn{2}} = \mathsf{M}(z_2)_{\tn{2}}\,  \mathsf{M}(z_1)_{\tn{1}} \, \mathsf{R} (z_1,z_2)_{\tn{12}}\,,
 \end{eqnarray}
 describing the commutation relations of the quantum monodromy matrix $\mathsf{M}(z)$ in terms of a quantum R-matrix $\mathsf{R} (z_1,z_2)_{\tn{12}}$.  These relations imply that the traces $\Tr\bigl( \mathsf{M}(z_1) \bigr)$ and $\Tr\bigl( \mathsf{M}(z_2) \bigr)$ commute, thus ensuring the quantum integrability of the theory. Moreover, they can be used to derive the spectrum of the model through the algebraic Bethe ansatz.

 In the case of a non-ultralocal theory, \textit{i.e.} when the $\Rc$-matrix is not skew-symmetric and the Maillet bracket thus contains terms proportional to $\partial_x\delta(x-y)$, the computation of the Poisson algebra of the monodromy is faced with ambiguities and the theory thus naively cannot be quantised within the Quantum Inverse Scattering Method. This is what is referred to as the problem of non-ultralocality~\cite{maillet_kac-moody_1985,maillet_new_1986}. Several approaches have been explored to circumvent this difficulty, although no general fully-consistent solution has been found so far. For instance, in the case where the symmetric part of the $\Rc$-matrix takes a specific form, the works~\cite{Freidel:1991jx,Freidel:1991jv} propose a certain prescription to resolve the ambiguities arising from non-ultralocality, eventually leading to commutation relations for the quantum monodromy $\mathsf{M}(z)$ in the form of the so-called Freidel-Maillet algebra. The latter takes a different, more complicated, form than the Yang-Baxter algebra \eqref{Eq:YBAlg}, but also defines a consistent integrable quantisation, for which $\Tr\bigl( \mathsf{M}(z_1) \bigr)$ and $\Tr\bigl( \mathsf{M}(z_2) \bigr)$ commute. However, the conditions on the $\Rc$-matrix that lead to this algebra are generally not satisfied for integrable $\sigma$-models.

 One thus has to find other approaches to try and solve the non-ultralocality problem for $\sigma$-models. For very specific examples, it turns out that a well-chosen gauge transformation of the Lax matrix leads to an ultralocal Poisson bracket~\cite{Bytsko:1994ae,Brodbeck:1999ib,Bazhanov:2017nzh,Delduc:2019lpe}, thus paving the way for applying the Quantum Inverse Scattering Method. However, we expect that the elliptic integrable $\sigma$-models considered in this paper do not admit such an ultralocal gauge. More recently, it was suggested in the work~\cite{Bazhanov:2018xzh} of Bazhanov, Kotousov and Lukyanov that despite its non-ultralocality, the Maillet bracket might sometimes still be consistent with having the Yang-Baxter algebra \eqref{Eq:YBAlg} in the quantum theory. This was proposed for a very specific integrable $\sigma$-model, the UV conformal point of the Bi-Yang-Baxter (or Klim\v{c}\'{i}k) model, and was further developped in~\cite{Kotousov:2022azm}. As mentioned in the first paragraph of this conclusion, we expect this Bi-Yang-Baxter model to admit an elliptic generalisation when the underlying Lie group is SU(2) or SL$_\R(N)$, which can alternatively be seen as a further integrable deformation of the elliptic PCM built in this paper. It is thus natural to wonder whether the quantisation of such a model can be approached using the proposal of~\cite{Bazhanov:2018xzh}. We stress however that the latter concerns the UV conformal point of the Bi-Yang-Baxter model: we expect that the UV limit of its elliptic generalisation sends the imaginary period of the torus to $i \infty$, thus degenerating the theory from elliptic to trigonometric. Therefore, we believe that the UV fixed-point of this elliptic deformation in fact coincides with the one of the Bi-Yang-Baxter model itself and thus fits into the framework of~\cite{Bazhanov:2018xzh,Kotousov:2022azm}. It would be interesting to study these aspects in more detail and for more general elliptic integrable $\sigma$-models.

 Finally, let us mention an alternative approach to circumvent the problem of non-ultralocality, proposed in~\cite{vicedo_integrable_2019} and based on the formalism of affine Gaudin models~\cite{feigin_quantization_2009}. Its starting point is the fact that classical affine Gaudin models possess an infinite number of Poisson-commuting higher-spin local charges~\cite{Evans:1999mj,Lacroix:2017isl}, which crucially are not extracted from the monodromy of their Lax matrix. The construction of these charges thus does not suffer from the ambiguities caused by the non-ultralocality problem. In particular, various progresses~\cite{feigin_quantization_2009,Frenkel:2016gxg,Lacroix:2018fhf,Lacroix:2018itd,Gaiotto:2020dhf,Kotousov:2021vih,Franzini:2022duf,Kotousov:2022azm} were made towards a quantisation of these charges into commuting local operators for conformal/chiral affine Gaudin models. We expect that the classical construction of local Poisson-commuting charges proposed in~\cite{Evans:1999mj,Lacroix:2017isl} will also apply to the elliptic integrable $\sigma$-models considered in the present paper and it would then be a natural perspective to study their quantisation.
	
	\paragraph{Geometry of the Spectral Parameter:} We end this conclusion with some general comments on the role of the spectral parameter and its geometry in the study of integrable $\sigma$-models. In principle, standard 4D Chern-Simons theory with disorder defects~\cite{costello_gauge_2019} allows for the construction of such models starting from the following geometric data: the choice of a Riemann surface $\ha{\C}$ in which the spectral parameter $z$ is valued and the choice of a meromorphic 1-form $\omega=\vp(z)\dd z$ on $\ha{\C}$. In the mathematical terminology, this data $(\ha{\C},\omega)$ corresponds to a point in the moduli space of abelian differentials (of the third kind). The recipe to build an integrable $\sigma$-model from the choice of $(\ha{\C},\omega)$ is now very well understood in the case where $\ha{\C}$ is the Riemann sphere $\mathbb{CP}^1$ (see for instance~\cite{costello_gauge_2019,delduc_unifying_2020,benini_homotopical_2022,lacroix_integrable_2021}). For $\ha{\C}$ of higher genus, it was discussed in~\cite[Section 15]{costello_gauge_2019} and \cite{derryberry_lax_2021}, at least for the cases where $\omega$ has only double poles. Formally, this construction characterises the target space of the $\sigma$-model as an appropriate moduli space of holomorphic bundles and constructs the metric and torsion in terms of a geometric object called the Szeg\"o kernel. However, in practice, it is difficult to obtain an explicit expression for these quantities when $\ha{\C}$ is of genus one or higher (see for instance the discussion in~\cite[Section 15]{costello_gauge_2019}).
	
	The elliptic integrable $\sigma$-models considered in the present paper are based on the same type of geometric data $(\ha{\C},\omega)$, with $\ha{\C}$ a genus-one Riemann surface, \textit{i.e.} a torus. However, they differ slightly from the models of~\cite{costello_gauge_2019,derryberry_lax_2021} as they are built from a 4D Chern-Simons theory with additional equivariance properties. This seemingly minor technical difference in fact has an important impact on the study of these models and in the end allows for a reasonably explicit construction of their target space, metric and B-field\footnote{Technically, this is due to the fact that in the equivariant 4D Chern-Simons theory, the $\ol{z}$-component of the gauge field can be written in the form $A_{\bar{z}} = - (\p_{\bar{z}} \hat{g})\hat{g}^{-1}$ (see Subsection \ref{subsec:paramA} for details). This is not the case in the setup of ~\cite[Section 15]{costello_gauge_2019} and \cite{derryberry_lax_2021}, making the analysis of the model more involved.}. We expect that this construction can be rephrased in terms of appropriately defined equivariant holomorphic bundles and Szeg\"o kernels, in the spirit of \cite{costello_gauge_2019,derryberry_lax_2021}. It would be interesting to investigate this geometric formulation in more detail. Moreover, one might wonder whether similar simplifications due to equivariance also exist for higher-genus surfaces.
	
	Let us also add a brief comment on the modular invariance of elliptic integrable $\sigma$-models obtained from the equivariant 4D Chern-Simons theory. A natural parametrisation of the geometric data defining such a model is given by the choice of the torus modulus $\tau$, together with the additional moduli describing the choice of the meromorphic 1-form $\omega$ on $\C/(\mathbb{Z}+\tau\mathbb{Z})$. For instance, in the simple case of the elliptic integrable DPCM, corresponding to $\omega$ having only a double pole (fixed at $z=0$ by translation invariance), we parametrised these additional moduli as $(\rho,\ha{z})$, describing an overall factor $\rho$ and the position of a zero $\ha{z}$ -- see Subsection \ref{sec:DPCM-HR}. We have argued around equation \eqref{eq:ModInv} that the theory is invariant under the modular group $\text{SL}(2,\Z)$, having the standard action by Möbius transformations on the torus modulus $\tau$ but also acting non-trivially on the remaining moduli $(\rho,\ha{z})$ and on the field of the theory. It would be interesting to study these properties of modular invariance for more general elliptic integrable $\sigma$-models and in particular understand how the modular group acts on the moduli space of abelian differentials (meromorphic 1-forms) with more complicated analytical structure than the one of the DPCM.
	
	To conclude, let us note that most of the discussion above concerned the construction of integrable $\sigma$-models at the classical level. It would be interesting to also explore whether other aspects of these models, such as their classical solutions and their quantisation, can be studied using the geometric language sketched above. As a first illustration, we note that the aforementioned conjecture of Costello in~\cite{derryberry_lax_2021} implies that an appropriate choice of parameterisation of the moduli space of abelian differentials (using the so-called absolute and relative periods of $(\ha{\C},\omega)$) allows to trivialise the 1-loop RG-flow of these theories. This suggests that the geometry of the spectral parameter also plays an important role in their quantisation.\vspace{-4pt}
	
	\section*{Acknowledgements} 
	
	We would like to thank Niklas Beisert, Johannes Broedel, Lewis Cole, Kevin Costello, Ryan Cullinan, Ben Hoare, Rob Klabbers, Shota Komatsu, Gleb Kotousov,  Nat Levine, Joaquin Liniado, Marc Magro, Ana Retore, Fiona Seibold, Konstantinos Siampos, Daniel Thompson and Beno\^it Vicedo for useful and interesting discussions. Moreover, we are grateful to Nat Levine and Beno\^it Vicedo for valuable comments on the draft. This work is partly based on the Master thesis of one of the authors (A.W.) prepared at ETH Z\"urich. We thank Niklas Beisert for acting as co-advisor for this thesis and for helpful interactions and comments during its preparation. The work of S.L. is supported by Dr. Max R\"ossler, the Walter Haefner Foundation and the ETH Z\"urich Foundation.
	
	\appendix
	
	\section{Elliptic Functions}\label{sec:EllipticAppendix}
	This appendix is a complement to the treatment of elliptic functions in the main text. We will expand on some of the properties of the Jacobi and Weierstrass families of elliptic functions and in particular show how to relate the two. This last point will allow us to identify the integrable elliptic $\text{SL}_\C(N)$-DPCM for $N=2$ with the Cherednik model.
	
	\subsection{The Jacobi Family of Elliptic Functions}\label{app:Jacobi}
	This family consists of $12$ functions, although they are not all independent. Every member is of the form $\text{ab}(z;m)$, where the role of $\text{a},\text{b}\in\{\text{s},\text{c},\text{d},\text{n}\}$ will be specified shortly and $m$ controls the periodicity of the functions. Specifically, the entire family of Jacobi functions is doubly-periodic with periods $4K(m)$ and $4iK'(m)$, where
	\begin{equation}
		K(m)=\int_0^1 \frac{\dd t}{\sqrt{(1-t^2)(1-mt^2)}} \qquad \text{ and } \qquad K'(m)=K(1-m)
	\end{equation}
	are the \textit{complete elliptic integrals of the first kind}. The letters a and b in the notation $\text{ab}(z;m)$ are distinct elements of $\{\text{s},\text{c},\text{d},\text{n}\}$ and are related to the location of the poles and zeroes of this function. More precisely, the first letter specifies a lattice $\Gamma_{\text{a}}$ of simple zeroes, while the second letter specifies a lattice $\Gamma_{\text{b}}$ of simple poles, in the following pattern:
	\begin{subequations}
		\begin{eqnarray}
			&\Gamma_{\text{s}} =\left\{2nK+2in'K'|n,n'\in\mathbb{Z}\right\}\,, \\[1pt]
			&\Gamma_{\text{c}} =\left\{K+2nK+2in'K'|n,n'\in\mathbb{Z}\right\}\,, \\[1pt]
			&\Gamma_{\text{d}} =\left\{iK'+2nK+2in'K'|n,n'\in\mathbb{Z}\right\}\,, \\[1pt]
			&\Gamma_{\text{n}} =\left\{K+iK'+2nK+2in'K'|n,n'\in\mathbb{Z}\right\}\,.
		\end{eqnarray}
	\end{subequations}
	Indeed, $\ab(z;m)$ is the unique meromorphic function that is doubly-periodic with periods $4K(m)$ and $4iK'(m)$ and has simple zeros at $z\in\Gamma_{\text{a}}$ and simple poles at $z\in\Gamma_{\text{b}}$. It then follows that
	\begin{equation}
		\text{ba}(z;m)^{-1}=\text{ab}(z;m)\,,\qquad \text{ac}(z;m)=\text{ab}(z;m)\,\text{bc}(z;m)\,,
	\end{equation}
	implying that one can reconstruct any of the 12 Jacobi elliptic functions using just three of them. Here, we will take these ``generators'' to be $\text{sc}(z;m)$, $\sd(z;m)$ and $\sn(z;m)$.
	
	On top of their periodicity, the Jacobi functions also have simple behaviours under shifts by the half-periods $2K(m)$ and $2iK'(m)$. In particular, the generators $\{\text{sc},\sd,\sn\}$ satisfy
	\begin{subequations}
		\begin{eqnarray}
			&\text{sc}(z+2K(m);m)=\text{sc}(z;m)\,,\qquad \text{sc}(z+2iK'(m);m)=-\text{sc}(z;m)\,,\\[1pt]&
			\text{sd}(z+2K(m);m)=-\text{sd}(z;m)
			\,,\qquad \text{sd}(z+2iK'(m);m)=-\text{sd}(z;m)\,,\\[1pt]&
			\text{sn}(z+2K(m);m)=-\text{sn}(z;m)\,,\qquad \text{sn}(z+2iK'(m);m)=\text{sn}(z;m)\,.
		\end{eqnarray}
	\end{subequations}
	These are the analogues of \eqref{eq:equivralpha} for the $\text{SU}(2)$ theory considered in section \ref{sec:su2}. The trigonometric limits of these three functions, corresponding to $m\ra 1$ or $m\ra 0$, are also simple, being given by
	\begin{subequations}
		\begin{eqnarray}
			&\text{sc}(z;1)=\sinh(z)\,,\qquad \sd(z;1)=\sinh(z)\,,\qquad \sn(z;1)=\tanh(z)\,,\\[1pt]
			&\text{sc}(z;0)=\tan(z)\,,\qquad \sd(z;0)=\sin(z)\,,\qquad \sn(z;0)=\sin(z)\,.
		\end{eqnarray}
	\end{subequations}
	Considering the right ratios and products of these, one can easily show that the $\text{SU}(2)$ elliptic DPCM \eqref{eq:EllipticD} correctly reproduces the trigonometric DPCM \eqref{eq:TrigD} in the limit $m\ra 1$. 
	
	\subsection{The Weierstrass Family of Elliptic Functions}\label{subsec:ralphafamily}
	
	The definition and the main properties of the Weierstrass functions have already been given in the subsection \ref{subsec:Belavin} of the main text. In this appendix, we will mainly concern ourselves with the proof of the two key features of the functions $\lbrace r^\alpha(z) \rbrace_{\alpha\in\mathbb{A}}$, namely their behaviour \eqref{eq:equivralpha} under shifts by $2\ell_i$ and their $\mathbb{A}$-compatibility \eqref{eq:rcompatible}.
	
	\paragraph{Property Under Shifts:} This relies on the definition \eqref{eq:defofralpha} of the family $\{r^\alpha(z)\}_{\alpha\in\mathbb{A}}$ in terms of Weierstrass $\sigma$-functions and the constants $q_\alpha$ and $Q_\alpha$ introduced in \eqref{eq:qalphadef}. Using the quasi-ellipticity properties \eqref{eq:zetasigmatransform} of $\sigma$, one finds
	\begin{equation*}
		r^\alpha(z+2\ell_i)=\exp(-2Q_\alpha\ell_i)\frac{\exp(2L_i\left[z+q_\alpha+\ell_i\right])}{\exp(2L_i\left[z+\ell_i\right])}r^\alpha(z)=\exp(\frac{4}{N}\left[L_i\,\alpha\times\bm{\ell}-\ell_i\,\alpha\times\bm{L}\right])r^\alpha(z)\,,
	\end{equation*}
	where we recall the definition of the cross-product in \eqref{eq:crossproduct}.
	One then notices that the square bracket can equivalently be written as $\alpha_i\,\bm{L}\times\bm{\ell}$. However, this expression simplifies using \textit{Legendre's relation}, which states that $\bm{L}\times\bm{\ell}=i\pi/2$. The above equation then becomes
	\begin{equation}
		r^\alpha(z+2\ell_i)=\exp(\alpha_i\frac{2i\pi}{N})r^\alpha(z)=\xi^{\alpha_i}r^\alpha(z)
	\end{equation}
	where $\xi=\exp(2 i\pi/N)$, proving the desired property \eqref{eq:equivralpha}.
	
	\paragraph{$\mathbb{A}$-Compatibility:}
	This property relies on the so-called \textit{Fay's identity} \cite{fay_theta_1973}, which takes the following form when expressed in terms of $\sigma$-functions\footnote{Fay's identity is usually written in terms of $\theta$-functions, but one can readily convert between the two.}:
	\begin{equation}
		\sigma(q)\sigma(q+w_1+w_2)=\sigma(q+w_1)\sigma(q+w_2)X(w_1,w_2;w_3,w_4)+\sigma(q+w_3)\sigma(q+w_4)X(w_3,w_4;w_1,w_2)\,,\label{eq:Fay}
	\end{equation}
	where the $w_i$'s satisfy $w_1+w_2=w_3+w_4$ and we have introduced the \textit{cross-ratio function} \cite{poor_fays_1992}
	\begin{equation}
		X(w_1,w_2;w_3,w_4)=\frac{\sigma(w_3)\sigma(-w_4)}{\sigma(w_3-w_1)\sigma(w_1-w_4)}\,.
	\end{equation}
	To match Fay's identity with \eqref{eq:rcompatible}, we identify
	\begin{equation*}
		q=z_1+q_\alpha\,,\qquad w_1=q_\beta\,,\qquad w_2=z_2-z_1-q_\alpha\,,\qquad w_3=-q_\alpha\,,\qquad w_4=z_2-z_1+q_\beta\,.
	\end{equation*}
	Inserting these identifications in \eqref{eq:Fay} and expanding the cross-ratio functions, we find
	\begin{align}
		\sigma(z_1+q_\alpha)\sigma(z_2+q_\beta) &=\sigma(z_1+q_\alpha+q_\beta)\sigma(z_2)\,\frac{\sigma(q_\alpha)\sigma(z_2-z_1+q_\beta)}{\sigma(q_\alpha+q_\beta)\sigma(z_2-z_1)} \\
		& \hspace{30pt} +\sigma(z_1)\sigma(z_2+q_\alpha+q_\beta)\,\frac{\sigma(q_\beta)\sigma(z_1-z_2+q_\alpha)}{\sigma(q_\alpha+q_\beta)\sigma(z_1-z_2)}\,, \notag
	\end{align}
	where we have used that $\sigma$ is odd. Dividing both sides by $\sigma(z_1)\sigma(z_2)\sigma(q_\alpha)\sigma(q_\beta)$, we rewrite this as
	\begin{equation}
		\frac{\sigma(z_1+q_\alpha)\sigma(z_2+q_\beta)}{\sigma(z_1)\sigma(q_\alpha)\sigma(z_2)\sigma(q_\beta)}=\frac{\sigma(z_1+q_\alpha+q_\beta)\sigma(z_2-z_1+q_\beta)}{\sigma(z_1)\sigma(q_\alpha+q_\beta)\sigma(z_2-z_1)\sigma(q_\beta)}+\frac{\sigma(z_2+q_\alpha+q_\beta)\sigma(z_1-z_2+q_\alpha)}{\sigma(z_2)\sigma(q_\alpha+q_\beta)\sigma(z_1-z_2)\sigma(q_\alpha)}\,.
	\end{equation}
	Multiplying by $\exp(-Q_\alpha z_1-Q_\beta z_2)$ on both sides and recalling the definition \eqref{eq:defofralpha} of $\lbrace r^\alpha(z) \rbrace$, this precisely simplifies to
	\begin{equation}
		r^\alpha(z_1)r^\beta(z_2)=r^{\alpha+\beta}(z_1)r^{\beta}(z_2-z_1)+r^{\alpha+\beta}(z_2)r^{\alpha}(z_1-z_2)\,,
	\end{equation}
	proving the desired property \eqref{eq:rcompatible}.
	
	\subsection{Relating the \texorpdfstring{$\text{SL}_{\mathbb{C}}$}{SLC}(2)-DPCM with the Cherednik model}\label{subsec:Nto2}
	
	Lastly, we want to show that for $N=2$, the elliptic DPCM in section \ref{sec:HigherRank} agrees with (the complexification of) the Cherednik model of section \ref{sec:su2}. First of all, one can readily convert between the basis of $\mathfrak{su}(2)$ introduced in \eqref{eq:su2basis} and the Belavin basis \eqref{eq:defofT} of $\mathfrak{sl}_\mathbb{C}(2)$\footnote{One checks that this agrees with the basis of $\mathfrak{su}(2)$ introduced in section \ref{subsec:Reality} to study reality conditions.}:
	\begin{equation}
		T_1=i\,T_{(0,1)}\,,\qquad T_2=T_{(1,1)}\,,\qquad T_3=i\, T_{(1,0)}\,.
	\end{equation}
	Thus, to match the deformation operators of sections \ref{sec:su2} and \ref{sec:HigherRank}, we need to show that
	\begin{equation}
		D_1=D_{(0,1)}\,,\qquad D_2=D_{(1,1)}\,,\qquad D_3=D_{(1,0)}\,.\label{eq:Dsumatch}
	\end{equation}
	The coefficients $D_a$ on the left-hand side are given by equation \eqref{eq:EllipticD}, in terms of Jacobi functions and three parameters $(h,\nu,m)$. In contrast, the ones $D_{(\alpha_1,\alpha_2)}$ on the right-hand side are defined by equation \eqref{eq:EllipticDeformationsHighRank}, using Weierstrass functions and four parameters $(\ell_1,\ell_2,\rho,\ha{z})$, subject to the dilation redundancy discussed in subsection \ref{sec:DPCM-HR}. 
	In order for the pole structures of the corresponding Lax connections \eqref{eq:EllipticL} and \eqref{eq:HighRankLax} to match, we will use the fixed-zero parametrisation, \textit{i.e.} use the dilation freedom to set $\ha{z}=1$. We then wish to find a choice of parameters $(\ell_1,\ell_2,\rho)$ expressed in terms of the $\text{SU}(2)$ parameters $(h,\nu,m)$ such that \eqref{eq:Dsumatch} holds.
	\newline
	
	Given the periods of the Jacobi functions and of the $\{r^\alpha\}$ functions \eqref{eq:LargeLattice}, it is clear that we must identify
	\begin{equation}
		\ell_1=\frac{K(m)}{\nu}\,,\qquad \ell_2=\frac{iK'(m)}{\nu}.
	\end{equation}
	Using this, $\wp$ and $\zeta$ can be expressed as
	\begin{equation*}
		\wp(z)=\frac{\nu^2}{\text{sc}^2(\nu z;m)}+\frac{\nu^2(2-m)}{3}\,,\qquad \zeta(z)=\nu\,\mathcal{E}(\nu z;m)+\frac{\nu\,\dn(\nu z;m)}{\text{sc}(\nu z;m)}+\frac{(m-2)\nu^2 z}{3}\,,
	\end{equation*}
	where $\mathcal{E}$ is the \textit{Jacobi epsilon function}. Moreover, one can identify $L_1$ and $L_2$ as
	\begin{equation}
		L_1=\nu\, E(m)+\frac{m-2}{3}\nu K(m)\,,\qquad L_2=-i\nu \, E'(m)+\frac{m+1}{3}i\nu K'(m)\,,
	\end{equation}
	where $E(m)=\mathcal{E}(K(m);m)$ is the \textit{complete elliptic integral of the second kind} and $E'(m)=E(1-m)$.
	
	We have yet to relate the parameter $\rho$ with $h$. This is most easily done by comparing the $O(1/z^2)$ term of the ``Weierstrass'' twist function \eqref{eq:higherrnaktwist} (using the identification of $\wp(z)$ given above) with that of the ``Jacobi'' twist function \eqref{eq:EllipticPhi}. We find
	\begin{equation}
		\rho=\frac{h}{\nu}\,\sd(\nu;m)\,\cn(\nu;m)\,.
	\end{equation}
	Using this, one can readily show that the two twist functions coincide.
	
	Lastly, we need to prove that the deformation coefficients also agree, \textit{i.e.} that \eqref{eq:Dsumatch} holds. This relies on a large amount of cancellation between different terms and the behaviour of the Jacobi elliptic functions under shifts\footnote{More precisely, we use the fact that any Jacobi elliptic function evaluated at $z+K(m)$ or $z+iK'(m)$ can be expressed in terms of another member of the family evaluated at $z$ (see for instance~\cite[Table 16.8]{10.5555/1098650}).} by $K(m)$ and $iK'(m)$. For brevity, we will not detail this computation here. In the end, one finds that \eqref{eq:EllipticD} and \eqref{eq:EllipticDeformationsHighRank} agree, with the identification \eqref{eq:Dsumatch}.
	
	\section{Relating The Trigonometric Limit to Yang-Baxter Deformations}\label{sec:YBAppendix}
	This appendix will give further details on the trigonometric limit considered in section \ref{subsec:MainYB}
	and its connection to Yang-Baxter deformations.
	
	\subsection{Elliptic Deformations in the Trigonometric Limit}
	As stated in section \ref{subsec:MainYB}, we will consider the limit $\ell_1\ra +\infty$ with $\ell_2=\frac{i\pi}{2\nu}$ held fixed, where $\nu>0$ is a positive parameter. Fixing $\ha{z}=1$ using the dilation freedom and changing variables from $\ell_2$ to $\nu$, we thus arrive at the operator ``tlim'' introduced in the main text, which acts by taking the limit $(\ell_1,\ell_2,\rho,\ha{z})\to (+\infty,\frac{i\pi}{2\nu},\rho,1)$. We now wish to calculate the trigonometric limit of the deformation coefficients $D_\alpha$ in \eqref{eq:EllipticDeformationsHighRank}. This requires us to consider the limit of the Weierstrass $\zeta$-function evaluated at various points. As noted in \eqref{eq:tlimzetawp}, if $z$ stays finite in the limit, one has
	\begin{equation}\label{eq:tlimZeta}
		\tlim \;\, \zeta(z)=\nu\,\coth(\nu z)-\frac{\nu^2 z}{3}\,.
	\end{equation}
	To obtain the limit of $D_\alpha$, we will now have to distinguish the cases $\alpha_2=0$ and $\alpha_2\neq 0$.
	
	\paragraph{Limit of $\bm{D_\alpha}$ for $\bm{\alpha_2=0}$:} We start with $\alpha_2=0$. In that case, $q_\alpha=q_{(\alpha_1,0)}$ does not depend on $\ell_1$ and thus stays finite in the trigonometric limit. Using \eqref{eq:tlimZeta}, one can then readily evaluate
	\begin{equation}
		\tlim \;\,\zeta\bigl(\ha{z}+q_{(\alpha_1,0)}\bigr)=\nu\coth(\nu+i\pi\frac{\alpha_1}{N})-\frac{\nu^2}{3}-\frac{i\pi\nu}{3}\frac{\alpha_1}{N}\,.
	\end{equation}
	Similarly, $Q_{(\alpha_1,0)}$ depends only on $L_2=\zeta(\ell_2)$, which has the finite limit
	\begin{equation}\label{eq:tlimL2}
		\tlim \;\, L_2 = -\frac{i\pi\nu}{6}
	\end{equation}
	by equation \eqref{eq:tlimZeta}. Combining these results, we finally find
	\begin{equation}
		\alpha_2=0: \qquad \tlim\;\, D_{(\alpha_1,0)}=\rho\nu\left\{\coth(\nu) - \coth(\nu+i\pi\frac{\alpha_1}{N})\right\}\,.\label{eq:TrigDalpha0appendix}
	\end{equation}
	
	\paragraph{Limit of $\bm{D_\alpha}$ for $\bm{\alpha_2\neq 0}$:} 
	The case with $\alpha_2\neq 0$ is more subtle since the argument $\ha{z}+q_\alpha$ of the relevant $\zeta$-function itself blows up. Indeed, a quantity of the form $\zeta(a+b\ell_1)$ (with finite $a,b\in\C$) will generally diverge in the trigonometric limit. To analyse this in more detail, let us consider the following series representation of the Weierstrass $\zeta$-function:
	\begin{equation}\label{eq:ZetaTrigSum}
		\zeta(z) = \frac{2\nu L_2}{i\pi}z + \nu\,\coth(\nu z) + \nu \sum_{n=1}^{+\infty} \Bigl[ \coth\bigl(\nu(z+2n\ell_1) \bigr) + \coth\bigl(\nu(z-2n\ell_1) \bigr) \Bigr]\,,
	\end{equation}
	where we used the parametrisation $\ell_2=\frac{i\pi}{2\nu}$, $\nu>0$, as above. We now want to determine the behaviour of $\zeta(a+b\ell_1)$ in the trigonometric limit $\ell_1\to +\infty$. A careful analysis shows that the terms
	\begin{equation}
		\coth\bigl(\nu(a+b\ell_1+2n\ell_1) \bigr) + \coth\bigl(\nu(a+b\ell_1-2n\ell_1) \bigr)
	\end{equation}
	all decay to zero exponentially fast when $\ell_1\to +\infty$ if $n\geq 1$ and $0<b<2$, in a way that makes their sum over $n\in\mathbb{Z}_{\geq 1}$ also converge to 0. Noting that $\coth(\nu\,(a+b\ell_1))$ converges to $1$ for the choice $0<b<2$ and recalling the finite limit \eqref{eq:tlimL2} of $L_2$, we can then easily extract the behaviour of $\zeta(a+b\ell_1)$ when $\ell_1\to +\infty$ from the series representation \eqref{eq:ZetaTrigSum}. More precisely, we confirm the fact that $\zeta(a+b\ell_1)$ diverges in this limit and find that
	\begin{equation}
		\tlim \;\,\left\lbrace \zeta(a+b\ell_1) - \frac{2b\nu L_2}{i\pi} \ell_1  \right\rbrace = \nu - \frac{a\nu^2}{3} \qquad \text{ for } \quad 0<b<2 \, .\label{eq:tlimZetal1}
	\end{equation}
	Crucially, the divergent part of $\zeta(a+b\ell_1)$ is exactly cancelled by the second term on the left-hand side, leaving a finite result in the end. Let us now consider the quantity
	\begin{equation}
		\zeta(a+b\ell_1) - b L_1 = \left\lbrace \zeta(a+b\ell_1) - \frac{2b\nu L_2}{i\pi} \ell_1  \right\rbrace - b\left\lbrace \zeta(\ell_1) - \frac{2\nu L_2}{i\pi} \ell_1  \right\rbrace\,,
	\end{equation}
	where we used $L_1=\zeta(\ell_1)$. By the identity \eqref{eq:tlimZetal1}, both terms on the right-hand side converge in the trigonometric limit and we get
	\begin{equation}\label{eq:tlimZetal1L1}
		\tlim \;\,\bigl\lbrace \zeta(a+b\ell_1) - b L_1 \bigr\rbrace =  \nu(1-b) - \frac{a\nu^2}{3}\,, \qquad \text{ for } \quad 0<b<2\,.
	\end{equation}
	Let us now come back to the coefficient $D_{\alpha}$ for $\alpha_2 \neq 0$. We choose to represent $\alpha_2$ as an integer with $0 < \alpha_2 < N$ and spell out the expression \eqref{eq:EllipticDeformationsHighRank} of $D_\alpha$ as
	\begin{equation}
		D_{(\alpha_1,\alpha_2)} = \rho \left\lbrace \zeta(\ha{z}) + \frac{2\alpha_1}{N}L_2 + \zeta\left( - \ha{z} - \frac{2\alpha_1}{N}\ell_2 + \frac{2\alpha_2}{N}\ell_1 \right) - \frac{2\alpha_2}{N}L_1 \right\rbrace\,,
	\end{equation}
	where we used the fact that $\zeta$ is odd. The trigonometric limit of the first two terms can be easily computed using equations \eqref{eq:tlimZeta} and \eqref{eq:tlimL2}, while the limit of the last two terms can be determined using the identity \eqref{eq:tlimZetal1L1} (whose applicability is ensured by the condition $0 < \alpha_2 < N$). After a few manipulations, we find
	\begin{equation}
		\alpha_2\neq 0: \qquad \tlim\;\, D_{(\alpha_1,\alpha_2)} = \rho \nu\left\{\coth(\nu) + 1 - \frac{2\alpha_2}{N} \right\}\,.\label{eq:TrigDalphaneq0apendix}
	\end{equation}
	
	\subsection{Yang-Baxter Deformations}
	
	In this subsection, we will use extensively the Belavin basis $\lbrace T_{(\alpha_1,\alpha_2)} \rbrace$, where $(\alpha_1,\alpha_2)$ runs through $\mathbb{Z}_N \times \mathbb{Z}_N \setminus \lbrace (0,0) \rbrace$ to conform with the notation above: we will always represent $\alpha_1$ and $\alpha_2$ as integers from $0$ to $N-1$.
	In the previous subsection, we showed that the trigonometric limit of the elliptic deformation \eqref{eq:EllipticDeformationsHighRank} defines an operator $\tlim \;\, D$, acting diagonally on this basis, with coefficients \eqref{eq:TrigDalpha0appendix} or \eqref{eq:TrigDalphaneq0apendix} depending on whether $\alpha_2$ vanishes or not. We now want to determine whether we can recreate this in the language of Yang-Baxter deformations~\cite{klimcik_yang-baxter_2002,klimcik_integrability_2009}, \textit{i.e.} as a deformation operator of the type
	\begin{equation}
		\ms{D}=h\frac{1-\eta^2}{1-\eta\ms{R}}\,,\label{eq:DYB}
	\end{equation}
	where $\ms{R}$ is a skew-symmetric solution of the modified classical Yang-Baxter equation \eqref{eq:mybe}. Here, we will take it to be an \textit{extended split Drinfel'd-Jimbo R-matrix}~\cite{Drinfeld:1985rx,Jimbo:1985zk}. Such a solution is defined with respect to a choice of Cartan subalgebra and root decomposition of $\mathfrak{sl}_\C(N)$. Here, we will take the Cartan subalgebra to be $\bigoplus_{\alpha_1=1}^{N-1}\C T_{(\alpha_1,0)}$\footnote{It is clear from the definition of $T_{(\alpha_1,0)}$ in the Belavin basis \eqref{eq:defofT} that these matrices generate a Cartan subalgebra, since they form a maximal set of independent diagonal matrices in $\slNC$.}. Moreover, we define the corresponding root vectors $E^{\pm}_{ij}$ as the matrices with a 1 in the entry $(i,j)$ and 0 in every other entry (the additional labels $+$ and $-$ correspond respectively to $i<j$ and $i>j$ and indicate whether these are positive or negative root vectors). The extended Drinfel'd-Jimbo R-matrix is then defined by
	\begin{equation}
		\ms{R}\left[E^\pm_{ij}\right]=\mp E^\pm_{ij}\,, \qquad \ms{R}\left[T_{(\alpha_1,0)}\right]=C_{\alpha_1}^{\phantom{\alpha_1}\beta_1}\,T_{(\beta_1,0)}\,,\label{eq:DrinfeldJimbo}
	\end{equation}
	where $(C_{\alpha_1}^{\phantom{\alpha_1}\beta_1})_{\alpha_1,\beta_1=1}^{N-1}$ are scalar coefficients (encoding the ``extended'' character of the R-matrix, also called a Reshetikhin twist~\cite{Reshetikhin:1990ep}). At this point, the only constraint on these coefficients is that the operator $\ms{R}$ should be skew-symmetric with respect to the bilinear form $\langle\cdot,\cdot\rangle$. Recalling the expression \eqref{eq:slcnkappa} of this form in the Belavin basis, one checks that it amounts to requiring
	\begin{equation}\label{eq:Rskewsym}
		C_{\alpha_1}^{\phantom{\alpha_1}N-\beta_1} = - C_{\beta_1}^{\phantom{\beta_1}N-\alpha_1}\,.
	\end{equation}
	We will now try to determine parameters $h,\eta$ and $C_{\alpha_1}^{\phantom{\alpha_1}\beta_1}$ such that the Yang-Baxter deformation operator $\ms{D}$ matches the trigonometric limit $\tlim\;D$.
	
	\paragraph{Action on the Cartan Subalgebra:} Since $\tlim\;D$ is diagonal in the basis $\lbrace T_{(\alpha_1,0)} \rbrace_{\alpha_1=1}^{N-1}$ of the Cartan subalgebra, it is clear that $\ms{R}$ needs to act diagonally on it as well in order to match $\tlim\;D$ with $\ms{D}$. We thus take
	\begin{equation}\label{eq:diagC}
		C_{\alpha_1}^{\phantom{\alpha_1}\beta_1}=\delta_{\alpha_1}^{\phantom{\alpha_1}\beta_1}C_{\alpha_1}\,, \qquad \text{ with } \qquad C_{\alpha_1}=-C_{N-\alpha_1}\,,
	\end{equation}
	where the last condition comes from the skew-symmetry requirement \eqref{eq:Rskewsym}. This leads to a very simple deformation operator on the Cartan subalgebra, being given by
	\begin{equation}
		\ms{D}\left[T_{(\alpha_1,0)}\right]=h\frac{1-\eta^2}{1-\eta\, C_{\alpha_1}}T_{(\alpha_1,0)}\,.\label{eq:DCartanid}
	\end{equation}
	
	\paragraph{Action on the Root Spaces:} We now want to determine the action of $\ms{D}$ on the remaining elements of the Belavin basis, \textit{i.e.} the matrices $T_{(\alpha_1,\alpha_2)}$ with $\alpha_2\neq 0$. These matrices are purely off-diagonal and are thus expressed only in terms of the positive and negative root-vectors $E_{ij}^\pm$ (and not the Cartan elements). Using equation \eqref{eq:DrinfeldJimbo}, we note that the action of $\ms{R}^2$ on these root-vectors is simply the identity. We can thus rewrite the action of the deformation operator \eqref{eq:DYB} on the non-Cartan generators as
	\begin{equation}
		\alpha_2\neq 0: \qquad \ms{D}\left[T_{(\alpha_1,\alpha_2)}\right]=h(1+\eta\ms{R})\frac{1-\eta^2}{1-\eta^2\ms{R}^2}\left[T_{(\alpha_1,\alpha_2)}\right]=h\bigl(1+\eta\ms{R}\bigr)\left[T_{(\alpha_1,\alpha_2)}\right].\label{eq:rootdeform}
	\end{equation} 
	To proceed, we then need to express $T_{(\alpha_1,\alpha_2)}$ with $\alpha_2\neq 0$ in the root-space basis $E_{ij}^\pm$. One finds
	\begin{equation}\label{eq:TE}
		T_{(\alpha_1,\alpha_2)}= \frac{1}{\sqrt{N}}\left(\sum_{i=1}^{\alpha_2}\xi^{\alpha_1\left[i-\alpha_2-1\right]} E^+_{i,i+N-\alpha_2} + \sum_{i=\alpha_2+1}^{N}\xi^{\alpha_1[i-\alpha_2-1]}E^-_{i,i-\alpha_2} \right)\,,
	\end{equation}
	where $\xi=\exp(2\pi i/N)$. The action of $\ms{R}$ on these generators is then given by
	\begin{align}
		\ms{R}\left[T_{(\alpha_1,\alpha_2)}\right]&=\frac{1}{\sqrt{N}}\left(-\sum_{i=1}^{\alpha_2}\xi^{\alpha_1\left[i-\alpha_2-1\right]} E^+_{i,i+N-\alpha_2}+\sum_{i=\alpha_2+1}^{N}\xi^{\alpha_1[i-\alpha_2-1]}E^-_{i,i-\alpha_2} \right)\\
		&=T_{(\alpha_1,\alpha_2)}-\frac{2}{\sqrt{N}}\sum_{i=1}^{\alpha_2}\xi^{\alpha_1\left[i-\alpha_2-1\right]} E^+_{i,i+N-\alpha_2}\,. \notag
	\end{align}
	The equation \eqref{eq:rootdeform} thus becomes
	\begin{equation}\label{eq:DTneq0}
		\ms{D}\left[T_{(\alpha_1,\alpha_2)}\right]=h\bigl(1+\eta\bigr)T_{(\alpha_1,\alpha_2)} -\frac{2h\eta}{\sqrt{N}}\,\sum_{i=1}^{\alpha_2}\xi^{\alpha_1\left[i-\alpha_2-1\right]} E^+_{i,i+N-\alpha_2}\,  .
	\end{equation}
	This action is not diagonal in the Belavin basis, due to the extra sum over positive root-vectors. It thus seems impossible to match $\ms{D}$ with the trigonometric operator $\tlim\;D$.
	To resolve this issue, let us consider adding to the action \eqref{eq:DPCMActionHighRank} a term proportional to the integral of $\left\langle j_+,\ad_{\ms{W}}\left[j_-\right]\right\rangle$, for some constant element $\ms{W}$ of $\slNC$. Using the ad-invariance of the bilinear form $\left\langle \cdot,\cdot\right\rangle$ and the Maurer-Cartan equation \eqref{eq:jflat}, this integrand can be recast as
	\begin{equation}
		\left\langle j_+,\ad_{\ms{W}}\left[j_-\right]\right\rangle=-\left\langle\left[j_+,j_-\right],\ms{W}\right\rangle=\left\langle\p_+j_--\p_-j_+,\ms{W}\right\rangle\,.\label{eq:Wdoesntmatter}
	\end{equation}
	This is a total derivative and it therefore changes the action only by a boundary term, without affecting the equations of motion. Adding such a term to the action amounts to considering a new deformation operator
	\begin{equation}
		\ms{D}' = \ms{D} + \mc{N}\,\ad_{\ms{W}}\,,
	\end{equation}
	where $\mc{N}$ is a proportionality constant, unfixed for the moment. Our strategy will now be to show that there exists a choice of $\ms{W}$ and $\mc{N}$ such that $\ms{D}'$ becomes diagonal in the Belavin basis. In the previous paragraph, we have chosen the coefficients $C_{\alpha_1}^{\phantom{\alpha_1}\beta_1}$ such that $\ms{D}$ acts diagonally on the elements $\lbrace T_{(\alpha_1,0)} \rbrace_{\alpha_1=1}^{N-1}$ of this basis, which generate the Cartan subalgebra. In order to keep this property for the shifted operator $\ms{D}'$, we will take $\ms{W}$ to be itself in this Cartan subalgebra, as $\ad_{\ms{W}}$ then acts trivially on $T_{(\alpha_1,0)}$. By construction, the equation \eqref{eq:DCartanid} thus also holds with $\ms{D}$ replaced by $\ms{D}'$.
	
	The right choice for $\ms{W}$ turns out to be the \textit{Weyl coweight} (see for instance \cite{kostant_principal_1959}), whose adjoint action on the root-vectors measures their height, \textit{i.e.}
	\begin{equation}
		\ad_{\ms{W}}\left[E_{ij}^\pm\right]=(j-i)E^{\pm}_{ij}\,.
	\end{equation}
	From the expression \eqref{eq:TE} of the generators $T_{(\alpha_1,\alpha_2)}$ ($\alpha_2\neq 0$), we then get
	\begin{align}
		\ad_{\ms{W}}\left[T_{(\alpha_1,\alpha_2)}\right]&=\frac{N-\alpha_2}{\sqrt{N}} \sum_{i=1}^{\alpha_2}\xi^{\alpha_1\left[i-\alpha_2-1\right]} E^+_{i,i+N-\alpha_2} - \frac{\alpha_2}{\sqrt{N}}\sum_{i=\alpha_2+1}^{N}\xi^{\alpha_1[i-\alpha_2-1]}E^-_{i,i-\alpha_2} \notag \\
		& = -\alpha_2\,T_{(\alpha_1,\alpha_2)} + \sqrt{N} \,\sum_{i=1}^{\alpha_2}\xi^{\alpha_1\left[i-\alpha_2-1\right]}E^+_{i,i+N-\alpha_2}\,. 
	\end{align}
	Combining this with equation \eqref{eq:DTneq0}, we observe that by choosing $\mc{N}=\frac{2h\eta}{N}$, we get a diagonal action of $\ms{D}'$ on the generators $T_{(\alpha_1,\alpha_2)}$ with $\alpha_2\neq 0$, given by
	\begin{equation}
		\ms{D}'\left[T_{(\alpha_1,\alpha_2)}\right]=\Big[h+h\eta\Big\{1-\frac{2\alpha_2}{N}\Big\}\Big]T_{(\alpha_1,\alpha_2)}\,.\label{eq:Drootid}
	\end{equation}
	
	\paragraph{Matching the Yang-Baxter Deformation With the Trigonometric Limit:}
	We have thus ensured that the shifted Yang-Baxter deformation operator $\ms{D}'$ acts diagonally on the Belavin basis, \textit{i.e.} $\ms{D}'[T_\alpha]=\ms{D}'_\alpha T_\alpha$. Explicitly, from \eqref{eq:DCartanid} and \eqref{eq:Drootid}, we read
	\begin{subequations}\label{eq:YBDalphaAppendix}
		\begin{equation}
			\alpha_2=0: \qquad \ms{D}'_{(\alpha_1,0)}=h\frac{1-\eta^2}{1-\eta \,C_{\alpha_1}}\,,
		\end{equation}
		\begin{equation}
			\alpha_2\neq 0: \qquad \ms{D}'_{(\alpha_1,\alpha_2)}=h+h\eta\Big\{1-\frac{2\alpha_2}{N}\Big\}\,.
		\end{equation}
	\end{subequations}
	We then wish to identify $h,\eta$ and $C_{\alpha_1}$ such that this matches \eqref{eq:TrigDalpha0appendix} and \eqref{eq:TrigDalphaneq0apendix}. We find
	\begin{equation}
		h=\rho\nu \coth(\nu)\,,\qquad \eta=\tanh(\nu)\,,\qquad C_{\alpha_1}=i\,\cot(\pi\frac{\alpha_1}{N})\,.
	\end{equation}
	Observe that the skew-symmetry requirement on $C_{\alpha_1}$ -- the second equation in \eqref{eq:diagC} -- is satisfied, using the trigonometric identity $\cot(\pi-z)=-\cot(z)$. This ends the proof that the trigonometric limit of the elliptic integrable DPCM coincides with a well-chosen Yang-Baxter deformation, up to a total derivative in the Lagrangian.
	
	\bibliographystyle{JHEP}

\begin{thebibliography}{10}
		
		\bibitem{maillet_kac-moody_1985}
		J.-M. Maillet,
		``\emph{Kac-Moody algebra and extended Yang-Baxter relations
			in the {O}({N}) non-linear $\sigma$-model}'',
		\href{https://doi.org/10.1016/0370-2693(85)91075-5}{Phys. Lett. B {\bf 162} (1985) 137--142}.
		
		\bibitem{maillet_new_1986}
		J.-M. Maillet,
		``\emph{New integrable canonical structures in two-dimensional
			models}'',
		\href{https://doi.org/10.1016/0550-3213(86)90365-2}{Nucl. Phys. B {\bf 269} (1986) 54--76}.
		
		\bibitem{belavin_solutions_1982}
		A.~A. Belavin and V.~G. Drinfel'd,
		``\emph{Solutions of the classical Yang-Baxter equation for simple Lie algebras}'',
		\href{https://doi.org/10.1007/BF01081585}{Funct. Anal. Its Appl. {\bf 16} (1982) 159--180}.
		
		\bibitem{hoare_integrable_2022}
		B.~Hoare, 
		``\emph{Integrable deformations of sigma models}'',
		\href{https://doi.org/10.1088/1751-8121/ac4a1e}{J. Phys. A {\bf 55} (2022) 093001} [\href{https://arxiv.org/abs/2109.14284}{{\ttfamily arXiv:2109.14284}}].
		
		\bibitem{cherednik_relativistically_1981}
		I.~V.~Cherednik, ``\emph{Relativistically Invariant Quasiclassical Limits of Integrable Two-dimensional Quantum Models}'',
		\href{https://doi.org/10.1007/BF01086395}{Theor. Math. Phys. \textbf{47} (1981) 422-425}
		
		\bibitem{sfetsos_anisotropic_2015}
		K.~Sfetsos and K.~Siampos,
		``\emph{The anisotropic $\lambda$-deformed {SU}(2)
			model is integrable}'',
		\href{https://doi.org/10.1007/BF01081585}{Phys. Lett. B {\bf 743} (2015) 160--165} [\href{https://arxiv.org/abs/1412.5181}{{\ttfamily arXiv:1412.5181}}].
		
		\bibitem{costello_gauge_2019}
		K.~Costello and M.~Yamazaki, ``\emph{{Gauge Theory And Integrability, III}}'' (2019), [\href{https://arxiv.org/abs/1908.02289}{{\ttfamily arXiv:1908.02289}}].
		
		\bibitem{bykov_sigma_2021}
		D.~V. Bykov, 
		``\emph{Sigma models as {Gross}–{Neveu} models}'',
		\href{https://doi.org/10.1007/BF01081585}{Theor. Math. Phys. {\bf 208} (2021) 993--1003} [\href{https://arxiv.org/abs/2106.15598}{{\ttfamily arXiv:2106.15598}}].
		
		\bibitem{derryberry_lax_2021}
		R.~Derryberry, ``\emph{{Lax formulation for harmonic maps to a moduli of bundles}}'' (2021), [\href{https://arxiv.org/abs/2106.09781}{{\ttfamily arXiv:2106.09781}}].
		
		\bibitem{bykov_cpn-1-model_2022}
		D.~V. Bykov, 
		``\emph{The $\mathbb{CP}^{n-1}$-model with fermions: a new look}'',
		\href{https://doi.org/10.1007/BF01081585}{Adv. Theor. Math. Phys. {\bf 26} (2022) no.~2, 295--324} [\href{https://arxiv.org/abs/2009.04608}{{\ttfamily arXiv:2009.04608}}].
		
		\bibitem{zakharov_relativistically_1978}
		V.~Zakharov and A.~Mikhailov, ``\emph{{Relativistically Invariant Two-Dimensional Models in Field Theory Integrable by the Inverse Problem Technique}}'', Sov. Phys. JETP \textbf{47} (1978) 1017-1027.
		
		\bibitem{maillet_hamiltonian_1986}
		J.M.~Maillet, ``\emph{{Hamiltonian Structures for Integrable Classical Theories From Graded Kac-Moody Algebras}}'', \href{https://doi.org/10.1016/0370-2693(86)91289-X}{Phys. Lett. 167B (1986) 401}.

  

\bibitem{Fateev:1996ea}
V.~A.~Fateev,  
``\emph{The sigma model (dual) representation for a two-parameter family of integrable quantum field theories}'',
\href{https://doi.org/10.1016/0550-3213(96)00256-8}{Nucl. Phys. B \textbf{473} (1996), 509-538}.

\bibitem{Lukyanov:2012zt}
S.~L.~Lukyanov,
``\emph{The integrable harmonic map problem versus Ricci flow}'',
\href{https://doi.org/10.1016/j.nuclphysb.2012.08.002}{Nucl. Phys. B \textbf{865} (2012), 308-329} [\href{https://arxiv.org/abs/1205.3201}{{\ttfamily arXiv:1205.3201}}].
		
		\bibitem{klimcik_yang-baxter_2002}
		C.~Klim\v{c}\'{i}k, ``\emph{{Yang-Baxter sigma models and dS/AdS T duality}}'',
		\href{https://doi.org/10.1088/1126-6708/2002/12/051}{JHEP {\bfseries   0212} (2002) 051} [\href{https://arxiv.org/abs/hep-th/0210095}{{\ttfamily arXiv:hep-th/0210095}}].
		
		\bibitem{klimcik_integrability_2009}
		C.~Klim\v{c}\'{i}k,
		``\emph{On integrability of the Yang-Baxter sigma-model}'',
		\href{https://doi.org/10.1063/1.3116242}{J. Math. Phys. \textbf{50} (2009) 043508}
		[\href{https://arxiv.org/abs/0802.3518}{{\ttfamily arXiv:0802.3518}}].

\bibitem{Delduc:2017fib}
F.~Delduc, B.~Hoare, T.~Kameyama and M.~Magro, ``\emph{Combining the bi-Yang-Baxter deformation, the Wess-Zumino term and TsT transformations in one integrable $\sigma$-model}'',
\href{https://doi.org/10.1007/JHEP10(2017)212}{JHEP \textbf{10} (2017), 212} [\href{https://arxiv.org/abs/1707.08371}{{\ttfamily arXiv:1707.08371}}].
		
		\bibitem{klimcik_integrability_2014}
		C.~Klim\v{c}\'{i}k,
		``\emph{Integrability of the bi-Yang-Baxter sigma-model}'',
		\href{https://doi.org/10.1007/s11005-014-0709-y}{Lett. Math. Phys. \textbf{104}  (2014) 1095}
		[\href{https://arxiv.org/abs/1402.2105}{{\ttfamily arXiv:1402.2105}}].
		
		\bibitem{Klimcik:1995ux}
		C.~Klim\v{c}\'{\i}k and P.~\v{S}evera,
		``\emph{Dual non-Abelian duality and the Drinfeld double}'',
		\href{https://doi.org/10.1016/0370-2693(95)00451-P}{Phys. Lett. B \textbf{351} (1995) 455--462} [\href{https://arxiv.org/abs/hep-th/9502122}{\ttfamily arXiv:hep-th/9502122}].
		
		\bibitem{Klimcik:1995dy}
		C.~Klim\v{c}\'{\i}k and P.~\v{S}evera,
		``\emph{Poisson-Lie T duality and loop groups of Drinfeld doubles}'',
		\href{https://doi.org/10.1016/0370-2693(96)00025-1}{Phys. Lett. B \textbf{372} (1996) 65--71} [\href{https://arxiv.org/abs/hep-th/9512040}{\ttfamily arXiv:hep-th/9512040}].

  \bibitem{Sfetsos:2013wia}
K.~Sfetsos, ``\emph{Integrable interpolations: From exact CFTs to non-Abelian T-duals}'',
		\href{http://dx.doi.org/10.1016/j.nuclphysb.2014.01.004}{Nucl. Phys. B \textbf{880} (2014), 225-246} 
		[\href{https://arxiv.org/abs/1312.4560}{{\tt arXiv:1312.4560}}].
		
		\bibitem{sfetsos_generalised_2015}
		K.~Sfetsos, K.~Siampos and D.~C.~Thompson,
		``\emph{Generalised integrable \ensuremath{\lambda} - and \ensuremath{\eta}-deformations and their relation}'',
		\href{http://dx.doi.org/10.1016/j.nuclphysb.2015.08.015}{Nucl. Phys. B \textbf{899} (2015) 489-512} 
		[\href{https://arxiv.org/abs/1506.05784}{{\tt arXiv:1506.05784}}].
		
		\bibitem{sklyanin_complete_1979}
		E.~K. Sklyanin, ``\emph{On complete integrability of the {Landau}-{Lifshitz}
			equation}'',  LOMI Preprint LOMI E3-79 (1979).
		
		\bibitem{belavin_discrete_1981}
		A.~A. Belavin, ``\emph{Discrete groups and the integrability of quantum systems}'',
		\href{https://doi.org/10.1007/BF01078301}{Funct. Anal. Its Appl. {\bf 14} (1981) no.~4,  260--267}.
		
		\bibitem{levin_hitchin_2003}
		A.~M.~Levin, M.~A.~Olshanetsky and A.~Zotov,
		``\emph{Hitchin systems\textendash{}symplectic hecke correspondence and two-dimensional version}'',
		\href{http://dx.doi.org/10.1007/s00220-003-0801-0}{Commun. Math. Phys. \textbf{236} (2003) 93} 
		[\href{https://arxiv.org/abs/nlin/0110045}{{\tt arXiv:nlin/0110045}}].
		
		\bibitem{feigin_quantization_2009}
		B.~Feigin and E.~Frenkel, ``\emph{{Quantization of soliton systems and Langlands duality}}'', in \emph{Exploration of
			New Structures and Natural Constructions in Mathematical Physics}, \href{https://doi.org/10.1142/e032}{Adv. Stud. Pure Math. 61 (2011) 185–274} [\href{https://arxiv.org/abs/0705.2486}{{\ttfamily arXiv:0705.2486}}].
		
		\bibitem{vicedo_integrable_2019}
		B.~Vicedo, ``\emph{{On integrable field theories as dihedral affine Gaudin  models}}'', \href{https://doi.org/10.1093/imrn/rny128}{Int. Math. Res. Not. {\bfseries 2020} (2020) 15} [\href{https://arxiv.org/abs/1701.04856}{{\ttfamily arXiv:1701.04856}}].
		
		\bibitem{delduc_assembling_2019}
		F.~Delduc, S.~Lacroix, M.~Magro and B.~Vicedo, ``\emph{{Assembling integrable $\sigma$-models as affine Gaudin models}}'',
		\href{https://doi.org/10.1007/JHEP06(2019)017}{JHEP {\bfseries 06}
			(2019) 017} [\href{https://arxiv.org/abs/1903.00368}{{\ttfamily
				arXiv:1903.00368}}].
		
		\bibitem{vicedo_4d_2021}
		B.~Vicedo, ``\emph{{4D Chern–Simons theory and affine Gaudin models}}'', \href{https://doi.org/10.1007/s11005-021-01354-9}{Lett. Math. Phys. \textbf{111} (2021) 24}
		[\href{https://arxiv.org/abs/1908.07511}{{\ttfamily arXiv:1908.07511}}].
		
		\bibitem{levin_2d_2022}
		A.~Levin, M.~Olshanetsky and A.~Zotov, ``\emph{{2D Integrable systems, 4D Chern\textendash{}Simons theory and affine Higgs bundles}}'', \href{https://doi.org/10.1140/epjc/s10052-022-10553-0}{Eur. Phys. J. C \textbf{82} (2022) no~7, 635} [\href{https://arxiv.org/abs/2202.10106}{{\ttfamily arXiv:2202.10106}}].
		
		\bibitem{costello_gauge_2018}
		K.~Costello, E.~Witten and M.~Yamazaki, ``\emph{{Gauge Theory and Integrability,  I}}'', 
		\href{https://doi.org/10.4310/ICCM.2018.v6.n1.a6}{ICCM Not. \textbf{06}, no~1 (2018), 46-119} [\href{https://arxiv.org/abs/1709.09993}{{\ttfamily arXiv:1709.09993}}].
		
		\bibitem{costello_gauge_2018-1}
		K.~Costello, E.~Witten and M.~Yamazaki, ``\emph{{Gauge Theory and Integrability, II}}'', 
		\href{https://doi.org/10.4310/ICCM.2018.v6.n1.a7}{ICCM Not. \textbf{06}, no.1 (2018), 120-146} [\href{https://arxiv.org/abs/1802.01579}{{\ttfamily arXiv:1802.01579}}].
		
		\bibitem{ToAppear:Gaudin}
		S.~Lacroix and A.~Wallberg, {\it {To appear}}.
		
		\bibitem{ToAppear:RG}
		S.~Lacroix and A.~Wallberg, ``\emph{Geometry of the spectral parameter and
renormalisation of integrable $\sigma$-models}'',\href{https://doi.org/10.1007/JHEP05(2024)108}{ JHEP \textbf{05} (2024) 108} [\href{https://arxiv.org/abs/2401.13741}{{\ttfamily arXiv:2401.13741}}]
		
		\bibitem{lozano_non-abelian_1995}
		Y.~Lozano,  ``\emph{Non-Abelian Duality and Canonical Transformations}'', 
		\href{https://doi.org/10.4310/ICCM.2018.v6.n1.a7}{Phys. Lett. B {\bf 355} (1995) 165--170} [\href{https://arxiv.org/abs/hep-th/9503045}{{\ttfamily arXiv:hep-th/9503045}}].
		
		\bibitem{alvarez_target_1996}
		O.~Alvarez and C.-H. Liu,  ``\emph{Target Space Duality between Simple Compact Lie Groups and Lie Algebras under the Hamiltonian Formalism: I. Remnants of Duality at the Classical Level}'', 
		\href{https://doi.org/10.4310/ICCM.2018.v6.n1.a7}{Commun. Math. Phys. {\bf 179} (1996) 185--213} [\href{https://arxiv.org/abs/hep-th/9503226}{{\ttfamily arXiv:hep-th/9503226}}].
		
		\bibitem{sfetsos_non--abelian_1996}
		K.~Sfetsos, ``\emph{Non--Abelian Duality, Parafermions and Supersymmetry}'', 
		\href{https://doi.org/10.1103/PhysRevD.54.1682}{Phys. Rev. D {\bf 54} (1996) 1682--1695} [\href{https://arxiv.org/abs/hep-th/9602179}{{\ttfamily arXiv:hep-th/9602179}}].
		
		\bibitem{kawaguchi_hybrid_2012}
		I.~Kawaguchi and K.~Yoshida, ``\emph{{Hybrid classical integrable structure of squashed sigma models -- a short summary}}'',
		\href{https://doi.org/10.1088/1742-6596/343/1/012055}{J. Phys. Conf. Ser.  {\bf 343} (2012) 012055}
		[\href{https://arxiv.org/abs/1110.6748}{{\ttfamily arXiv:1110.6748}}].
		
		\bibitem{kawaguchi_classical_2012}
		I.~Kawaguchi, T.~Matsumoto and K.~Yoshida, ``\emph{{On the classical equivalence of monodromy matrices in squashed sigma model}}'',
		\href{https://doi.org/10.1007/JHEP06(2012)082}{JHEP \textbf{06} (2012) 082}
		[\href{https://arxiv.org/abs/1203.3400}{{\ttfamily arXiv:1203.3400}}].
		
		\bibitem{fay_theta_1973}
		J.~D. Fay, ``\emph{Theta Functions on Riemann surfaces}'',
		\href{https://doi.org/10.1007/BFb0060090}{Springer-Verlag (1973)}, Berlin, Heidelberg .
		
		\bibitem{eichler_theory_1985}
		M.~Eichler and D.~Zagier, ``\emph{The Theory of Jacobi Forms}'', \href{https://doi.org/10.1007/978-1-4684-9162-3}{Volume 55 of Progress in Mathematics (1985)},  Birkhäuser, Boston, MA.
		
		\bibitem{Drinfeld:1985rx}
		V.~G. Drinfeld,  ``\emph{{Hopf algebras and the quantum Yang-Baxter equation}}'',
		\href{https://doi.org/10.1142/9789812798336_0013}{Sov. Math. Dokl. {\bf 32} (1985) 254--258}.
		
		\bibitem{Jimbo:1985zk}
		M.~Jimbo, ``\emph{{A q-difference analog of U(g) and the Yang-Baxter equation}}'',
		\href{https://doi.org/10.1007/BF00704588}{Lett. Math. Phys. {\bf 10} (1985) 63--69}.
		
		\bibitem{nekrassov_four_1996}
		N.~Nekrasov, ``\emph{{Four Dimensional Holomorphic Theories}}'',  PhD thesis, Princeton University (1996). \href{http://media.scgp.stonybrook.edu/papers/prdiss96.pdf}{Available online}.
		
		\bibitem{costello_supersymmetric_2013}
		K.~Costello, ``\emph{{Supersymmetric gauge theory and the Yangian}}'' (2013),   [\href{https://arxiv.org/abs/1303.2632}{{\ttfamily arXiv:1303.2632}}].
		
		\bibitem{lacroix_4-dimensional_2022}
		S.~Lacroix, 
		``\emph{4-dimensional Chern-Simons theory and integrable field theories}'',
		\href{https://doi.org/10.1088/1751-8121/ac48ed}{J. Phys. A {\bf 55} (2022) 083001} [\href{https://arxiv.org/abs/2109.14278}{{\ttfamily arXiv:2109.14278}}].
		
		\bibitem{schmidtt_symmetric_2021}
		D.~M.~Schmidtt, ``\emph{{Symmetric space $\lambda$-model exchange algebra from 4d holomorphic Chern-Simons theory}}'', \href{https://doi.org/10.1007/JHEP12(2021)004}{JHEP \textbf{21} (2020) 004} [\href{https://arxiv.org/abs/2109.05637}{{\ttfamily arXiv:2109.05637}}].
		
		\bibitem{delduc_unifying_2020}
		F.~Delduc, S.~Lacroix, M.~Magro and B.~Vicedo, ``\emph{{A unifying 2d action for  integrable $\sigma$-models from 4d Chern-Simons theory}}'', \href{https://doi.org/10.1007/s11005-020-01268-y}{Lett. Math. Phys. \textbf{110} (2020) 1645–1687}
		[\href{https://arxiv.org/abs/1909.13824}{{\ttfamily arXiv:1909.13824}}].
		
		\bibitem{benini_homotopical_2022}
		M.~Benini, A.~Schenkel and B.~Vicedo, ``\emph{{Homotopical analysis of 4d Chern-Simons theory and integrable field theories}}'',
		\href{https://doi.org/10.1007/s00220-021-04304-7}{Commun. Math. Phys. (2022) 1417--1443} [\href{https://arxiv.org/abs/2008.01829}{{\ttfamily arXiv:2008.01829}}].
		
		\bibitem{prokofev_elliptic_2023}
		V.~Prokofev and A.~Zabrodin, ``\emph{Elliptic Cauchy matrices}'' (2023), [\href{https://arxiv.org/abs/2305.02837}{{\ttfamily arXiv:2305.02837}}].
		
		\bibitem{Wess:1971yu}
		J.~Wess and B.~Zumino,  ``\emph{{Consequences of anomalous Ward identities}}'',
		\href{https://doi.org/10.1016/0370-2693(71)90582-X}{Phys. Lett. B {\bf 37} (1971) 95--97}.
		
		\bibitem{Novikov:1982ei}
		S.~P.~Novikov,  ``\emph{{The Hamiltonian formalism and a many valued analog of Morse theory}}'',
		\href{https://doi.org/10.1070/RM1982v037n05ABEH004020}{Usp. Mat. Nauk \textbf{37N5} (1982) no.5, 3-49}.
		
		\bibitem{Witten:1983ar}
		E.~Witten,
		``\emph{Nonabelian Bosonization in Two-Dimensions}'',
		\href{https://doi.org/10.1007/BF01215276}{Commun. Math. Phys. \textbf{92} (1984) 455-472}.
		
		\bibitem{Zotov:2010kb}
		A.~V. Zotov, ``\emph{{1+1 Gaudin Model}}'',
		\href{https://doi.org/10.3842/SIGMA.2011.067}{SIGMA {\bf 7} (2011) 067} [\href{https://arxiv.org/abs/1012.1072}{{\ttfamily arXiv:1012.1072}}].
		
		\bibitem{lacroix_integrable_2021}
		S.~Lacroix and B.~Vicedo, 
		``\emph{Integrable $\mathcal{E}$-Models, 4d Chern-Simons Theory and Affine Gaudin Models. I.~Lagrangian Aspects}'',
		\href{https://doi.org/10.3842/SIGMA.2021.058}{SIGMA \textbf{17} (2021) 058} [\href{https://arxiv.org/abs/2011.13809}{\ttfamily arXiv:2011.13809}].

  \bibitem{Young:2005jv}
C.~A.~S.~Young, 
		``\emph{Non-local charges, Z(m) gradings and coset space actions}'',
		\href{https://doi.org/10.1016/j.physletb.2005.10.090}{Phys. Lett. B \textbf{632} (2006), 559-565} [\href{https://arxiv.org/abs/hep-th/0503008}{\ttfamily arXiv:hep-th/0503008}].

  \bibitem{Ke:2011zzb}
S.~M.~Ke, X.~Y.~Li, C.~Wang and R.~H.~Yue, ``\emph{Classical exchange algebra of the nonlinear sigma model on a supercoset target with Z(2n) grading}'',
		\href{https://doi.org/10.1088/0256-307X/28/10/101101}{Chin. Phys. Lett. \textbf{28} (2011), 101101}.
		
		\bibitem{delduc_classical_2013}
		F.~Delduc, M.~Magro and B.~Vicedo, ``\emph{{On classical $q$-deformations of
				integrable $\sigma$-models}}'',
		\href{https://doi.org/10.1007/JHEP11(2013)192}{JHEP {\bfseries 1311}
			(2013) 192} [\href{https://arxiv.org/abs/1308.3581}{{\ttfamily arXiv:1308.3581}}].
		
		\bibitem{delduc_affine_2017}
		F.~Delduc, T.~Kameyama, M.~Magro and B.~Vicedo,  ``\emph{{Affine q-deformed symmetry and the classical Yang-Baxter $\sigma$-model}}'',
		\href{https://doi.org/10.1007/JHEP03(2017)126}{JHEP {\bf 1703} (2017) 126} [\href{https://arxiv.org/abs/2010.07879}{{\ttfamily arXiv:2010.07879}}].
		
		\bibitem{delduc_rg_2021}
		F.~Delduc, S.~Lacroix, K. Sfetsos and K. Siampos, ``\emph{RG flows of integrable sigma-models and the twist function}'',   \href{https://doi.org/10.1007/JHEP02(2021)065}{JHEP \textbf{02} (2021) 065} [\href{https://arxiv.org/abs/2010.07879}{{\ttfamily arXiv:2010.07879}}].
		
		\bibitem{hassler_rg_2021}
		F.~Hassler, ``\emph{RG flow of integrable $\mathcal{E}$-models}'',   \href{https://doi.org/10.1016/j.physletb.2021.136367}{Phys. Lett. B \textbf{818} (2021) 136367} [\href{https://arxiv.org/abs/2012.10451}{{\ttfamily arXiv:2012.10451}}].
		
		\bibitem{Hassler:2023xwn}
		F.~Hassler, S.~Lacroix, and B.~Vicedo, ``\emph{The Magic Renormalisability of
			Affine Gaudin Models}'' (2023), [\href{https://arxiv.org/abs/2310.16079}{{\ttfamily arXiv:2310.16079}}].
		
		\bibitem{levine_universal_2023}
		N.~Levine, ``\emph{Universal 1-loop divergences for integrable sigma models}'',   \href{https://doi.org/10.1007/JHEP03(2023)003}{JHEP {\bf 2023} (2023) 3} [\href{https://arxiv.org/abs/2209.05502}{{\ttfamily arXiv:2209.05502}}].
		
		\bibitem{Levine:2023wvt}
		N.~Levine, ``\emph{Equivalence of 1-loop RG flows in 4d Chern-Simons and integrable 2d sigma-models}'' (2023), [\href{https://arxiv.org/abs/2309.16753}{{\ttfamily arXiv:2309.16753}}].
		
		\bibitem{Valent:2009nv}
		G.~Valent, C.~Klimcik and R.~Squellari, ``\emph{One loop renormalizability of the Poisson-Lie sigma models}'', \href{https://doi.org/10.1016/j.physletb.2009.06.001}{Phys. Lett. B \textbf{678} (2009) 143-148}  [\href{https://arxiv.org/abs/0902.1459}{\ttfamily arXiv:0902.1459}].
		
		\bibitem{sfetsos_renormalization_2010}
		K.~Sfetsos, K.~Siampos and D.~C.~Thompson,
		``\emph{Renormalization of Lorentz non-invariant actions and manifest T-duality}'', \href{https://doi.org/10.1016/j.nuclphysb.2009.11.001}{Nucl. Phys. B \textbf{827} (2010) 545-564}  [\href{https://arxiv.org/abs/0910.1345}{\ttfamily arXiv:0910.1345}].
		
		\bibitem{Faddeev:1985qu}
		L.~D. Faddeev and N.~Y. Reshetikhin, ``\emph{Integrability of the Principal Chiral Field Model in (1+1)-dimension}'', \href{https://doi.org/10.1016/0003-4916(86)90201-0}{Annals Phys. {\bf 167} (1986) 227}.
		
		\bibitem{Appadu:2017bnv}
		C.~Appadu, T.~J. Hollowood, D.~Price, and D.~C. Thompson, ``\emph{Yang Baxter and Anisotropic Sigma and Lambda Models, Cyclic RG and Exact S-Matrices}'', \href{https://doi.org/10.1007/JHEP09(2017)035}{JHEP {\bf 09} (2017) 035}  [\href{https://arxiv.org/abs/1706.05322}{\ttfamily arXiv:1706.05322}].
		
		\bibitem{Zamolodchikov:1979ba}
		A.~B. Zamolodchikov, ``\emph{Z(4) Symetric Factorized S-matrix in two space-time dimensions}'', \href{https://doi.org/10.1007/BF01221446}{Commun. Math. Phys. {\bf 69} (1979) 165--178}.

  \bibitem{Sklyanin:1979pfu}
E.~K.~Sklyanin, L.~A.~Takhtadzhyan and L.~D.~Faddeev, ``\emph{Quantum inverse problem method. I}'', \href{https://doi.org/10.1007/BF01018718}{Theor. Math. Phys. \textbf{40} (1979) no.2, 688-706}.

\bibitem{Freidel:1991jx}
L.~Freidel and J.~M.~Maillet, ``\emph{Quadratic algebras and integrable systems}'', \href{https://doi.org/10.1016/0370-2693(91)91566-E}{Phys. Lett. B \textbf{262} (1991), 278-284}.

\bibitem{Freidel:1991jv}
L.~Freidel and J.~M.~Maillet, ``\emph{On classical and quantum integrable field theories associated to Kac-Moody current algebras}'', \href{https://doi.org/10.1016/0370-2693(91)90479-A}{Phys. Lett. B \textbf{263} (1991), 403-410}.

\bibitem{Bytsko:1994ae}
A.~G.~Bytsko, ``\emph{The Zero curvature representation for nonlinear o(3) sigma model}'', \href{https://doi.org/10.1007/BF02355322}{J. Math. Sci. \textbf{85} (1994), 1619-1628}  [\href{https://arxiv.org/abs/hep-th/9403101}{\ttfamily arXiv:hep-th/9403101}].

\bibitem{Brodbeck:1999ib}
O.~Brodbeck and M.~Zagermann, ``\emph{Dimensionally reduced gravity, Hermitian symmetric spaces and the Ashtekar variables}'', \href{https://doi.org/10.1088/0264-9381/17/14/310}{Class. Quant. Grav. \textbf{17} (2000), 2749-2764}  [\href{https://arxiv.org/abs/gr-qc/9911118}{\ttfamily arXiv:gr-qc/9911118}].

\bibitem{Bazhanov:2017nzh}
V.~V.~Bazhanov, G.~A.~Kotousov and S.~L.~Lukyanov, ``\emph{Quantum transfer-matrices for the sausage model}'', \href{https://doi.org/10.1007/JHEP01(2018)021}{JHEP \textbf{01} (2018), 021}  [\href{https://arxiv.org/abs/1706.09941}{\ttfamily arXiv:1706.09941}].

\bibitem{Delduc:2019lpe}
F.~Delduc, T.~Kameyama, S.~Lacroix, M.~Magro and B.~Vicedo, ``\emph{Ultralocal Lax connection for para-complex $\mathbb{Z}_T$-cosets}'', \href{https://doi.org/10.1016/j.nuclphysb.2019.114821}{Nucl. Phys. B \textbf{949} (2019), 114821}  [\href{https://arxiv.org/abs/1909.00742}{\ttfamily arXiv:1909.00742}].

\bibitem{Bazhanov:2018xzh}
V.~V.~Bazhanov, G.~A.~Kotousov and S.~L.~Lukyanov, ``\emph{On the Yang\textendash{}Baxter Poisson algebra in non-ultralocal integrable systems}'',   \href{https://doi.org/10.1016/j.nuclphysb.2018.07.016}{Nucl. Phys. B \textbf{934} (2018), 529-556} [\href{https://arxiv.org/abs/1805.07417}{{\ttfamily arXiv:1805.07417}}].

\bibitem{Kotousov:2022azm}
G.~A.~Kotousov, S.~Lacroix and J.~Teschner, ``\emph{Integrable sigma models at RG fixed points: quantisation as affine Gaudin models}'',   \href{https://doi.org/10.1007/s00023-022-01243-4}{Annales Henri Poincare \textbf{25} (2024) no.1, 843-1006} [\href{https://arxiv.org/abs/2204.06554}{{\ttfamily arXiv:2204.06554}}].

\bibitem{Evans:1999mj}
J.~Evans, M.~Hassan, N.~MacKay et A.~Mountain, ``\emph{{Local conserved charges
  in principal chiral models}}'',
  \href{http://dx.doi.org/10.1016/S0550-3213(99)00489-7}{Nucl. Phys.
  {\bf B561} (1999) 385--412},
  [\href{https://arxiv.org/abs/hep-th/9902008}{{\tt hep-th/9902008}}].

\bibitem{Lacroix:2017isl}
S.~Lacroix, M.~Magro et B.~Vicedo, ``\emph{{Local charges in involution and
  hierarchies in integrable sigma-models}}'',
  \href{http://dx.doi.org/10.1007/JHEP09(2017)117}{JHEP {\bf 09} (2017)
  117}, [\href{https://arxiv.org/abs/1703.01951}{{\tt 1703.01951}}].
  
\bibitem{Frenkel:2016gxg}
E.~Frenkel et D.~Hernandez,  ``\emph{{Spectra of quantum KdV Hamiltonians, Langlands duality, and affine opers}}'', \href{https://doi.org/10.1007/s00220-018-3194-9}{Commun. Math. Phys. \textbf{362} (2018) no.2, 361-414},
  [\href{https://arxiv.org/abs/1606.05301}{{\tt 1606.05301}}].

\bibitem{Lacroix:2018fhf}
S.~Lacroix, B.~Vicedo et C.~Young, ``\emph{{Affine Gaudin models and
  hypergeometric functions on affine opers}}'',
  \href{http://dx.doi.org/10.1016/j.aim.2019.04.032}{Adv. Math. {\bf 350} (2019) 486-546},
  [\href{https://arxiv.org/abs/1804.01480}{{\tt 1804.01480}}]

\bibitem{Lacroix:2018itd}
S.~Lacroix, B.~Vicedo et C.~A.~S. Young, ``\emph{{Cubic hypergeometric integrals
  of motion in affine Gaudin models}}'', \href{http://dx.doi.org/10.4310/ATMP.2020.v24.n1.a5}{Adv. Theor. Math. Phys. {\bf 24} (2020) 1, 155-187},
  [\href{https://arxiv.org/abs/1804.06751}{{\tt 1804.06751}}].
  
\bibitem{Gaiotto:2020dhf}
D.~Gaiotto, J.~H.~Lee, B.~Vicedo and J.~Wu, ``\emph{{Kondo line defects and affine Gaudin models}}'', \href{https://doi.org/10.1007/JHEP01(2022)175}{JHEP \textbf{01} (2022), 175},
  [\href{https://arxiv.org/abs/2010.07325}{{\tt 2010.07325}}].
  
\bibitem{Kotousov:2021vih}  
G.~A.~Kotousov and S.~L.~Lukyanov, ``\emph{{ODE/IQFT correspondence for the generalized affine $ \mathfrak{sl} $(2) Gaudin model}}'', \href{https://doi.org/10.1007/JHEP09(2021)201}{JHEP \textbf{09} (2021), 201},
  [\href{https://arxiv.org/abs/2106.01238}{{\tt 2106.01238}}].
    
\bibitem{Franzini:2022duf}
T.~Franzini and C.~A.~S.~Young, ``\emph{{Quartic Hamiltonians, and higher Hamiltonians at next-to-leading order, for the affine $\mathfrak{sl}_2$ Gaudin model}}'',  [\href{https://arxiv.org/abs/2205.15815}{{\tt 2205.15815}}].



		\bibitem{poor_fays_1992}
		C.~Poor, ``\emph{Fay's Trisecant Formula and Cross-Ratios}'', \href{https://doi.org/10.2307/2159386}{Proceedings of the American Mathematical Society {\bf 114} (1992) no.~3, 667--671}.
		
		\bibitem{10.5555/1098650}
		M.~Abramowitz and I.~Stegun, ``\emph{Handbook of Mathematical Functions, With Formulas, Graphs and Mathematical Tables}'', Dover Publications (1974), Inc., USA.
		
		\bibitem{Reshetikhin:1990ep}
		N.~Reshetikhin, ``\emph{Multiparameter quantum groups and twisted quasitriangular  Hopf algebras}'', \href{https://doi.org/10.1007/BF00626530}{Lett. Math. Phys. {\bf 20} (1990) 331--335}.
		
		\bibitem{kostant_principal_1959}
		B.~Kostant, ``\emph{The Principal Three-Dimensional Subgroup and the Betti Numbers of a Complex Simple Lie Group}'', \href{https://doi.org/10.2307/2372999}{American Journal of Mathematics {\bf 81} (1959) no.~4, 973--1032}.
		
	\end{thebibliography}
	
	\begingroup\raggedright\endgroup
	
\end{document}